# Contents

# Part 4: Gravitational Microlensing

Joachim Wambsganss

Zentrum für Astronomie der Universität Heidelberg (ZAH), D-69120 Heidelberg, Germany
email: jkw@ari.uni-heidelberg.de

Gravitational Microlensing can be thought of as a version of strong gravitational lensing in which the image separation is too small to be resolved. Multiple images are formed, but their typical separation – $\Delta\theta \approx 2\,\theta_{\rm E}$ – is far below the limiting resolution determined by observational constraints. Given the dependence of the Einstein radius on lens mass and geometry, it is clear that microlensing will occur for sufficiently small masses and sufficiently distant lenses and sources. In very general terms, microlensing deals with the lensing effects of compact objects in the mass range $10^{-6} \leq m/M_\odot \leq 10^6$. This translates into Einstein radii/angular separations of a milli-arcsecond or smaller for the two main distance regimes: "galactic" – lens/source distances of order 10 kpc, and "extragalactic/cosmological" – lens/source distances of order Gpc. Both regimes will be discussed at length in the subsequent sections.

The mathematical possibility of microlensing – lensing effects of stellar mass objects on background stars – was already discussed many decades ago (Chwolson 1924, Einstein 1936). Right after the discovery of the first gravitational lens system in the late 70s (Walsh et al. 1979), interest in lensing by stellar mass objects was revived: Chang & Refsdal (1979), Gott (1981), Paczyński (1986a,b), Kayser et al. (1986). Bohdan Paczyński was the first to use the term "microlensing" for light deflection by stellar masses. The first observational detection of the microlensing effect came in 1989 (Vanderriest et al. 1989; Irwin et al. 1989) when individual stars in the lensing galaxies of gravitationally lensed quasars QSO 0957+561 and QSO 2237+0305 altered the magnification of one of the quasar images relative to other(s). The first Galactic microlensing events were reported in 1993: Alcock et al. (1993), Aubourg et al. (1993), and Udalski et al. (1993).

Most of the sections in this part of the proceedings deal with stellar or Galactic microlensing, the last one treats quasar or cosmological microlensing. There exist a number of good and more detailed reviews on microlensing, e.g. Paczyński (1996), Mao (2001), Mollerach & Roulet (2002), Courbin, Saha & Schechter (2002).

# 1 Lensing of Single Stars by Single Stars

## 1.1 Brief History

Commonly it is assumed that light deflection is a modern phenomenon. However, more than 200 years ago scientists started to think about it. In the beginning of the 19th century, Johann Georg Soldner wrote an article entitled "Über den Einfluß der Schwerkraft auf die Ausbreitung des Lichtes"[1], in which he investigated the possibility that a light ray be attracted by the gravitation of a "heavenly body" (Soldner 1801). He even derived the deflection angle for a light ray passing close to the solar limb, arriving at half the correct value. In 1911, Einstein had thought about light deflection as well and published the same value (Einstein 1911). Only with the completion of the General Theory of Relativity, Einstein found the value that was later confirmed by the famous solar eclipse expedition. Chwolson (1924) mentioned a "fictitious double" star, an apparent illusion due to the light deflection of a foreground star by a background star, even considering a ring-like image for perfect alignment between lensing and lensed star. He was uncertain whether this might ever be observable. Years later, Einstein published again a letter on star-star lensing, initiated by a visit of the Czech engineer Mandl (Einstein 1936). He mentions the appearance of a luminous circle for perfect alignment between source and lens and derives the magnification properties. But he was skeptical regarding the observability: "of course, there is not much hope of observing this phenomenon directly". Renn et al. (1997) report that Einstein had dealt with the same question already as early as 1911/1912: in his notebooks he had derived the relations regarding double images, magnification, separation of images etc., but apparently had never bothered to publish it. Link (1937, 1967) had treated lensing of stars by stars as well, and produced tables for the magnification of finite sized background stars. With the seminal papers in the 1960s (Klimov 1963, Liebes 1964, Refsdal 1964a,b) lensing was put on firm theoretical footing and applicable to interesting astrophysical goals. Chang & Refsdal (1979, 1984) suggested that the lensing action of individual stars affect the apparent brightness of multiply imaged quasars. Gott (1981) suggested that lensing by stellar-mass objects in halos of lensing galaxies can be used to detect compact dark matter, and finally Paczyński (1986a,b) introduced the term "microlensing", both for the action of stars in distant lensing galaxies on quasar images and for stars or dark matter objects in the Milky Way acting on background stars in the bulge or in the Large/Small Magellanic Clouds.

## 1.2 Theoretical Background

This section considers the mathematically simplest case: the lensing effect of a single foreground star on a single background star in the Milky Way or Local

---

[1] "On the influence of gravity on the propagation of light"

Group. It is typical and representative for most Galactic lensing systems. In this regime, lens and source distances are of the order kpc, the lenses have roughly stellar masses. This results in angular Einstein ring radii which are on the order of $10^{-3}$ arcsec, and $D_s$ is small enough that we may reasonably set $D_{ds} = D_s - D_d$.

**Point Source – Point Lens**

Over a large range of current astrophysical interest, the stellar source subtends a considerably smaller angle than $\theta_E$, so that it may be approximated as a point. Symmetry allows us the freedom to choose the origin as the position of the lens and a position along the positive $\theta_1$ axis for the point source. The characteristic length scale is $\theta_E$, by which we scale the one-dimensional lens equation

$$y = x - \frac{1}{x} \quad , \tag{1}$$

where $x \equiv \theta/\theta_E$ is a (normalized) image position corresponding to the (normalized) source position $y \equiv \beta/\theta_E$. The two solutions

$$x_\pm = \frac{1}{2}\left(y \pm \sqrt{y^2 + 4}\right) \tag{2}$$

to the quadratic lens equation correspond to positions which straddle the lens on the sky, with the positive parity image (+) on the source side of the lens, always magnified and further away from the lens than the negative parity image (−), which is the less magnified of the two.

Indeed, we can formulate the magnification of the two images as (cf. Introduction):

$$\mu_\pm = \frac{1}{\det \mathcal{A}_\pm} = \left(1 - \frac{1}{x_\pm^4}\right)^{-1} = \pm\frac{1}{4}\left[\frac{y}{\sqrt{y^2+4}} + \frac{\sqrt{y^2+4}}{y} \pm 2\right] . \tag{3}$$

Note that the image separation is $\Delta x \equiv |x_+ - x_-| = \sqrt{y^2 + 4}$. Relations for the total magnification $\mu$, and the sum and ratio of the individual image magnifications can then be derived (however, only the total magnification is observable in (photometric) microlensing):

$$\mu \equiv \mu_+ + |\mu_-| = \mu_+ - \mu_- = \frac{1}{2}\left[\frac{y}{\sqrt{y^2+4}} + \frac{\sqrt{y^2+4}}{y}\right] = \frac{y^2 + 2}{y\sqrt{y^2+4}} \tag{4}$$

$$\mu_+ + \mu_- = 1 \tag{5}$$

$$\left|\frac{\mu_-}{\mu_+}\right| = \left(\frac{y - \sqrt{y^2+4}}{y + \sqrt{y^2+4}}\right)^2 = \left(\frac{x_-}{x_+}\right)^2 . \tag{6}$$

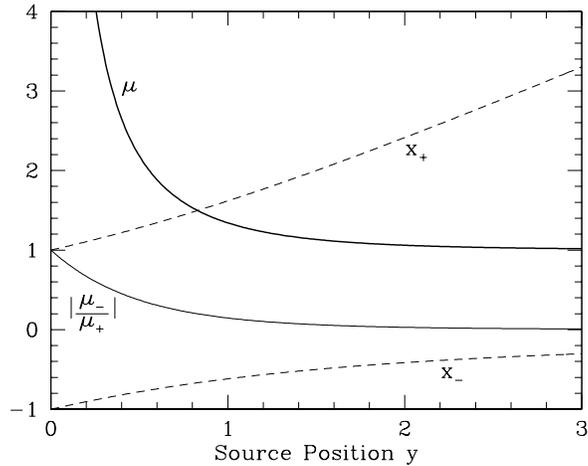

**Fig. 1.** Total magnification $\mu$, normalized image positions $x_\pm$, and ratio of image magnifications for a point lens as a function of the normalized source position $y$ (Figure courtesy Penny Sackett)

For a point lens, the two-dimensional magnification distribution in the source plane – the magnification pattern – consists of circular contours of constant total magnification: the circles are centered on the lens position with a magnification value $\mu \to 1/y$ for small impact parameters $y$, and $\mu \to 1$ for large $y$. For source positions $y > 0$, the $x_+$ image is responsible for the lion's share of the total magnification $\mu$ for all source positions $y$ (Fig. 1). At the fiducial source position $y = 1$ where $\mu = 1.34$, e.g., the positive parity image contributes 87% of the total magnification.

### Microlensing "Events"

If the instantaneous magnification of a microlensing event were the only measurable quantity in a static stellar microlensing scenario, then the science of microlensing would be much less rich and its literature much less voluminous[2]. However, stars move around the Galactic center (and have an additional random velocity component with respect to one another). The relative velocities are such, that the time scale of the relative change of lens and source positions is of order of weeks or shorter. Hence this motion introduces a temporal component into the lensing geometry, causing the impact parameter and thus the normalized bending angle and the magnification to vary measurably as a function of time.

---

[2] In fact, without independent knowledge of the unlensed flux emanating from a source, it would not be possible at all to ascertain whether, or by how much, a background star is lensed.

In general, the observer, the lens, and the source are all in motion, all with a certain three-dimensional velocity vector. The temporal behavior of the magnification — the "light curve" of the source — is dictated by the *relative* motion of the lens across the observer-source line-of-sight. The distances $D_\mathrm{d}$, $D_\mathrm{s}$, and $D_\mathrm{ds}$, and thus the relative scalings of $\beta$, $\theta$, and $\alpha$, may also change with time, but in astrophysical situations these changes have a negligible effect on the lensing equation compared to changes in the impact parameter due to the projected relative motion.

The characteristic time scale for these changes is given by the Einstein time

$$t_\mathrm{E} \equiv D_\mathrm{d}\, \theta_\mathrm{E}/v_\perp, \tag{7}$$

where $v_\perp$ is the transverse speed of the lens relative to the source-observer line-of-sight. For $y < 1$, the total magnification of a point source can be expected to change appreciably over a time $t_\mathrm{E}$. Stellar lenses in our own Galaxy are associated with typical $t_\mathrm{E}$ on the order of a month, and thus the change in the observed brightness of the source they induce are referred to as microlensing "events".

Since the magnification depends only on the source position in units of $\theta_\mathrm{E}$ (cf. eq. 4), we need only describe how the source moves relative to the lens as a function of time to obtain a description of the light curve of the microlensed source. Assuming that for the duration of the observable event (say, several $t_\mathrm{E}$) this motion is rectilinear, the relative motion of the source on the sky differs from that of the lens only by the sign. Taking the time $t_0$ to be that at which the source-lens separation $y$ takes on its smallest value $y_0$, the trajectory of the source can be represented by

$$y(t) = \sqrt{y_0^2 + \left(\frac{t-t_0}{t_\mathrm{E}}\right)^2}\;, \tag{8}$$

since the line joining the lens to the source at time $t_0$ is perpendicular to the lens-source relative motion[3]. The corresponding light curve $F(t) = \mu(t)F_s$ of the source, examples of which are shown in Fig. 2, is then obtained by substituting $y(t)$ into eq. (4) and multiplying by the unlensed source flux $F_\mathrm{s}$.

---

[3] It may be worth noting that in the microlensing literature, the normalized source-lens separation on the sky, **y**, is often alternatively denoted as **u**. This convention is used below as well.

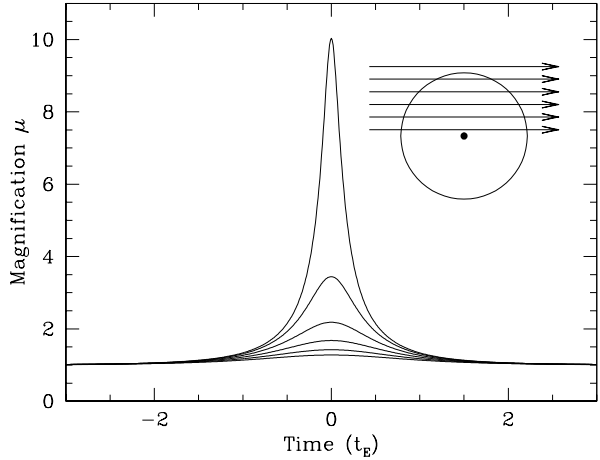

**Fig. 2.** Point-lens, point-source light curves for minimum impact parameters $y_0 = 0.1$ (top), 0.3,...,1.1 (bottom) and the corresponding trajectories across the Einstein ring (Figure courtesy Penny Sackett)

## Observables

A simple, point-source, point-lens microlensing light curve is thus described by four parameters: unlensed flux $F_0$, $t_0$, $y_0$ and $t_E$. Of these, $F_0$ can be measured in the absence of microlensing, $t_0$ sets an arbitrary time scale, and $y_0$ depends on the random placement of lens and source on the sky. Only $t_E = D_d \theta_E / v_\perp$ contains physical information about the lensing system, albeit in a degenerate combination. Assuming that the source distance $D_s$ can be determined from its properties (membership in a stellar system, spectral type and apparent magnitude), we are left with three physical parameters, lens mass $M$, lens distance $D_d$, and lens relative transverse velocity $v_\perp$, to determine from one observable. Other additional information is needed, if we want to learn more about the lens system. Unlike other forms of gravitational lensing, a microlens is not observed directly in general (however, cf. Subsection 'Direct Lens Detection' in Section 5).

As Fig. 1 illustrates, significant magnification occurs when the source lies within one angular Einstein radius of the lens. The microlensing event itself gives us very little possibility to measure or estimate the lens mass $M$, the lens distance $D_d$ or the transverse velocity $v_\perp$ independently. To understand the severity of this degeneracy, consider a Galactic microlensing system in which the source is known to lie in the Galactic Bulge at, say, precisely 8 kpc and
the Einstein time $t_{\rm E}$ has been precisely determined to be (a rather typical)
40 days. Assuming the lens to be bound to the Galaxy, it is likely that $0 <
v_\perp < 600$ km$^{-1}$ with values nearer to the middle of the range statistically
favoured. Fig. 3 shows the resulting degeneracy in the mass and distance of
the lens. The distribution of lens masses ranges from those massive brown
dwarfs ($\lesssim 0.1 {\rm M}_\odot$) to that of a heavy stellar black hole ($\sim 10 {\rm M}_\odot$).

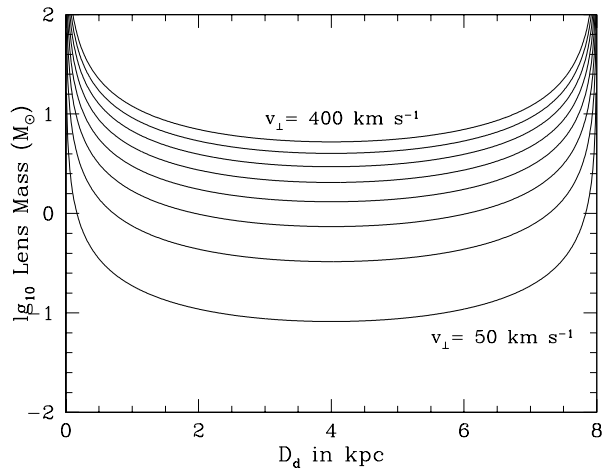

**Fig. 3.** Degeneracy between mass $M$ and distance $D_{\rm d}$ for a Galactic microlens for
different assumptions for its transverse speed ($50 \leq v_\perp \leq 400$ km s$^{-1}$, in steps of
$50$ km s$^{-1}$) across the line-of-sight to the source (Figure courtesy Penny Sackett)

### 1.3 How good is the point lens - point source approximation ?

Information about individual image positions, magnifications and shape is
entirely lost in a standard microlensing situation. This results in degeneracies
in the lens-source combinations that can lead to the same observables. In order
to understand what information remains, and how it is related to the physical
parameters of the system, we begin with a simple model and gradually add
complexity.

To a very good approximation, we can assume that the mass distribution
of a star is spherically symmetric, hence the projected mass distribution is axisymmetric, independent of the direction. Since the mean free path of photons is quite short in all but the extreme outer layers of stellar atmospheres, we also assume that stellar lenses are not transparent. Taken together, this implies that we can model a single stellar lens as a point mass $M$ as long as we consider only impact parameters larger than the stellar radius of the lens. The ratio of the angular size of the source compared to

$$\theta_\mathrm{E} = \sqrt{(4\,G\,M D_\mathrm{ds})/(c^2\,D_\mathrm{d}\,D_\mathrm{s})} \qquad (9)$$

will determine whether it is appropriate to use the point source approximation.

Although a normal star acting as a lens emits light that could yield information about lens properties, this light is mixed with that of the source and any other luminous object lying in the same resolution element. However, most sources are bright giants, and most lenses are presumably faint (M-)dwarfs: Since bright stars are easier to detect at great distances, giants are much more likely sources. And since low mass stars are much more abundant than massive ones in the Milky Way disk (consequence of the the Initial Mass Function of stars), they are much more likely lenses. Combined with the luminosity-mass relation this means: in a microlensing event, the source light dominates lens light dramatically. Even if the lens contributes substantially to the total flux received during a microlensing event, the lens mass-distance degeneracy may not be reduced. The microlensing light curve is now a function of five parameters: $F_\mathrm{s}$ (actual source flux at baseline), $t_0$, $y_0$, $t_\mathrm{E}$, and the blending flux $F_\mathrm{b}$ (constant flux contribution of some unrelated star) with the result that $t_E$ is more difficult to determine.

The symmetry of the point-source, point-lens light curve aids in determining $t_0$ and the total baseline flux $F_0 = F_\mathrm{s} + F_\mathrm{b}$ can be measured well after the lensing event is over. The fraction of $F_0$ that is contributed by the source, $f_\mathrm{s} \equiv F_\mathrm{s}/F_0$, must be determined from subtleties in the light curve itself (e.g., color changes, or astrometric information), and thus can be strongly degenerate with $y_0$ and $t_\mathrm{E}$.

Most of the measured microlensing events are indeed well fitted and described by the simple point lens - point source approximation and linear relative motion. Occasionally, some events were very well covered with hundreds of data points with small error bars. For some of them, the point lens - point source approximation did not produce satisfactory fits. In binary microlensing events with caustic crossings, the point source approximation breaks down. This can be used to determine source size and even source profile information. These cases and effects will be discussed in Section 5.

### 1.4 Statistical Ensembles

Gravitational microlensing offers the opportunity to measure the density and total mass of a population of objects - bright or dark - between a background population of sources and the observer on Earth. Paczyński (1986b) worked out this idea quantitatively, and applied it to objects potentially making up the dark matter halo of the Milky Way. If such objects had masses in the stellar range (very roughly from $10^{-6} \leq M/M_\odot \leq 10^2$), they would produce time-variable magnification of background stars in the Large or Small Magellanic Clouds. Quantitatively important in such a situation are: probability and duration of such events.

The optical depth to gravitational microlensing is equal to the ratio of surface mass density of microlensing objects to the critical mass density (cf. Introduction). For a variable mass density, the optical depth is an integral expression along the line-of-sight (Paczyński 1986b):

$$\tau = \int_0^{D_s} \frac{4\pi GD}{c^2} \rho(D_d) dD_d, \tag{10}$$

where $\rho(D_d)$ is the average microlensing matter density at distance $D_d$ from the observer, and $D = (D_d D_{ds}/D_S)$. The resulting optical depth depends a bit on the exact direction and parameters of the isothermal halo. Assuming a simple isothermal sphere model ($M(R) = V_{\text{rot}}^2 R/G$, $\rho(R) = V_{\text{rot}}^2/4\pi G R_{\text{GC}}^2$) for the dark halo of the Milky Way ($R_{\text{GC}}$ is the distance to the Galactic Center), Paczyński derived the numerical value of the optical depth turns out to be of order

$$\tau_0 = 5 \times 10^{-7}. \tag{11}$$

This means that roughly one out of a million stars in the nearby galaxies will be strongly lensed, i.e. the source is located within the Einstein radius of the lens and hence magnified by at least $\mu \geq 3/5^{0.5} \approx 1.34$. The concept of optical depth can easily be visualized in the following way: if all the lenses would be represented as dark disks with their respective Einstein radii, then the sum of the areas of all these disks would cover exactly the fraction $\tau_0$ of the sky.

The event duration (defined as the time it takes to cross the Einstein radius) depends on the transverse velocity of the object and its mass (cf. equation 7). A typical value for an object at a distance of $D_d = 10$ kpc and a tangential velocity of $v_\perp = 200$ km/sec is

$$t_0 \approx 6 \times 10^6 \text{ sec } \left(\frac{M}{M_\odot}\right)^{0.5} \approx 0.2 \text{ yr } \left(\frac{M}{M_\odot}\right)^{0.5} \tag{12}$$

However, even assuming that all the lenses had the same mass and the same (randomly oriented) three-dimensional velocity, the event durations would cover a wide range. In particular, there should be a tail of relatively long events, because a fraction of lenses my have a three dimensional velocity vector close to radial. Also, since the actual lens population most likely consists of a range of masses and velocities and distances, this would broaden the duration distribution even more.

If all events had the same time scale $t_0$, then the event rate $N$ would be given as (cf. Paczyński 1996):

$$N = \frac{2}{\pi} \, n \, \tau \, \frac{\Delta t}{t_0}, \tag{13}$$

where $n$ is the total number of sources monitored, $\tau$ is the optical depth, and $\Delta t$ is the time interval of the monitoring campaign. In his review article, Paczyński (1996) derives the probability distribution of event durations; a more detailed analysis can be found in Mao & Paczyński (1996).

## 2 Binary Lenses

After treating the case of a single lens, the logical next step is the binary lens scenario. In the lens equation, the only change is that the deflection angle now consists of the sum of two point lenses:

$$\boldsymbol{\alpha}(\boldsymbol{x}) = \frac{4G}{c^2} \left( \frac{M_A(\boldsymbol{x} - \boldsymbol{x}_A)}{(\boldsymbol{x} - \boldsymbol{x}_A)^2} + \frac{M_B(\boldsymbol{x} - \boldsymbol{x}_B)}{(\boldsymbol{x} - \boldsymbol{x}_B)^2} \right), \tag{14}$$

where $M_A$, $M_B$ are the masses of the two lenses and $\boldsymbol{x}_A$, $\boldsymbol{x}_B$ are their positions. Due to the non-linearity of the lens equation, the effect of replacing a single lens by two separate lenses does not at all have the effect of a simple sum or superposition of two single lens cases: the caustics and the two-dimensional magnification distributions in the source plane look *very* different compared to the single-lens case, and so do the lightcurves.

The major new phenomenon of a binary (or any asymmetric) lens, compared to the isolated point lens is the occurence of extended *caustics* in the source plane (cf. Introduction), due to the astigmatism of the lens. Caustics separate regions of different image multiplicities: when a source crosses a caustic, a new image pair is created or destroyed. Due to the very small image separation for stellar mass lenses (compared to the resolution of the telescope), these new images cannot be observed directly. However, since these new images are very highly magnified, the combined magnification of all images (which is an observable) is dominated by these bright new images: the lightcurve of a source undergoing a caustic crossing exhibits high peaks. Formally, a point source would even be infinitely magnified. Due to the realistic finite source size, the actual magnification remains finite but can get very high (events with magnifications of more than five magnitudes have been detected).

In a binary lens scenario, the second lens introduces three new parameters:

- the mass ratio $q = m_1/m_2$,
- the binary separation $d$ (in units of the Einstein radius for the total mass $m = m_1 + m_2$),
- the angle $\phi$ between the source trajectory and the line connecting the two lenses.

This allows for a *very* large variety of binary lens lightcurves: "A double lens is vastly more complicated than a single one" (Paczyński 1996).

## 2.1 Theory and Basics of Binary Lensing

The properties of a system consisting of two point lenses have been explored in great detail in a seminal paper by Schneider & Weiss (1986). They derive analytically the critical curves and caustics for the binary lens with equal masses: $m_1 = m_2$, i.e. $q = 1$. They found three regimes for binary lensing: when the two lenses are widely separated, they act like two single lenses which are slightly perturbed: the degenerate point caustic of an isolated lens is slightly deformed into a small asymmetric asterisk with four cusps, and the circular critical line, the Einstein ring, is slightly deformed into an oval (see top left panel of Fig. 4 for a separation of $d = 1.2$). Once the separation of the two lenses approaches one Einstein radius, the two critical lines merge, forming the "infinity" sign, and the two separate caustics merge accordingly (cf. top right panel of Fig. 4). For further decreasing separation between the binary components, there is now one closed critical line and one closed six-cusp-caustic (middle panel of Fig. 4). When the separation reaches $d = 8^{-0.5} \approx 0.35355$, another change of topology occurs: two regions inside the main critical line detach, and the caustic divides up into three parts as well: two triangular shaped caustics and one four-cusp asterisk (bottom left panels of Fig. 4). When the two lenses approach each other even further, the two triangular caustics move away from the main caustic very rapidly (bottom right panel of Fig. 4).

Fig. 4 (a reproduction of Fig. 2 from Schneider & Weiss (1986)) also indicates the parity of the images in the lens plane with a plus or minus sign (the explanation of the additional labels can be found in the original paper). Accordingly, the image configuration for a binary lens can be very diverse, even for given mass ratio and separation. This was illustrated by Schneider & Weiss (1986) as well for an extended source and is reproduced here in Fig. 5 for equal masses ($q = 1$) and a separation of $d = 0.5$: A source inside the caustic (inset) has five images (top left panel in Fig. 5). When the source touches a caustic (other three panels), two or three images merge, respectively. In the panels, the size/area of an image is proportional to its magnification.

Schneider & Weiss (1986) studied also the effect of the source size on the magnification during a caustic crossing. In Fig. 6 (reproduced from their Fig. 9) the "lightcurve" of a variety of circular sources with different radii and constant surface brightness is shown: the maximum magnification as well as the exact shape of the lightcurve depend strongly on the source radius, here shown for a range from $\theta/\theta_E = 0.05, 0.03, 0.01, 0.005$ and for point-like source.

The study of Schneider & Weiss (1986) was generic in the sense that it was applicable to close pairs of galaxies as well as double stars. Mao & Paczyński (1991) concentrated entirely on microlensing of binary stars. Based on the observational fact that more than 50% of all stars are members of binaries or multiple star systems, they predicted that it is unavoidable that microlensing lightcurves of binary stars and star-plus-planet systems will be observed. However, due to the large range of separations, most of the physical binary stars will either act as two individual lenses (for large separation) or

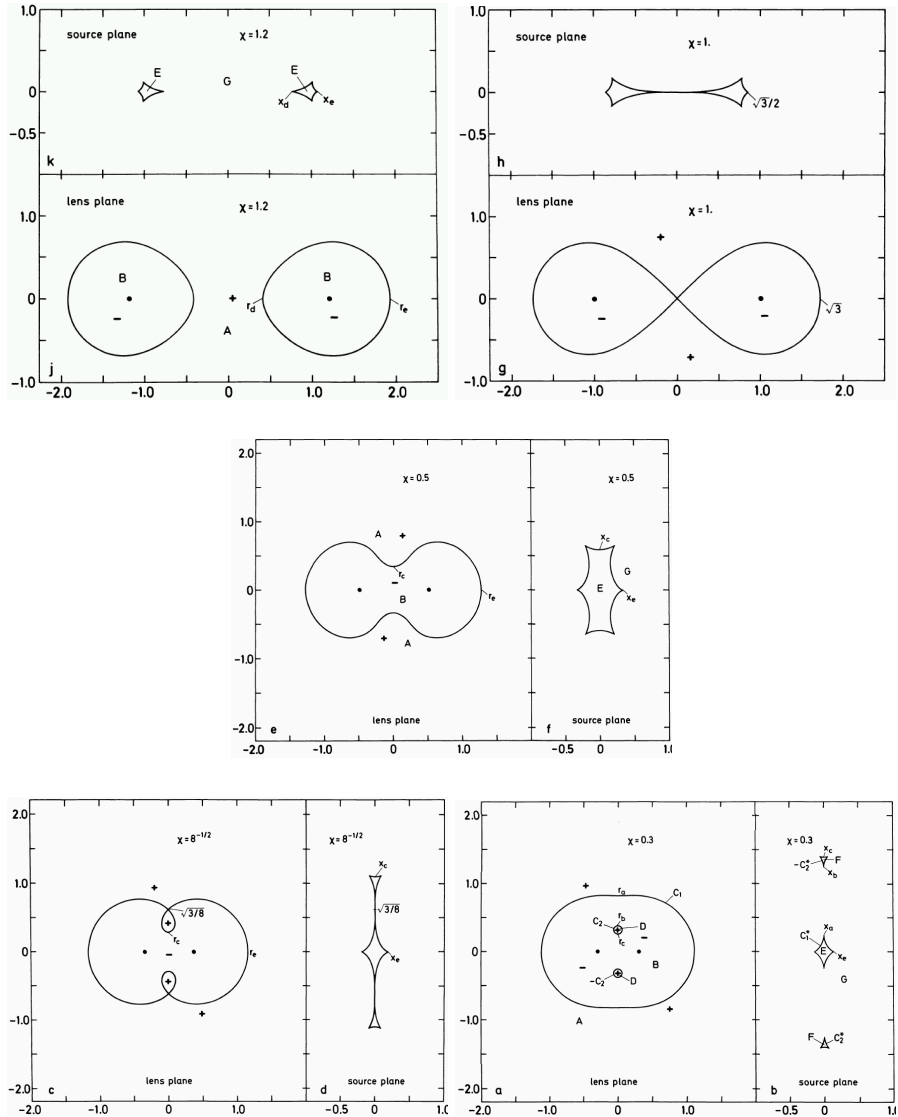

**Fig. 4.** The critical curves in the lens plane (lower part of each panel) and the caustics in the source plane (upper part of each panel) are depicted for a binary lens situation with equal masses ($M_A = M_B$) and decreasing separation: from 1.2 $R_E$ (top left) to 0.3 $R_E$ (bottom right). In particular the "transition cases" (separation 1.0 $R_E$ and $8^{-1/2} R_E$) are of interest (from Schneider & Weiss (1986), Fig. 2)

468

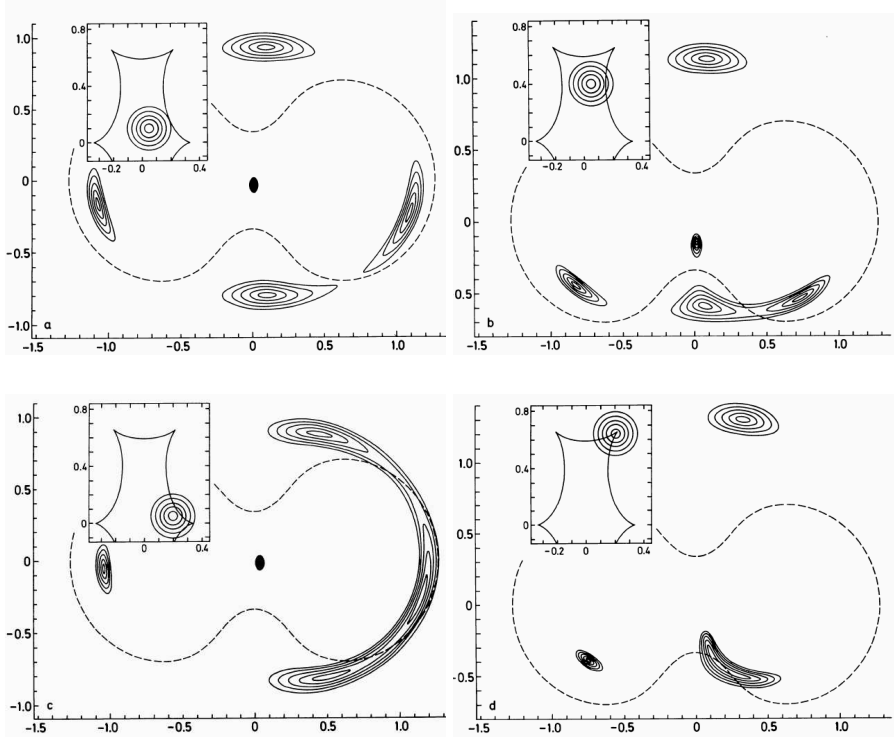

**Fig. 5.** Image configurations for an extended source and a binary lens with equal masses ($M_A = M_B$) and separation 0.5 $R_E$. The critical curves are shown as dashed lines, the solid contours indicate the image shapes. The inset at the top left corners of the four panels indicate the respective source position with respect to the binary lens caustic (from Schneider & Weiss (1986), Fig. 6)

as a single lens (for very small separation). If the (projected) separation is of order the Einstein radius of the combined mass, it gets "interesting" for lensing (cf. also Figs. 4 and 5), i.e. deviations from single-lens lightcurves are to be expected. They concluded that about 10% of all observed stellar microlensing events should show signatures of the binarity of the lens.

Witt & Mao (1995) explored the binary lens further and found that the minimum total magnification for a source inside the caustic is three. They suggested that for an observed lightcurve in which this is apparently not the case, there are two possibilities: either there is light from another component ("blending"), which could be the lens itself or an unrelated background star, or the lens system consists of more than two stars (triple lens).

An illustration of the variety of lightcurves for a binary lens with separation $d = 0.5$ (and mass ratio $q = 1$) is shown in Fig. 7 for 5 parallel tracks and a finite source size. At http://www.sns.ias.edu/∼gaudi/movies.html, Scott Gaudi's website, he provides a full suite of animations of binary microlensing scenarios with variable mass ratio and separation, indicating the critical lines, the caustics, the individual micro-images and the lightcurve for relative motion. A static example is shown in Fig. 8.

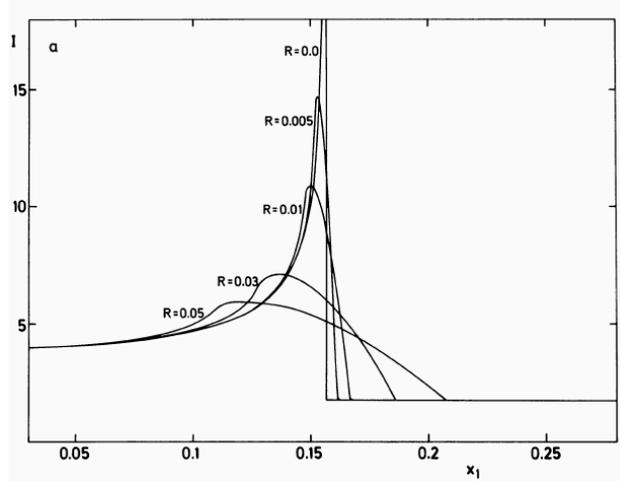

**Fig. 6.** Lightcurves of a caustic crossing (fold) for a point source (R = 0.0) and extended sources with radii varying from $R = 0.005 R_E$ to $R = 0.05 R_E$ (from Schneider & Weiss (1986), Fig. 9a)

### 2.2 First Microlensing Lightcurve of a Binary Lens: OGLE-7

The first microlensing lightcurve of a binary lens to be detected was OGLE-7 (more on the OGLE-team in Section 3.3), for which two peaks were measured in the course of the 1993 season (see Fig. 9, top panel). This lightcurve was originally classified as "unusual", because the star had brightened by more than 2 magnitudes but deviated from the expected single-lens-single-source lightcurve: The flux as a function of time displayed a double-peak structure, following a completely flat and constant lightcurve at a low level in the previous season. This relatively bright phase lasted for about 60 days. The OGLE team (Udalski et al. 1994) found a simple binary lens solution (Fig. 9, bottom panel) with the following parameters: mass ratio $q = 1.02$, projected separation $a = 1.14 R_E$, impact parameter $b = 0.050 R_E$, angle $\theta = 48.3$ degrees, time scale $t_E = 80$ days, baseline magnitude $I_0 = 18.1$, and fraction of "blended" light $f = 56\%$.

The last value indicates that the measured apparent brightness must consist of two contributions, and so Udalski et al. (1994) concluded that the lensed star should have a composite spectrum: the lens model required light from an additional (unresolved) star.

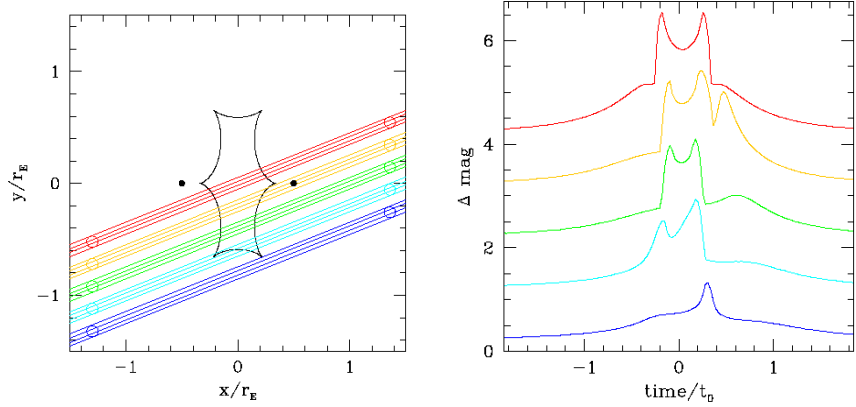

**Fig. 7.** The left hand panel indicates a binary lens caustic, the positions of the two (equal mass) lenses with separation $1.0 R_E$, and the parallel tracks of an extended source (width of track corresponds to two source radii: $R = 0.05 R_E$). On the right hand panel, the five corresponding lightcurves are displayed, offset by one magnitude for easier visibility (from Sackett [private communication], after Paczyński 1996)

As mentioned above, Witt & Mao (1995) showed a few months later that the minimum magnification between the peaks of a double caustic crossing is three (for OGLE-7 it was factors of 2.2 and 2.4, respectively, in R and I filters). So in hindsight this fact showed unambigously that either there is blended light contributing to the lightcurvce, or the event OGLE-7 is caused by lensing of a system with more than two components.

### 2.3 Binary Lens MACHO 1998-SMC-1

In the following years, the data reduction systems of the microlensing teams were dramatically improved in order to allow detection of microlensing events while they were still ongoing, with the goal of real-time detection. MACHO-98-SMC-1 was the first caustic crossing binary event towards the Magellanic Clouds which was 'caught in action' in this way (Alcock et al. 1999). This allowed very good coverage of the lightcurve: Once an anomaly is recognized "on the run", the observing strategy can be changed immediately with a much more frequent monitoring of the active event. As a consequence, a prediction for the time of the second caustic crossing became possible.

The event was originaly recognized on May 25.9, 1998 (UT), when it had brightened by 0.9 mag. At this time the first "alert" was activated (see Fig. 10,

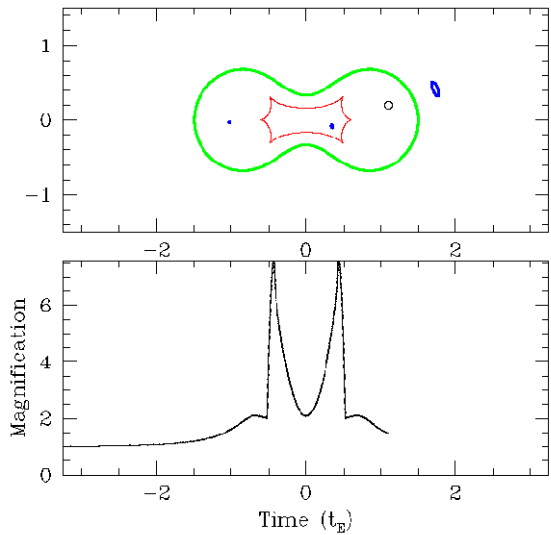

**Fig. 8.** Snapshot of a binary microlensing animation made available on the web by Scott Gaudi, under http://www.sns.ias.edu/∼gaudi/movies.html. This particular situation illustrates the case of a binary lens with equal masses: in the top panel the critical line (thick solid line), the caustics (thin line with 6 cusps), the source size and position (little circle) and the three images (thick circles/ellipses) are shown for one particular instant of time; the source moves along a horizontal line from left to right. The bottom panel indicates the lightcurve (sum of all the micro-images) as it develops during the animation, with the current instant displayed at the top panel corresponding to the end of the black line

top panel). On June 6.5, 1998 (UT), a sudden brightening by another 1.5 mag was detected. This caused a level-2 alert, which meant that MACHO-98-SMC-1 was a likely caustic crossing event. An accurate prediction of the timing of the second caustic crossing was then a very important task (later it was shown by Jaroszyński & Mao (2001) that a reliable prediction of the exact timing of the second caustic crossing is intrinsically difficult and possible only relatively late). The first prediction for the second caustic crossing was for UT June $(19.3 \pm 1.5)$, issued on June 15.3 (see Fig. 10, bottom panel). This value was revised on June 17 to UT June $(18.2 \pm 1.5)$, when it actually happened. The possibility to analyse the lightcurve while it is still ongoing made it hence possible to react quickly. This resulted in an amazing 1598 data points for this double peak microlensing event within about 50 days (Fig. 10) !

Measuring the caustic crossing time can help break the degeneracy between the lensing parameters. The method is very simple: assuming that locally the

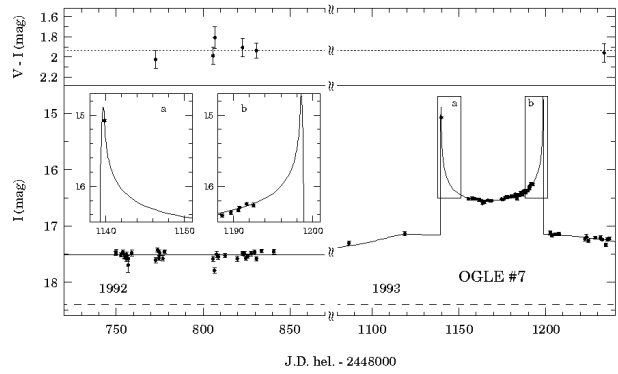
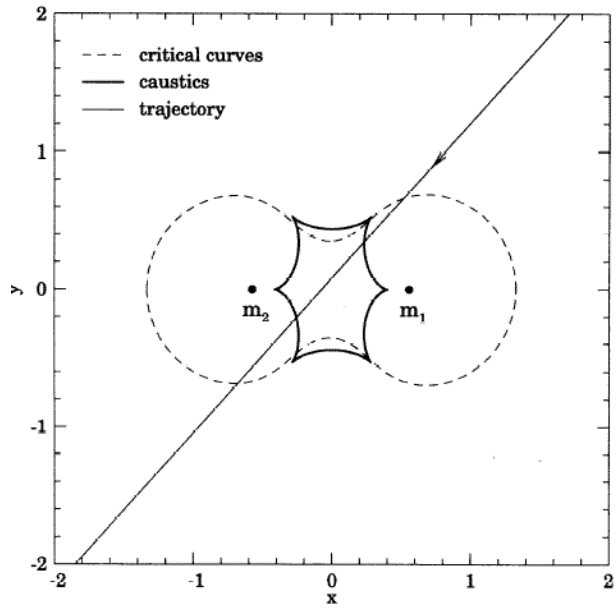

**Fig. 9.** Lightcurve obtained by the OGLE team of the first double-lens microlensing event OGLE-7 (top) and corresponding double lens configuration with caustics, critical lines and relative track (from Udalski et al. 1994)

473

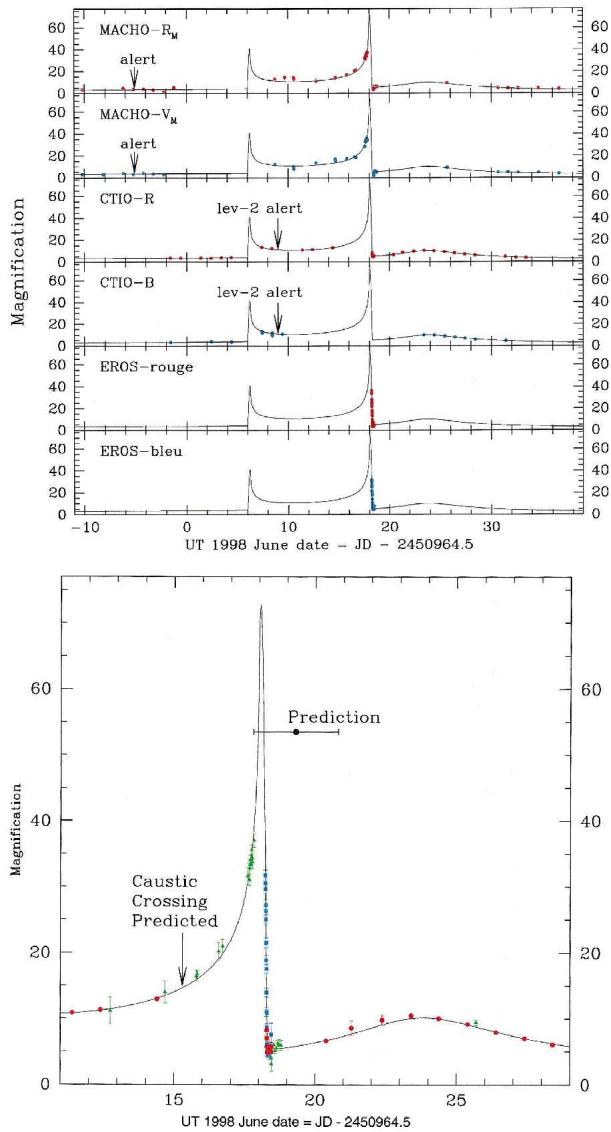

**Fig. 10.** Lightcurve obtained by the MACHO team of the first double-lens microlensing in the Small Magellanic Cloud, SMC-1 (top). The six panels represent data from three observing sites and two filters, respectively. The times of the first and second alerts are indicated by arrows. The bottom panel is a zoom around the second caustic crossing, indicated is the predicted epoch of the caustic crossing, and the time when it was announced (from Alcock et al. 1999)

474

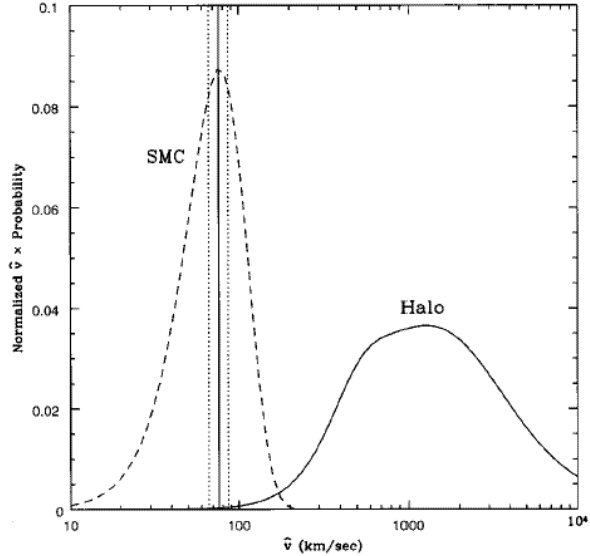

**Fig. 11.** Probability distribution for certain values of the transverse velocities of the lens in MACHO-SMC-1 (from Alcock et al. 1999), assuming that the lens resides in the LMC (dashed) or in the Milky Way halo (solid). The vertical solid (dotted) line shows the measured value (including the error bars)

(fold) caustic is a straight line in the source plane, the duration of the caustic crossing is then just the time it takes the star to move its own stellar diameter, $2\,t_*$. As described in Alcock et al. (1999), they could make a very accurate measurement of the duration of the caustic crossing: $t_* = (0.116 \pm 0.010)$ days. With the knowledge of the physical size of the star, $R_* = (1.1 \pm 0.1) R_\odot$, from its spectral type ($T_{\text{eff}} = 8000 K$), it became possible to determine the proper motion of the lens with respect to the source. Alcock et al. (1999) estimated the transverse velocity of the lens projected to the SMC distance to be $v = (76 \pm 10)$ km/s. This in turn allows to estimate the distance to the lens, one of the very interesting parameters which cannot be obtained in "normal" microlensing cases, due to the degeneracy of the parameters lens mass, distance and transverse motion (cf. Section 1.2).

In Fig. 11, the expected distribution of relative lens velocities projected to the source (SMC) plane are shown for two potential lens populations: if the lenses are in the Milky Way Halo, the typical (projected) velocities are in the range of about 1000 km/sec, whereas for a lens population in the SMC, it is rather around 60 km/sec.

The measured value of the projected transverse velocity hence clearly favours a lens position in the (foreground of the) SMC. Alcock et al. (1999) analysed this quantitatively as well: The probability for a halo star to have such a low velocity is only 0.12%, whereas 38% of the SMC stars would have such a value or smaller. Hence Alcock et al. (1999) concluded that the lensing system responsible for MACHO-99-SMC-1 is much more likely to reside in the SMC rather than in the Galactic halo, hence it is a case of "self lensing" (Sahu 1994).

### 2.4 Binary Lens MACHO 1999-BLG-047

The well covered microlensing event MACHO 1999-BLG-047 displays a "nearly normal" lightcurve with a small but highly significant deviation close to the peak (Fig. 12). Since such small-amplitude deviations near the peak of a lightcurve can be produced by planetary lenses (cf. Griest & Safizadeh 1998), this event attracted a lot of interest. However, roughly equal mass binary lenses with either very small or very wide separation can introduce very similar features in the lightcurve. The analysis of Albrow et al. (2002) showed that these two cases can be distinguished with a high quality data set, and that in particular the event MACHO 1999-BLG-047 is produced by an extreme binary event. However, the analysis yielded two "islands" in the mass ratio versus separation diagram which both satisfied the observational data equally well (Fig. 13): it was not possible to find a unique solution. The two best fit models require the binary lens to be either a close binary with parameters $d/R_E = 0.134 \pm 0.009$ and $q = 0.340 \pm 0.041$ or $d/R_E = 11.31 \pm 0.96$ and $q = 0.751 \pm 0.193$ (more details see in Albrow et al. 2002).

### 2.5 Binary Lens EROS BLG-2000-005

The triple-peak microlensing event EROS BLG-2000-005 (Fig. 14, An et al. 2002) became one of the most spectacular examples of stellar microlensing (An et al. 2002). Originally detected by the EROS team, an alert was issued on May 5, 2000 for a possible microlensing event. On June 8, 2000, the MPS team (Microlensing Planet Search) sent an anomaly alert, stating that the star has changed its brightness by 0.5 mag compared to the previous night, and that it was still brightening at the remarkable rate of 0.1 mag per 40 minutes (!). The PLANET team increased the monitoring frequency of this event and kept it at a high level until about a year later.

The I-band data of the PLANET team are presented in Fig. 14, containing 1286 data points (cf. An et al. 2002). The different symbols indicate the four PLANET telescopes in South Africa (SAAO), Tasmania (Canopus), Chile (YALO) and West Australia (Perth). The two main maxima in the lightcurve are more than 3.5 mag above the baseline and show very steep rising or dropping flanks, indicative of caustic crossings. The third peak is slightly less steep and shows the characteristics of a cusp passage. The inset in the top right part

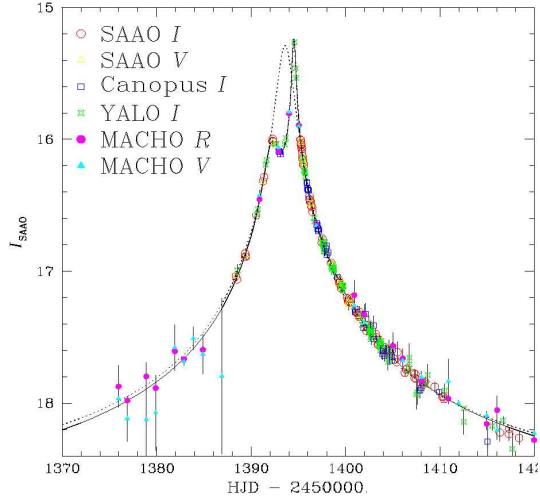

**Fig. 12.** Lightcurve of the binary microlensing event MACHO 99-BLG-47 obtained by the MACHO and PLANET teams (from Albrow et al. 2002). The symbols denote the data points from the various observatories. The solid line is the best fit, the dotted line is the best single lens fit, clearly not at all reproducing the high magnification data points

of Fig. 14 shows the track of the source star relative to the binary lens caustic. The two caustic crossings are labelled 'A' and 'B', whereas the cusp passage has the label 'C'.

The quality of the data is very good and the duration of the event long enough, in order to measure the parallax due to the motion of the Earth around the Sun. In Fig. 15, the geometry of the event as projected on the sky is shown (from An et al. 2002). It shows very clearly the difference of the relative paths as seen from the Earth (solid track) and from the Sun (dashed track), i.e. the parallax effect.

In Fig. 16, a close-up of the previous figure is shown for the time of the cusp passage. The circle represents the source at the time of closest approach to the cusp (see also the inset panel).

The system could not be modelled satisfactorily without including the orbital motion of the binary (cf. Figs. 15 and 16). This made it possible to measure the projected Einstein radius $\tilde{r}_E = (3.61 \pm 0.11)$ AU. The angular Einstein radius, on the other hand, could be determined from the finite source effects on the lightcurve to $\theta_E = (1.38 \pm 0.12)$ mas with an estimate of the physical source size from its position in the color-magnitude-diagram, these two measurements result in a determination of the lens mass: $M_{\rm lens} = (0.612 \pm 0.057) M_\odot$. This is the first time that a microlens parallax was measured for a caustic crossing event, and also the first time that the lens mass degeneracy could be broken and that the mass of a microlens could be derived from photometric measurements alone.

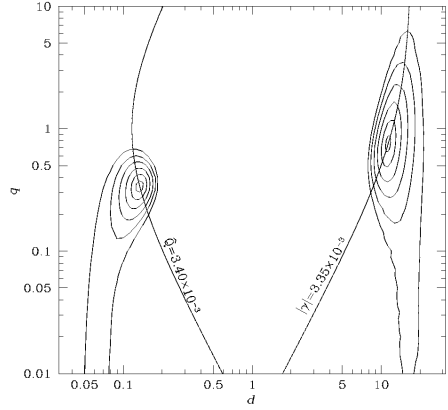

**Fig. 13.** Mass ratio versus separation for the binary lens parameters of the PLANET analysis of the binary microlensing event MACHO 99-BLG-47: shown are contour of good fits, based on PLANET and MACHO data. The binary separation d is in units of the Einstein radius of the combined mass, and the mass ratio q is the ratio of the farther component to the closer component to the source trajectory (i.e., q > 1 means that the source passes by the secondary). Contours are shown for $\Delta\chi^2 = 1, 4, 9, 16, 25, 36$ (with respect to the global minimum). It is obvious that there are two well separated minima. Also drawn are the curves of models with the same quadrupole moment $\hat{Q}$ as the best-fit close-binary model and the same shear $\gamma$ as the best-fit wide-binary model (from Albrow et al. 2002)

# 3 Microlensing and Dark Matter: Ideas, Surveys and Results

### 3.1 Why we need dark matter: flat rotation curves (1970s)

Since the 1970s, measurements of the rotation curves of galaxies showed that the (rotational) velocity as a function of radius is roughly constant: galaxies have flat rotation curves (e.g., Bosma 1978, Rubin 1983), see also Fig. 17, top left panel. This is a non-trival result: in the solar system, as a contrast, the planets follow the Kepler law: velocity decreases with the square root of the radius (bottom left panel in Fig. 17). In general, for a stable circular orbit, gravitation is balanced by the centrifugal forces:

$$\frac{GM(r)}{r^2} = \frac{v(r)^2}{r}.$$

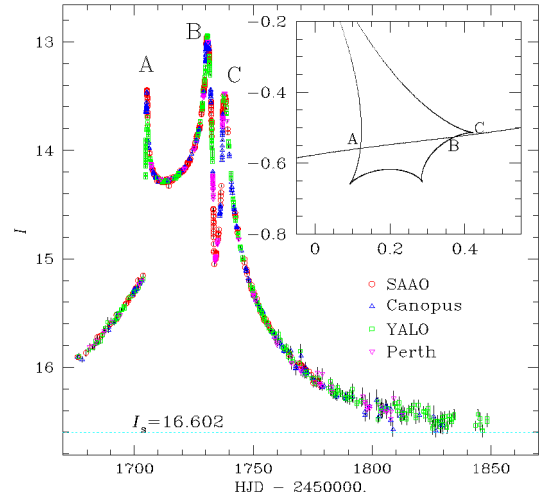

**Fig. 14.** Lightcurve of the binary microlensing event EROS BLG-2000-5 (from An et al. 2002). The I-band data points are displayed, the baseline magnitude $I_s$ is given, the three peaks are labelled A, B and C. The inset at the top right indicates the geometry of the stellar track crossing the caustic, with the three caustic crossings labelled according to the maxima in the lightcurve

If $v(r) = $ const for a broad range of radii $r$, this implies: $M(r) \propto r$. In other words: flat rotation curves mean that the mass of the galaxy increases linearly with radius. In Fig. 17, three idealized rotation curves are shown for solid body rotation (top right), Keplerian rotation (bottom left) and a relation in which mass increases linearly with radius: $M(r) \propto r$.

Using the 21cm Hydrogen line, the rotation velocity of spiral galaxies could even be measured far beyond the visible stellar part: The interesting - and very unexpected - result was: galactic rotation curves remain flat even outside the regions in which stars exist. These observations imply: more than 90% of the mass of a galaxy must be in an unknown and invisible form, soon to be called "dark matter". From its presumed roughly spherical distribution around the visible galaxies, the concept of "dark matter halos" was established[4]

For some time, an alternative explanation for the flat rotation curves was put forward: the concept that Newton's law of gravity (and also Einstein's General Theory of Relativity) changes on large length scales. Two of the theories in the latter categories are the "MOdified Newtonian Dynamics", or MOND (Milgrom 2001), and the "conformal gravity" (Mannheim 1992). These theories change the relation between gravity and distance and try to avoid the concept of "dark matter" altogether. We cannot go into any more detail here, but rather refer the interested reader to a recent review on alternative theories of gravity (Sanders & McGaugh 2002). However, in the light of the latest results of the Wilkinson Microwave Anisotropy Probe (WMAP, Spergel et al. 2003), these alternative explanations seem not to be viable any more.

---

[4] To this day it is not really understood what dark matter is. The concept of "Dark Matter Halos", however, is so ubiquitous inside and outside physics, that it made it even into art/literature, see Reza (2000).

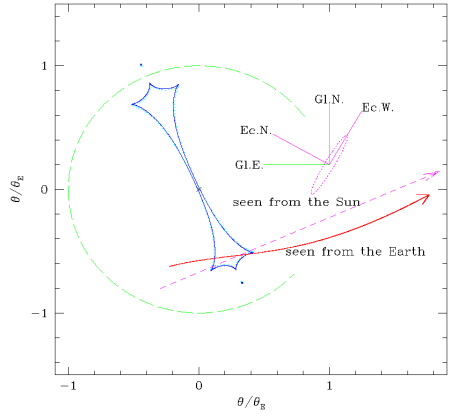

**Fig. 15.** Geometry of microlensing event EROS BLG-2000-5 as projected on the sky (from An et al. 2002). The origin (marked with a small cross) is the center of mass of the binary lens. The path of the source relative to the lens as seen from the Earth is shown as the solid curve, whereas the relative proper motion as seen from the Sun is indicated as the short-dashed line (length of both trajectories corresponds to six months). The circle (long-dashed line) is the Einstein ring, and the lines within (solid and dotted) are the caustics of the binary system at two different epochs. The labels "Gl.E." and "Gl.N." indicate the directions East and North in galactic coordinates, "Ec.N." and "Ec.W." stand for North and West in ecliptic coordinates, respectively (from An et al. 2002)

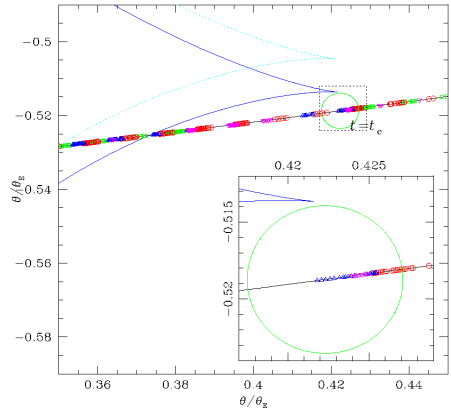

**Fig. 16.** Relative track of the source in the binary microlensing event EROS BLG-2000-5 during the time of closest approach to the cusp (from An et al. 2002). The symbols indicate the positions of the source center during data taking at the various observatories (cf. Fig. 14)

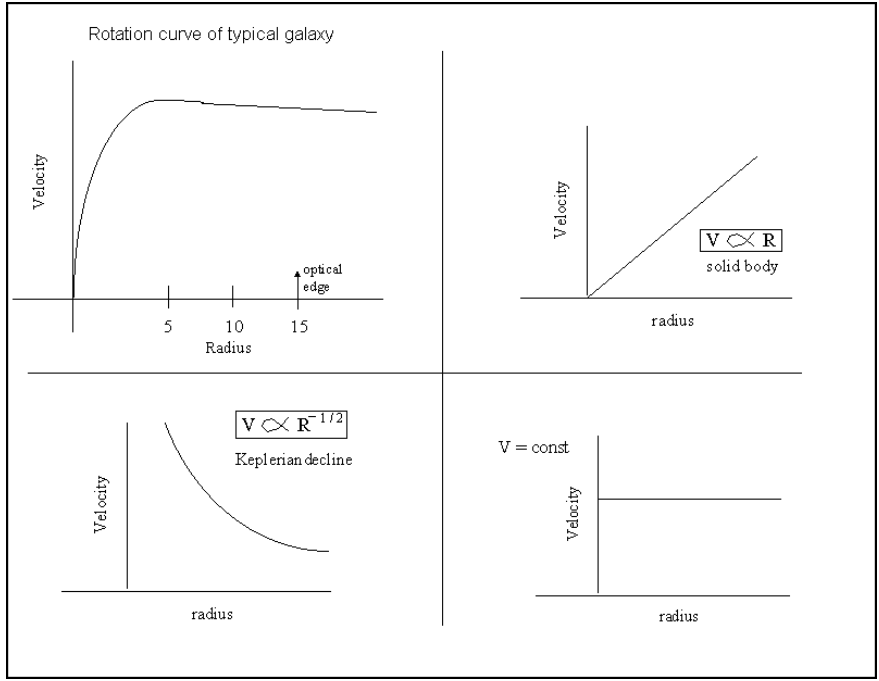

**Fig. 17.** Different types of rotation curves (schematically): observed flat rotation curve of spiral galaxies (top left); solid body rotation curve (top right): $v_{\rm rot}(r) \propto r$; Keplerian rotation curve (bottom left): $v_{\rm rot}(r) \propto r^{-0.5}$; rotation curve for constant velocity (bottom right): $v_{\rm rot}(r) = const$ (in other words: $\propto r^0$)

Though experimentalists have tried very hard for many decades (Rees 2003), no physical candidate for dark matter was detected. Two main dark matter candidate types were proposed: massive elementary particles and astrophysical compact objects. The list of elementary particle candidates for dark matter comprises many dozen candidates, among them many hypothetical ones: neutralino, Higgs particle, WIMPs (weakly interacting massive particles), axions. The suggested astrophysical candidates were stellar or sub-stellar mass black holes, neutron stars, white dwarfs, brown dwarfs, or planets.

### 3.2 How to search for compact dark matter (as of 1986)

In 1986, Bohdan Paczyński proposed a clean experiment for testing whether the latter type, compact astrophysical objects of roughly stellar mass, can make up the dark matter halo of the Milky Way (Paczyński 1986b). His idea was simple[5] and brilliant at the same time: if a class of compact objects in the mass range of very roughly $10^{-6} \leq m/M_\odot \leq 10^6$ exists in the Milky Way halo and makes up a fair fraction of the dark matter, then occasionally one of these objects must pass very close to the line-of-sight to a background star in the Large Magellanic Cloud (LMC). As a consequence, the apparent brightness of this background star is magnified temporarily, in exactly the way that is explained in Section 1: single lens, single source.

Paczyński determined the fraction of background stars that would be within the Einstein radius of these MACHOs[6], the so called optical depth, to be in the range $p_{\text{MACHO}} = 10^{-6}...10^{-7}$. This is a remarkably small number: it means that the apparent brightness of a few million stars has to be monitored very frequently, in order to find the handful of candidate lightcurves[7].

### 3.3 Just do it: MACHO, EROS, OGLE et al. (as of 1989)

What sounded like science fiction at the time (Paczyński even refers to it this way in his original article), soon became reality, due to four developments:

1. Optical CCD chips got bigger, and it became possible to build cameras consisting of an array of such CCDs. This way one could determine the apparent brightness of many stars in "one shot".
2. Software could be developed for automatic data reduction pipelines, so that a large number of objects could be treated and analysed with no or little human interaction.
3. Computer power kept increasing according to Moore's law, i.e. speed doubling roughly every 18 months, as well as data storage became available in sufficient amounts, so that by the mid 1990s literally tens of millions of stars could be monitored frequently enough with lightcurves being produced.
4. Scientists realized that the normal procedure of applying for a certain chunk of time at a certain telescope did not make much sense: they needed (and succeeded in getting !) dedicated telescopes.

---

[5] It was in fact so simple that the referee first rejected the paper; only after some discussion between author, referee and editor, the paper was published; and at the time of this writing, it has collected more than 500 references.

[6] MACHO - MAssive Compact Halo Object, an acronym coined for these dark matter candidates originally by Kim Griest (1991)

[7] Paczyński's optimistic suggestion that this might be measurable was in stark contrast to Einstein's pessimistic view exactly 50 years earlier: he had derived the basic equation and estimated the probabilites and written that there is 'no hope of observing such a phenomenon directly' (Einstein 1936).

In the years following Paczyński's article, three teams formed and started to address the scientific question posed. Later on, a number of additional collaborations followed:

- the MACHO Team (USA/Australia), MAssive Compact Halo Objects: http://wwwmacho.anu.edu.au/
- the EROS Team (France), Expérience pour la Recherche d'Objets Sombres: http://eros.in2p3.fr/
- the OGLE Team (Poland/USA), The Optical Gravitational Lensing Experiment: http://www.astrouw.edu.pl/∼ogle/
- the MOA Team (NZ/Japan), Microlensing Observations in Astrophysics: http://www.physics.auckland.ac.nz/moa/

They determined the apparent brightnesses of stars in the direction to the LMC/SMC and to the Galactic Bulge a few times per week, constructed lightcurves, identified the variable stars, and searched for the rare "needle-in-the-haystake" microlensing signal among the millions of stars.

### 3.4 "Pixel"-Lensing: advantage Andromeda !

In 1992, Arlin Crotts had suggested to use the Andromeda Galaxy as a "unique laboratory for gravitational microlensing". M31 is roughly 15 times as distant as the LMC/SMC, hence individual stars cannot be resolved any more: only the combined flux of many stars can be measured in any resolution element of the CCD camera. This means that a possible microlensing event would be "buried" among an ensemble of constant or variable unrelated stars. The magnification consequently is diluted: "blending" dominates the lightcurve very heavily. Only very high magnification events would be detectable. However, Crotts (1992) pointed out that M31 does have a number of advantages compared to LMC/SMC searches: smaller angular size of source stars, (much) greater total mass, favourable geometry and foreground/background asymmetry, which should statistically allow to distinguish microlensing events due to Milky Way halo objects from those produced by M31-halo objects. This method was subsequently somewhat improperly called "pixel-lensing" and became popular under this name.

Subsequently, a number of teams jumped on the pixel-lensing train:

- AGAPE: Andromeda Galaxy Amplified Pixel Experiment (later POINT-AGAPE): http://cdfinfo.in2p3.fr/Experiences/Agape/
- MEGA: Microlensing Exploration of the Galaxy and Andromeda http://www.astro.columbia.edu/∼arlin/MEGA/
- WeCAPP: Wendelstein Calar Alto Pixellensing Project, http://www.usm.uni-muenchen.de/people/fliri/wecapp.html.

For the "Theory of Pixel Lensing", see Gould (1996).

## 3.5 Current interpretation of microlensing surveys with respect to halo dark matter (as of 2004)

In the more than ten years which have passed since the first publication of stellar microlensing events towards the LMC (Alcock et al. 1993, Aubourg et al. 1993), many more microlensing events have been discovered (and some of the first discovered events were retracted because they were later classified as misinterpretations of a rare kind of variable stars, so-called blue bumpers [Tisserand & Milsztajn, private communication 2004]). With a baseline of five years or longer, statistically quantitative results have been obtained. In the meantime, both MACHO and EROS have ended their campaigns. Final results and/or conference summaries have been published. The two robust results of these two experiments are:

- A certain relatively small number of lightcurves of LMC (and SMC) stars have been obtained which were definitely produced by microlensing of an intermediate single or binary star. "Intermediate" between source star and observer could mean in the Galactic Halo, in the disk of the Milky Way, or in the foreground of the LMC/SMC.
- The total number of these microlensing lightcurves (fewer than two dozen) is definitely far too small to explain ALL the dark matter in the Galactic halo by compact objects, even if all the lenses were objects in the Galactic halo. What fraction of the halo dark matter could still be explained by MACHOs is a matter of debate. The estimates range from about 20% to zero.

Here the results are summarized:

**MACHO**

The MACHO team ended their operation in 1999. Results are summarized in Alcock et al. (2000b): They were in operation for 5.7 years and had monitored 11.7 million stars in the LMC. The identification of microlensing events is not trivial: They applied two criteria and found 13 or 17 events, respectively. The time scales of these events range from $t_E = 34$ to 230 days. According to their modelling, they expected 2 - 4 events from known stellar populations in the Milky Way.

Their analysis results in an optical depth of

$$\tau_{\text{LMC(MACHO)}} = 1.2^{+0.4}_{-0.3} \times 10^{-7},$$

plus an estimated systematic error of 20%. Their interpretation is that about 20% of the Milky Way halo could be made of dark matter objects in the mass range $0.15 \leq m/M_\odot \leq 0.9$, with a 90% confidence interval of 8% - 50%.

The lack of long duration events allows them to put limits on more massive objects, in particular potential black holes: they conclude that objects in the mass range $0.3 \leq m/M_\odot \leq 30$ cannot make up the entire dark matter halo (Alcock et al. 2001).

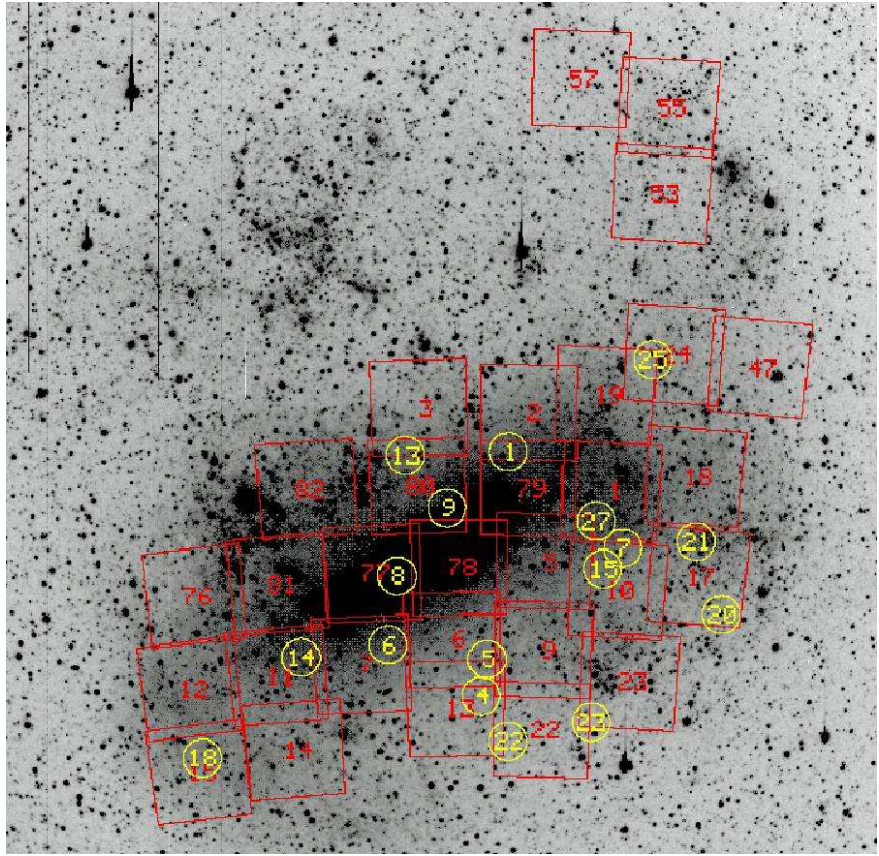

**Fig. 18.** MACHO fields with events indicated, from Alcock et al. (2000b): R-band LMC, 8.2 degrees at a side, 30 MACHO fields (squares), with 17 microlensing events (numbers in circles)

In Fig. 18, the 30 central monitoring fields of the MACHO team are indicated on an R-band image, including the location of the 17 identified events[8]. The detection efficiencies of the MACHO team – defined as the fraction of events of a certain duration that would have been identified in the data, given the actual sampling and data quality – for microlensing events of certain duration are shown in Fig. 19 as a function of increasing coverage: 1 year, 2 years

---

[8] The monitoring data of the MACHO team for 73 million stars in the LMC, the Small Magellanic Cloud and in the Galactic Bulge are available to the general public at http://wwwmacho.mcmaster.ca or http://wwwmacho.anu.edu.au.

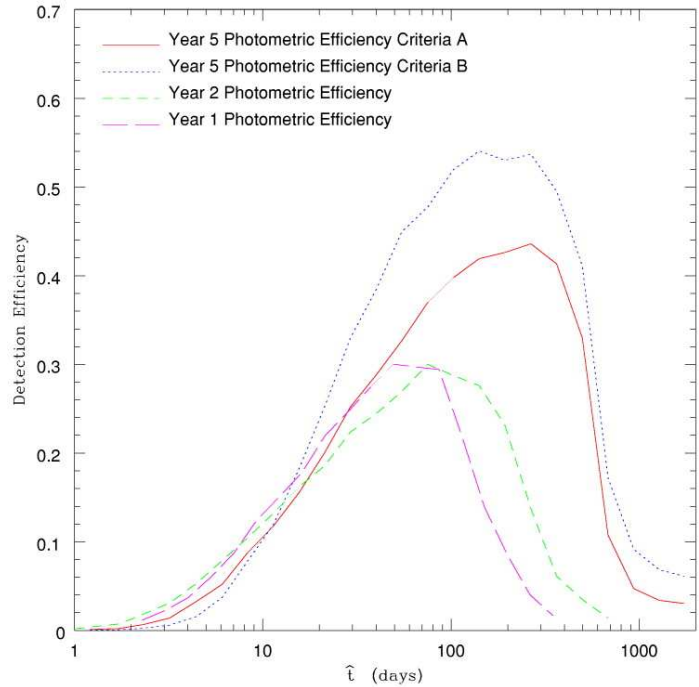

**Fig. 19.** Detection efficiencies of the MACHO experiment for one year, two years, and five years of operation, as a function of event duration (from Alcock et al. 2001)

and 5 years (and two selection criteria for the latter). More information and detailed analyses can be found in Alcock et al. (2000a,b).

**EROS**

The EROS team ended their operation in February 2003. Preliminary results were already published in Lasserre et al. (2000): They had ruled out sub-solar mass dark matter objects as an important component of the Galactic Halo. In an analysis of 5 years of EROS data towards the Small Magellanic Cloud, Afonso et al. (2003a) concluded: Objects in the mass range from $2 \times 10^{-7} M_\odot$ to $1 M_\odot$ cannot contribute more than 25% of the total halo. They derived an upper limit on the optical depth:

$$\tau_{\text{SMC(EROS)}} \leq 10^{-7}$$

(stating in addition that the long duration of all the EROS SMC candidates may point to the fact that they are more likely due to unidentified variable stars or self-lensing within the SMC, rather than due to halo objects). A preliminary analysis of the full 6.7 year EROS data set on the LMC strengthens this result (Glicenstein private communication 2003; Tisserand, private communication 2004): some of the formerly claimed EROS candidates turned out to be variable stars, and the derived optical depth towards the LMC is in the range

$$\tau_{\text{LMC(EROS)}} \approx 10^{-7}.$$

A graphic depiction of the EROS mass exclusion range for both the SMC and LMC directions can be found in Fig. 20 (from Afonso et al. 2003a). Similar results were obtained in an analysis of the microlensing experiments by Jetzer et al. (2004).

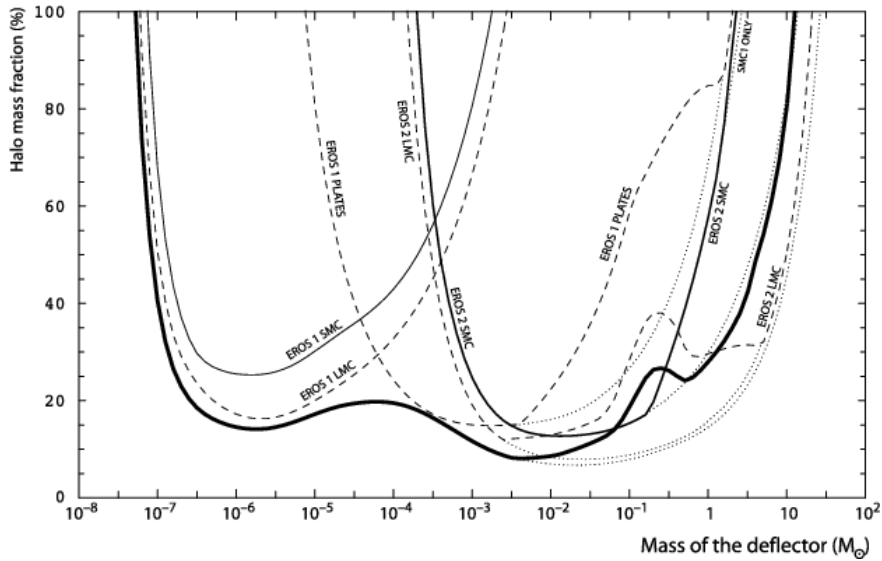

**Fig. 20.** Limits on Halo Mass fraction of EROS team: the 95% exclusion probability for the standard halo model are shown. Dashed lines: limits towards the LMC from EROS-1 and EROS-2; thin line: limit towards the SMC; thick line: combination of five EROS sub-experiments. The dotted line indicates the limits that would have been obtained without any detected events, it illustrates the overall sensitivity of the EROS experiment (from Afonso et al. 2003a)

Some additional results of EROS are: they had identified 4 long duration events, which are most likely not produced by a halo population. The main EROS result concerning the dark matter fraction is: less than 25% of the standard Milky Way Halo can be in objects with masses between $2 \times 10^{-7} M_\odot$ and $1 M_\odot$.

**Where are the lenses ?**

Combining the EROS and MACHO data results in roughly 20 microlensing events in the directions towards the LMC and the SMC. Originally, the experiments were set up to find compact dark matter objects in the Galactic Halo. So the big question is: *are the lenses* that caused the microlensing events *dark matter objects* ? This is difficult to answer, because this would involve to uncover the nature of an invisible object, which is almost impossible to do. However, what may help answer this question is a related one: *where are the lenses ?* And the answer to this latter question may be easier to obtain. From a purely observational point of view, the lenses could be at three distinct locations:

- in the Milky Way (thin/thick) disk: the lenses – in that case presumably normal stars – might become visible a few years after the event, once they have moved away from the bright background LMC/SMC star.
- in the Milky Way Halo: then they could be the searched for dark matter population. The density of events should be proportional to the density of stars in the LMC.
- inside the LMC/SMC: foreground stars could act as lenses on background stars: the density of events should be proportional to the star density squared.

Jetzer et al. (2004) conclude that the microlensing events are produced by various populations: a combination of self-lensing in the LMC, thick disk, spheroid, plus some "true machos" in the halo of the Milky Way and the LMC itself. Taking advantage of the apparent near-far asymmetry of the spatial distribution of the LMC events, Mancini et al. (2004) re-analysed the possibility of self-lensing. Their main conclusion is that even considering this, self-lensing cannot account for all the observed microlensing events towards the LMC.

In an earlier conference proceedings contribution, Kailash Sahu (2003) discussed the issue: "Microlensing towards the Magellanic Clouds: Nature of the Lenses and Implications for Dark Matter" in some detail. In particular, Sahu investigated the question of the distance of the 17 MACHO events towards the Magellanic Clouds. As a first step he summarizes:

- for one of them, a binary-lens event, the distance could be determined securely via its caustic crossing time scale: it is within the SMC (cf. Fig.10).

- for three more, the lens location could be estimated. This estimate is less certain than for the SMC event, but the evidence suggests that it is very likely that the lenses are located within the Magellanic Clouds as well.

As an independent second step, Sahu (2003) mentions that – assuming that most of the events are dark matter objects in the Galactic Halo – the time scales of the events towards the LMC would imply that masses are of the order of 0.5 $M_\odot$ (cf. Alcock et al. 2000a). However, with the same line of thought, the most likely masses for the events towards the SMC would be in the range 2 - 3 $M_\odot$. Could the mass distribution of objects with different masses be different from each other ? No model of the Galaxy is consistent with such an inhomogeneous mass distribution. On the other hand, if one assumes that most of the events are caused by foreground objects in the LMC/SMC, then the expected masses would be of order 0.2 $M_\odot$ for both LMC and SMC.

A third line of argument uses the frequency of binary lenses. Two of the 17 events are caused by binary lenses. In both cases, the lenses are most likely objects located within the LMC/SMC. Assuming that roughly 50% of the potential lenses in the LMC/SMC are in binary systems (similar to the stars in the solar neighbourhood), one would expect that 10% of all microlensing events would show binary characteristics (cf. Mao & Paczyński 1991 and Section 2.1). This implies that of order 20 events are expected to be caused by single stars within the LMC/SMC. So this would be perfectly consistent, if most of the observed microlensing events are caused by foreground stars within LMC/SMC.

A fourth argument of Sahu (2003): If the microlensing events are caused by 0.5 $M_\odot$ objects in the Galactic Halo (as claimed from the LMC observations), one would have expected to detect about 15 events in the direction towards the SMC, with time scales of about 40 days. Not a single event of this kind was detected: in fact, both SMC events are shown to be due to self-lensing.

Although each individual of these four arguments is not very strong, the combination of them provides relatively firm evidence against them being interpreted as mostly due to halo objects.

The conclusions in Sahu (2003) are: *"Close scrutiny of the microlensing results towards the Magellanic Clouds reveals that stars are major contributions as lenses, and the contribution of MACHOs to dark matter is 0% to 5%."*. This view might not be shared by everyone working in the field. However, it is certainly a viable one[9].

### 3.6 Microlensing towards the Galactic Bulge

As originally suggested by Paczyński (1986b), monitoring stars in the Galactic Bulge turned out to be a very fruitful enterprise. Originally meant as a safety measure[10], in the mean time the Bulge microlensing turned out to be a source of exciting astrophysical results in itself.

---

[9] In an earlier independent analysis, Graff (2001) had concluded: "Occam's razor suggests ... that microlensing experiments have simply found a background of ordinary stars".

[10] Microlensing experiments towards the Bulge would produce microlensing events with certainty due to the known population of disk stars, which could be used as a test of the experimental setup; otherwise, the potential lack of microlensing events towards the LMC/SMC could always have two reasons: there are no MACHOs, or the experiment does not work properly.

The angular distribution of the microlensing events lead to the re-discovery of the galactic bar (see Stanek et al. 1994, Paczyński et al. 1994). The microlensing optical depth in this direction turned out to be higher than expected: the original results by MACHO/OGLE were roughly

$$\tau_{\text{Bulge(MACHO/OGLE)}} \approx 3 - 4 \times 10^{-6}$$

(Udalski et al. 1994b, Alcock et al. 1997). This caused some kind of problems of our understanding of the Galactic dynamics: the high optical depth for microlensing implies much more mass than people had thought there is in the inner part of the Galaxy. Quite a number of papers dealt with this issue and tried to solve the discrepancy.

Recently, EROS published their analysis for the optical depth towards the Galactic bulge, based on the identification of 16 microlensing events with clump giants from a region of 15 contiguous one-square-degree fields with a total of $1.42 \times 10^6$ clump giants. The distribution of the time scales of their microlensing events is displayed in Fig. 21: in a logarithmic presentation, most events were found with Einstein time scales of 10 to 30 days. EROS found a much lower value than what was previously favoured:

$$\tau_{\text{Bulge(EROS)}} = (0.94 \pm 0.29) \times 10^{-6}$$

(Afonso et al. 2003b), which lead to the remark "The issue of the optical depth to the bulge is solved" by one of the experts in the Galactic microlensing community (Andy Gould, private communication March 2003). Considering the error bars of the published values, the problem was never really severe: the new EROS result still agrees with most of the previous results at the $2\sigma$ level. But it is closer to the many predicted values. For a detailed discussion of the differences between the various observational and theoretical analyses, see Afonso et al. (2003b).

## 4 Microlensing Surveys in Search of Extrasolar Planets

The very first time that microlensing by planets was mentioned in the literature was the paper by Shude Mao and Bohdan Paczyński from 1991: "Gravitational microlensing by double stars and planetary systems". This seminal paper with more than 150 citations by now (Dec. 2003) states the situation and explores the possibilities. Experiments for the detection of compact objects of stellar mass in the halo or the disk of the Milky Way via microlensing were planned and prepared at that time. Mao and Paczyński (1991) figured out that binary signatures of the lenses should be visible in some of the lightcurves. In addition, they stated that this microlensing technique will be able to detect planetary systems ultimately as well.

In this section, the current state of microlensing searches for extrasolar planets is summarized. The basics of the method are explained, the advantages and disadvantages are discussed and compared with other planet-search

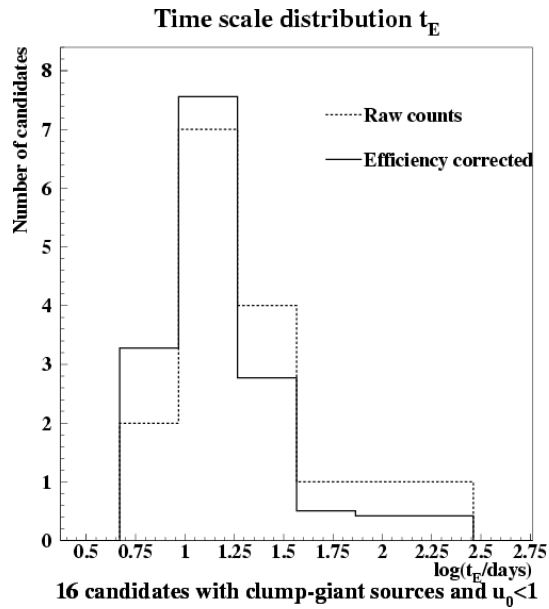

**Fig. 21.** Time scales of the 16 EROS bulge candidate events of clump giants: dashed line corresponds to raw data, the solid line shows the (rescaled) distribution, corrected for the detection efficiency (from Afonso et al. 2003b)

techniques. The teams active in the microlensing searches (OGLE, MOA, PLANET, MicroFUN) are presented. A number of recent observational and theoretical results on planet microlensing are mentioned. Good descriptions of the basics of planet microlensing can be found, e.g., in Paczyński (1996), Sackett (2001) and Gaudi (2003).

### 4.1 How does the microlensing search for extrasolar planet work ? THE METHOD

Only a few years after the original idea proposed by Bohdan Paczyński (1986b) to use gravitational microlensing as potential test for stellar mass objects in the Galactic halo, Mao & Paczyński (1991) calculated that roughly 10% of all lensing events *had* to show the signature of a binary companion. So it only was a quantitative question: One had to monitor the apparent brightness of a very large number of stars in the Milky Way bulge, with the goal to detect the passage of a binary star or star-plus-planet system in the line-of-sight to one of these background stars, producing a very characteristic magnification lightcurve.

Compared to the situation of a single stellar lens, there are three additional parameters in a situation of a star-plus-planet lens (as shown in the binary lens case, Section 2.1): the mass ratio $q = M_{\rm PL}/M_*$, the projected separation

491

between planet and star $d$, and the angle between the relative source track and the connecting line between star and planet. The binary-lens nature of the star-plus-planet system affects the observed lightcurves most strongly if the separation is in a certain range, the so-called lensing zone: $0.6 \leq d/R_E \leq 1.6$. In this case, the planet caustic(s) are within the Einstein radius of the host star. Due to a coincidence, this lensing zone corresponds to a projected distance range of order 1 AU. This means that the microlensing method is in principle capable of detecting planets at distances overlapping with the habitable zone[11].

In Fig. 22, six magnification patterns are shown for planet distances very close to the Einstein radius, so called "resonant lensing": $d/R_E = 1.11, ..., 0.91$ (Wambsganss 1997). The lightcurves on the right hand side (displayed is the "difference lightcurve" between the star-plus-planet lightcurve and the star-only lightcurve) show that the deviations are typically of small amplitude (few percent) and short duration (few percent of the Einstein time, i.e. order a day or shorter).

Very nice animations of planet microlensing showing relative tracks, individual images and magnification as a function of time for various mass ratios $q$ and separations $d$ are provided by Scott Gaudi at:
http://cfa-www.harvard.edu/~sgaudi/movies.html.

### 4.2 Why search for extrasolar planets with microlensing ? – ADVANTAGES and DISADVANTAGES

Searching for extrasolar planets is a tough astrophysical enterprise. There are a number of different techniques being pursued: radial velocity variations or doppler wobble, transits, astrometric variations, pulsar timing, or direct detection. Each of those methods is used by a number of groups (more than 20 different teams, e.g., for transit searches alone, see review by Horne (2003)). So it is a fair question to ask: why bother applying yet another technique ?

In this subsection, the microlensing method for planet searching is compared to the other indirect methods. It will be shown that microlensing is indeed a complementary method with different strengths, and that it is very worthwhile pursuing this search technique. As the starting point, here follows a list of commonly mentioned "disadvantages" of the microlensing planet searching technique (with a few comments added in parentheses):

1. The probability for an individual planet-lensing event is very small *(yes indeed, the chance for detecting a planet-microlensing event by monitoring*

---

[11] 'Habitable zone' is defined as the distance range around a central star which would allow life to develop on a planet; because of lack of better criteria – and based on life as we know it – what is chosen in the simplest version is a temperature range between 0 and 100 degrees Celsius (centigrade), which allows water to be in the fluid phase.

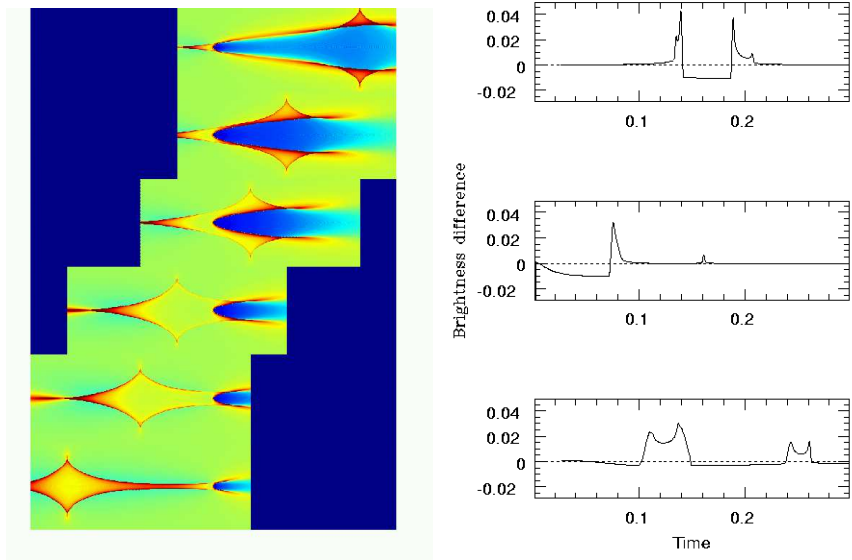

**Fig. 22.** Left: Caustics of a saturn-mass planet (mass ratio $q = 10^{-4}$) with a projected separation close to one Einstein radius: the six panels from top to bottom show parts of the magnification patterns for separations $d/R_E = 1.105, 1.051, 1.025, 0.975, 0.951, 0.905$; right: typical (difference) lightcurves obtained from the second, fourth and sixth panel; time scale is in units of the Einstein time $t_E$. The amplitude is give in magnitudes (after Wambsganss 1997)

   an arbitrary background star in the galactic bulge is very roughly of order $10^{-8}$ or smaller).
2. The duration of the planet-induced deviation in the microlensing lightcurve is very short *(yes, estimated typical durations for planet deviations are of order hours to days)*.
3. The planets – once found – will be very distant *(true, most likely distance is a few kpc)*, and even worse: the exact distance determination will turn out to be very difficult or close to impossible *(true, unless we get additional information about the event)*.
4. It is close to impossible to do subsequently more detailed investigations of the planet *(fair enough)*.
5. The lightcurve shapes caused by extrasolar planets are diverse, occasionally there might be a parameter degeneracy when modelling the event, with no unique relation between lightcurve and planet parameters *(yes)*.
6. Even when unambiguously detected, what can be determined is not the *mass* of the planet, but only the mass ratio between host star and planet *(true)*.

7. No independent confirmation will be possible *after* the detection: it is a once-and-only event *(yes)*.

These are fair points of critique towards using microlensing as a planet search technique. So, why bother anyway ? Firstly I would like to emphasize and recall that almost all these arguments were put forward already more than a decade ago, then used against the "normal" stellar/dark matter microlensing which had been proposed by Paczyński (1986b) and produced the first results a few years later (Alcock et al. 1993, Aubourg al. 1993, Udalski et al. 1993). Today no one has any doubts any more about the reality of the many stellar microlensing events, despite, say, their non-repeatability. Secondly, I now try to present one by one good reasons why the above arguments – though true to a large degree – are not really arguments *against* using the microlensing technique for planet searching:

1. <u>*Small probability:*</u> The probability for "normal" microlensing events in the galactic halo or disk (i.e., directions to the LMC/SMC or the galactic bulge) is already very small (of order $10^{-6}...10^{-7}$). Nevertheless, more than a dozen microlensing events have been found towards the LMC/SMC (Alcock et al. 2000) and more than 1000 events (!) have been detected in the direction of the galactic bulge (see, e.g., on the OGLE web page http://www.astrouw.edu.pl/∼ogle/ogle3/ews/ews.html). This shows: small probability in itself is certainly not a strong argument against using this technique. It is just a matter of statistics: even today it is possible to monitor of order $10^7$ stars on a regular basis with sampling every few days on comparably small operational cost. Doubtlessly, this number will increase by an order of magnitude every few years.
2. <u>*Short duration:*</u> In the current "mode-of-operation", the planet-searching teams take advantage of the relatively coarse sampling in the time domain of the microlensing monitoring teams (in particular OGLE and MOA), they work "piggy-back": once a deviation indicative of a stellar microlensing event is detected by these monitoring teams, the planet-searching teams follow those alerted events with a very dense coverage in time. This can result in lightcurves with an average sampling of many data points per hour. A number of events with more than 1000 data points (An et al. 2002) with photometric accuracy of 1% or better have been observed. Due to a set-up of telescopes in Australia, South Africa and Chile, lightcurve coverage around the clock is possible, weather permitting (see 'The 24-Hour Night Shift', Sackett 2001). So even planetary deviations in the lightcurve lasting only a couple of hours can be covered very well with many data points.
3. <u>*Large (and unknown) distance to the planet in general:*</u> The distances to the microlensing planets will be larger by one or two orders of magnitude than those found with the conventional techniques. This is true, too, for the pulsar planets (Wolscszan 1994) and not a disadvantage in itself. The "not-well-determined" aspect can be treated in a statistical way for a sample of events. If there is additional information available (parallax, astrometric signatures), the distance can be determined for the invididual events (cf. Alcock et al. 2001, Gould 2001).

4. *More detailed investigation impossible:* Indeed, a more detailed study of the planet candidate will turn out to be very difficult. However, we may be able to get more information about the star which the planet is circling: Alcock et al. (2001) show, that due to the relative proper motion, the projected positions of source star and lens star will move away from each other, so that we may be able to detect and measure the parent star and the relative proper motion of the star-plus-planet system, a few years after the event.

5. *Parameter degeneracy:* Lightcurves covering only the central caustic or only the outer caustic are likely to have two sets of solutions. However, there is a wide range of planetary lightcurves which will result in unique solutions/fits, if the data sampling and quality is good enough.

6. *Only mass ratios determinable:* Most stars in the disk of the Milky Way are low mass main sequence stars, M-dwarfs. Hence there is a relatively narrow range of absolute masses for most of the planets. Statistically, the planet mass distribution from microlensing can be determined to the same accuracy to which we know the mass function of the (host) stars. Furthermore, the most successful exoplanet search method to date – the radial velocity technique – also cannot determine the individual planet mass to better than a factor $1/\sin i$, due to the unknown inclination $i$ of the orbital plane of the planetary system relative to the line-of-sight.

7. *Once-and-only event, no independent confirmation:* Most star-plus-planet microlensing events will not repeat, this is true. But whether the event is "believable" or not is just a question of signal-to-noise: once there are enough data points with small enough error bars, this is convincing. A lightcurve consisting of more than 1000 data points with accuracy of order of 1 % or better (cf. PLANET team caustic crossing data of event EROS-BLG 2000-005, An et al. 2002) is beyond any reasonable doubt. In addition, lightcurves are often collected by two or more separate teams, which is a good independent confirmation. Furthermore, supernovae or gamma-ray bursts also do not repeat; no one takes this as an argument against them being real.

So all the arguments commonly used against microlensing as a useful planet search technique can be refuted or weakened. If the sampling and the photometric accuracy are good enough, planet microlensing deviations will be believed by the astronomical community. Occasionally there might still be model degeneracies. The most significant ones, though, just concern the projected separation between planet and host star: for each solution with separation $d$ there is usually also one with separation $1/d$. We have to live with this, as well as we do with the unknown $\sin i$ of the radial velocity planets.

After having discussed in detail the potential or perceived disadvantages, let us now come to the positive aspects of planet searching with the microlensing technique, compared to the other methods:

- *No bias for nearby stars:* Almost all the conventional planet search techniques concentrate their efforts on nearby stars, mainly because the signals are stronger, the closer the host stars are. The solar neighbourhood, however, might not be representative for the galactic planet population. Microlensing searches for planets are sensitive to stars anywhere along the line-of-sight to the source star in the galactic bulge at a distance of about 8.5 kpc, most sensitive for a lens position roughly half-way in between.
- *No bias for planets around solar-type stars/main sequence stars:* Almost all the conventional planet search techniques select and target the host stars. The very successful radial velocity technique cannot be applied to all stellar types, in particular not to active stars with broad and/or variable lines, so it has limited applications. Microlensing searches are "blind" for the characteristics of their host stars. Planet and host star will be found in proportion to their actual frequency in the Milky Way disk. The host stars of the microlensing planets will represent fair samples of the planet-carrying stars in the Milky Way. Planet microlensing is not constrained to any spectral type of host star, nor does it exclude any early type or active stars.
- *No strong bias for planets with large masses:* All conventional techniques are most sensitive to massive planets, with sensitivity strongly declining with decreasing planet mass. To first order, the microlensing signal – the amplitude of the lightcurve deviation – is independent of the planet mass. The duration and hence the probability for detection decreases, though, with decreasing planet mass. However, the size of the source star is important, and the lightcurve signal will be affected/smoothed by the finite source diameter, resulting in a lower amplitude signal (compared to a point source) and hence a lower detection probability.
- *Earth-bound method sensitive down to (almost) Earth-masses:* In principle, it is possible to detect even Earth-mass planets with ground based monitoring via microlensing. In practise, however, this would mean extremely high monitoring frequency and photometric accuracy. It is certainly true, though, that currently microlensing is able to reach down to lower planet masses than any other technique.
- *Most sensitive for planets in lensing zone, overlapping with habitable zone:* In the current mode-of-operation ("alerted" microlensing events being followed by dedicated planet-search groups), the most likely range of projected separations is the so-called lensing zone, roughly corresponding to a projected separation between 0.6 AU and 1.6 AU (Bennett & Rhie 1996). For low mass main sequence stars, this region overlaps with the habitable zone. This coincidence makes microlens-detected planets particularly in-teresting with regard to the question whether and how many planets exist in the habitable zone.

- *Multiple planet systems detectable:* There are two "channels", in which microlensing can even detect multiple planet systems: well sampled, very high magnification events have such small impact parameters that they pass the central caustic, which carries the signature of all the planets. Another channel would be the chance passage through two or more planet caustics, in case they happen to lie along the path of the background source star.

- *"Instantanous" detection of large semi-major axes:* The detection of long period planets is a long lasting process with the radial velocity or astrometry or transit techniques (years, decades ?): ideally it takes at least one full period for confirmation, better two or three. Microlensing will find large-separation planets basically instantaneously. The measured (projected !) distance between planet and host-star is, though, only a lower limit to the real semi-major axis (statistically, the 3-dimensional distribution can be inferred under the assumption that there is no preferred direction of the planetary orbital planes in the Milky Way).

- *Detection of free-floating planets ("isolated bodies of planetary mass"):*
  The next generation of microlensing searches for planets most likely will not work in the two-step mode-of-operation described below, with one team sampling lightcurves coarsely and then follow-up teams sampling selected candidate frequently. Rather, they will do very massive photometry ground-based (cf. Sackett 1997), or potentially even continuously from space, as the satellite project "Microlensing Planet Finder" (MPF, formerly called GEST) promises to do (Bennett & Rhie 2002, Bennett et al. 2003). Once such an experiment is implemented, microlensing will also detect a potential population of free-floating planets, by the microlensing signature of single lenses with small mass, i.e. very short duration (Han & Kang 2003).

- *Ultimately best statistics of galactic population of planets:* Gravitational microlensing will ultimately provide the best statistics for planets in the Milky Way; it is not without biases, but the biases in the microlensing search technique are very different from those of all other methods and can easier be quantified.

So gravitational microlensing is a very powerful and promising method for the search for extrasolar planets. It is largely complementary to other planet search techniques and has relatively little sensitivity to the planet mass. It also has a number of not-so-favourable aspects, which, however, are more than balanced by the advantages listed above.

### 4.3 Who is searching ? THE TEAMS: OGLE, MOA, PLANET, MicroFUN

The search for planets with the microlensing technique is currently done in a two-step process with shared tasks:

1. Stellar microlensing events have to be discovered while they are still in progress. This task is being done by two monitoring teams which measure the apparent brightness of a few million stars every few days:
   - MOA ("Microlensing Observations in Astronomy"; New Zealand/Japan, 60cm telescope on Mt. John, NZ): covers about 20 square degrees few times per night; geared to high magnification events (Bond & Rattenbury et al. 2002): 10 events expected per season with $A_{\max} > 100$. In total, 74 alerts in the whole 2003 bulge season. MOA alert page: http://www.massey.ac.nz/$sim$iabond/alert/alert.html.
   From 2005 on, the MOA team will use a dedicated 1.8m telescope for their microlensing searches, which will improve their efficiency dramatically.
   - OGLE ("Optical Gravitational Lens Experiment", Poland/USA; 1.3m telescope on Las Campanas, Chile): monitor 170 million stars regularly (Udalski 2003). In total 462 alerted events in the 2003 bulge season. OGLE alert page:
   http://www.astrouw.edu.pl/$\sim$ogle/ogle3/ews/ews.html

   These monitoring teams use the image subtraction technique (Alard & Lupton 1998) for accurate photometry and do basically online data reduction (Wozniak et al. 2001). Once they have discovered an ongoing microlensing event, these teams alert the community for follow-up observations, which involves the second step:

2. Two specialized teams concentrate only on follow-up monitoring of currently ongoing microlensing events:
   - PLANET ("Probing Lens Anomaly NETwork"; international team, various telescopes in Australia, South Africa and Chile): monitor selected on-going events around the clock. PLANET home page:
   http://planet.iap.fr
   - MicroFUN ("MICROlensing Follow-Up Network", US/SA/Israel/ Korea; 1.3m telescope, Cerro Tololo): informal consortium of observers dedicated to photometric monitoring of interesting microlensing events in the Galactic Bulge. MicroFUN home page:
   http://www.astronomy.ohio-state.edu/$\sim$microfun/.

   Both follow-up teams monitor only *alerted events* with high frequency (ideally few times per hour) and high photometric accuracy. At any given time there are usually a few dozen interesting events being followed up.

### 4.4 What is the status of microlensing planet searches so far ?
### THE RESULTS

At the time of the 33rd Saas Fee Advanced Course on Gravitational Lensing (April 2003), there were no definitive results on the detection of planets with the microlensing technique. A few candidates had been proposed, however, they remain controversial. Here a selected number of recent observational and theoretical results with respect to planet microlensing are presented:

### PLANET results

The PLANET team has put limits on Jupiters orbiting Galactic M-dwarfs (Gaudi et al. 2001, Gaudi et al. 2002): Analysis of 5 years of PLANET monitoring data towards the bulge with respect to short-duration events from single-lens light curves yielded a well defined sample of 43 intensely monitored events. The search for planet perturbations over a densely sampled region of parameter space (two decades in mass ratio and projected separation) resulted in no viable planetary lensing candidates. This analysis found that less than 25% of the primary lenses can have companions with mass ratio $q = 10^{-2}$ and separations in the "lensing zone": $0.6 \leq d/R_E \leq 1.6$. With a model for the mass, velocity, and spatial distribution of the stars/lenses in the bulge, astronomical limits could be obtained: less than 33% of the M dwarfs in the Galactic bulge can have companions with $M_{\text{Jupiter}}$ between 1.5 AU and 4 AU; and, less than 45% of the M dwarfs in Galactic bulge can have companions with $M_{\text{Jupiter}}$ between 1 AU and 7 AU.

### Event OGLE-2002-BLG-055: possibly planetary ?

The microlensing event OGLE-2002-BLG-055 was investigated by Jaroszynski & Paczyński (2002). The lightcurve contains one data point which lies $\Delta m = 0.6$ mag above the "single-lens, single-source" fit (Fig. 23, left). There is only this one deviant point, but it is very reliable. The authors argue correctly, that there is no reason to ignore it. The simplest interpretation for the lightcurve is: a binary lens with parallax and mass ratio $q = 0.001 - 0.01$. The lower $q$-value would correspond to roughly a Jupiter-mass planet (depending on the exact mass of the primary). The authors caution, however: with a single deviant point, it is impossible to fit a unique model (cf. Fig. 23, right) ! Instead, they conclude: In order to make sure that for similar events in the future, more data points in the relevant epoch will be obtained, the OGLE observing strategy should be modified in the following sense:

1. instant verification of deviant points in a microlensing lightcurve,
2. in positive case (deviation confirmed): change observing strategy, follow this particular event by frequent time sampling to make unique model possible.

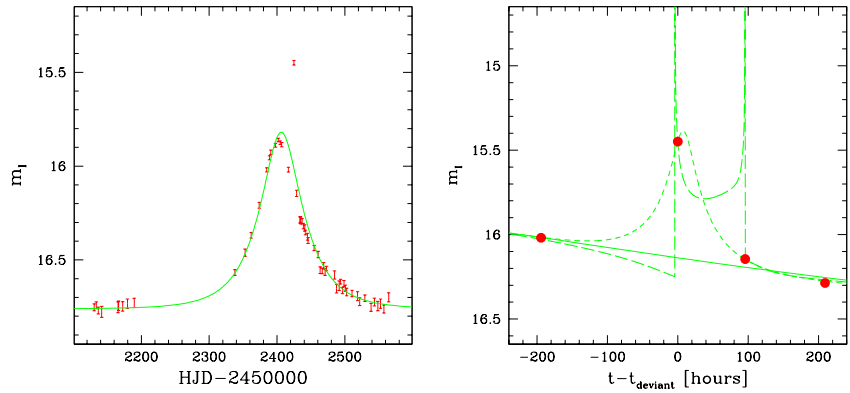

**Fig. 23.** Left: OGLE data of the microlensing event OGLE-2002-BLG-055 with the best fit single-lens-lightcurve including all data points. Right: zoom around the deviating data point with two well fitting binary lens models of mass ratios $q = 0.01$ (short-dashed) and 0.001 (long-dashed). The solid line indicates the best-fit single-lens lightcurve (from Jaroszynski & Paczyński, 2002)

As a very fast consequence of this suggestion, Andrzej Udalski implemented a very fast check-and-verification system, the OGLE Early Early Warning System (EEWS). It uses the automatic data reduction and searches for deviations "on the fly", i.e. recognize and verify possible planetary disturbance in real time with instant follow up (Udalski 2003). Very impressively, OGLE is now able to verify or falsify a deviant data point within 5 minutes ! As shown by events OGLE-2003-BLG-170 and OGLE-2003-BLG-194 in June 2003 (cf. OGLE web page at http://www.astrouw.edu.pl/∼ogle): it works ! This new developement is very promising indeed.

### Limits on number/orbits of exoplanets from 1998-2000 OGLE data

Tsapras et al. (2003) analysed the OGLE data base of the years 1998 to 2000. They put limits on the number and orbits of extrasolar planets. They focused on the frequency of "cool" Jupiters at a few astronomical units separation, based on 145 OGLE events. They used a maximum likelihood technique and found $n \leq 2$ realistic candidate events for a mass ratio of $q = 10^{-3}$. Their result: less than $21 \times n\%$ of all the lensing stars have Jupiter-mass planets within $1 < a/\mathrm{AU} < 4$. An additional result of their analysis is: it is more efficient to observe *many events less densely* in time, than intensively monitoring only a small number of events

### New Theoretical Results

In a recent analysis, Gould, Gaudi, & Han (2003, 2004) looked into the question, how the different planet searching techniques fare in terms of sensitivity to Earth mass planets. Playing every method "to its strength", they found that only microlensing provides a realistic prospect with high signal-to-noise values for Earth-mass companions. In particular for orbital periods of order one year or larger, microlensing fares very well. Their conclusion is: microlensing has the best chances of all the methods studied for realistically detecting Earth-mass planets, with the above mentioned limitation that only the mass ratio is determined, which leaves the mass itself uncertain to within a factor of a few.

### New channels for planet detection

Already a few years ago, Di Stefano & Scalzo (1999a,b) had pointed out two other ways of finding planetary systems with microlensing: They showed that the microlensing signature of planets in wide orbits ($d > 1.5 R_E$) could be seen as an isolated event of short duration. They figured that a distribution of events by stars with wide-orbit planets is necessarily accompanied by a distribution of shorter events. What is very important: very accurate photometry is necessary ! Since the size of the star is comparable to the Einstein radius $R_E$ of the planet, the amplitude $\Delta m$ in the lightcurve will be low, the shape of the event is distorted and broader than the point source approximation (cf. Han et Kang, 2003). In addition, for very wide planetary orbits, there could also be *repeating* single-lens events, in case the track of the source relative to the lens passed both within an Einstein radius of the stars *and* the planet. These events will be rare, but they must occur, and hence previous stellar microlensing events should be monitored with higher frequency in the following observing seasons !

### Additional future ways for planet detection with microlensing

Recently, one additional aspect of planet microlensing was discussed: Ashton & Lewis (2001) looked into the question whether planets accompanying the source stars can be detected. They found that during a caustic crossing, the (reflected) light of the planet can be *very* highly magnified due to the very small size of this secondary source, and hence potentially be detected. The deviation is proportional to $f \times \theta_P$, where $f$ is the fraction of star light reflected by the planet, and $\theta_P$ is the angular radius of the planet in Einstein radii. They figured that even rings, satellites and atmospheric features on planets are detectable this way. Even in an optimistic scenario, though, it will take quite a number of years until such a measurement will be possible. But it is a very exciting possibility.

### 4.5 When will planets be detected with microlensing ? THE PROSPECTS

Considering that ...

- ... both OGLE and MOA have improved their alert efficiencies considerably, so that already now there are of order 1000 events per year measured,

- ... OGLE has implemented their early early warning system (EEWS),
- ... the PLANET team has improved their priority scheme for selecting between the events going on at the same time,
- ... PLANET and MicroFUN keep doing follow-up photometry with high sampling
- ... MOA follows very/extremely high magnification events,
- ... the new MOA 1.8m dedicated telescope is under construction, (first light planned for 2004/05),
- ... microlensing IS sensitive down to Earth masses,

... the question is WHEN rather then WHETHER planets will be detected with the microlensing technique. My answer is: SOONER rather than LATER. I am very optimistic that within the next 2 to 3 years at the latest, the first convincing planet will have been detected with the microlensing technique.

### 4.6 Note added in April 2004 (about one year after the 33rd Saas Fee Advanced Course)

During the 33rd Sass Fee Advanced Course on Gravitational Lensing – which took place in April 2003 – the author (J.W.) had offered a bet (which was accepted by one of the student participants) that the first convincing detection of an extrasolar planet with the microlensing technique would take place within 12 months time.

Indeed, on a NASA press conference in April 2004, it was announced that MOA/OGLE/MICROfun had detected a microlensing event which can be explained only with a very low mass companion to the primary star: OGLE 2003-BLG-235 or MOA 2003-BLG-53. The result is published meanwhile as Bond et al. (2004), see also Fig. 24. In the original words of the authors:

> "A short-duration (∼7 days) low-amplitude deviation in the light curve due to a single-lens profile was observed in both the MOA and OGLE survey observations. We find that the observed features of the light curve can only be reproduced using a binary microlensing model with an extreme (planetary) mass ratio of $0.0039^{+11}_{-07}$ for the lensing system. If the lens system comprises a main-sequence primary, we infer that the secondary is a planet of about 1.5 Jupiter masses with an orbital radius of ∼3 AU."

The author considers this a very convincing planet microlensing event.

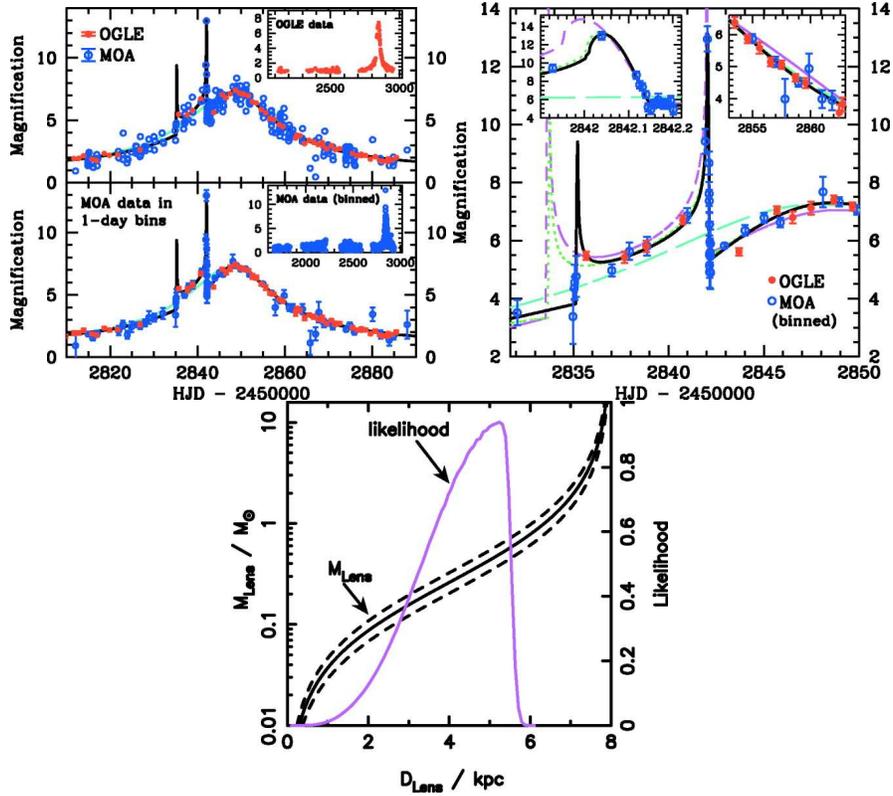

**Fig. 24.** Top left: Lightcurve of microlensing event OGLE 2003-BLG-235/MOA 2003-BLG-53: open (filled) symbols are MOA (OGLE) data points. Data points are shown individually in the top level, and binned in one-day intervals in the bottom panel. Top right: Data points and models covering about 18 days around the planetary deviation: long-dashed line – single lens case; short-dashed line – double lens with $q \geq 0.03$; solid line – best fit with $q = 0.004$. Bottom: Constraints and likelihood for the distance and mass of the lens: the thick solid line with the accompanying dashed lines as error limits shows the constraints on lens mass and distance from the measurement of the Einstein radius. The thin line (likelihood function) assumes the lens to be main sequence star (from Bond et al. 2004)

### 4.7 Summary

Microlensing as a planet search technique has stepped out of its infancy. It is a viable method which is complementary to other techniques. There is one unambiguous microlensing planet detection (Fig. 24; Bond et al. 2004) as of yet (April 2004): A star-plus-planet system with a mass ratio of $q = 0.004$. Furthermore, microlensing monitoring has put limits on the frequency

503

of Jupiter-like planets at semi-major axes between 1 AU and 4 AU around M-dwarfs: PLANET results show that less than one third of M-dwarfs host them (Gaudi et al. 2001). Soon these limits will be pushed further down, maybe to the few percent level. Stellar/binary microlensing lightcurves with > 1000 data points have been obtained: binary/planetary signatures can be covered with very high signal-to-noise: non-repeatability is no problem. With improved detection software: OGLE/MOA produce routinely more than 1000 alerts per year of events caught "in action". Implementation of EEWS (OGLE) guarantees verification of any signification deviation "on-the-fly", within 5 minutes ! Microlensing remains the most promising method for the detection of Earth-mass companions, either ground- or space-based.

# 5 Higher Order Effects in Microlensing:

As originally worked out by Einstein (1936) and Paczyński (1986b), the magnification of a point source by a point lens is a very simple function of impact parameter or time (see also Section 1.2). The first observed microlensing lightcurves were well fit by this functional form. However, the point-lens-point-source ansatz with a linear relative motion between source, lens and observer is clearly a mathematical idealisation.

In one sense, realistic situations are more complicated. In another sense, this helps us measure more/additional parameters and sometimes even break some of the degeneracies mentioned in previous sections. Some of the real world effects will be discussed here, for example:

- Blending - due to the dense star fields which are studied and the (very) large number of faint stars, often more than one star contributes to the light within the seeing disk. As a consequence, the measured microlensing lightcurve consists of two parts: a more-or-less constant background contribution (the "blending"), and the source star which is being microlensed.

- Parallax - For microlensing events with a duration of many months or longer (i.e. comparable to the orbital period of the Earth), the relative motion cannot be treated strictly as a straight line, but rather the changing observer position has to be considered. This leads to a modulation of the point source - point lens (PSPL)-lightcurve: it is not symmetric any more.

- Binary Lens - a binary lens clearly provides the most dramatic deviation from a PSPL-lightcurve: the caustic crossings are distinctly different features; this situation was already treated in Section 2. For a binary lens in a short-period orbit, the caustic configuration may change both its shape and position in the course of the microlensing event.

- Finite Source Effects, Limb Darkening, Star Spots - for small impact parameter/high magnification microlensing events and for caustic crossing in binary events, the finite size of the source has to be considered. It can strongly affect the lightcurve. During caustic crossings, the one-dimensional surface brightness profile of a background source can be determined from a careful analysis/fitting of the well-sampled lightcurve. In principle, even star spots can be determined this way.

- Direct Detection of the Lens - most (micro-)lenses presumably are stellar objects; due to the relative motion between lens and source (which can be measured in units of angular Einstein radii per time), the lens will move away from the position of the source. By selection, the light contributing to the lightcurve is dominated by the background (giant) star. However, if the lens is a faint/low-mass star, at some angular distance from the source star, it may become visible. From the spectrum and the stellar type, the mass and maybe even the distance of the lens can be estimated, and hence the degeneracies can be broken.

- Binary Source - the source can also be a physical binary system. The measured lightcurve is then a superposition of two PSPL-curves, in general with different impact parameters, different colors and different times of closest approach.

**Blending**

The microlensing monitoring programs need to cover as many stars as possible on one CCD frame. Hence they select dense star fields, e.g. towards the Galactic bulge. The typical angular separation between stars in such fields is much smaller than the seeing disk. Hence it is unavoidable that flux of more than one star is contributing to the light measured for the light curve. The blending can be due to a physical companion of the source star, due to the lens itself, or due to a random superposition of a star along the line-of-sight (which is too far away in units of Einstein angles to affect the point-lens lightcurve, but still within the seeing disk).

The source stars for the lensing events are usually giants in the Galactic bulge, they dominate the light. However, the additional 'blending' light cannot be entirely neglected. Di Stefano & Esin (1995) investigated this question. They concluded that the optical depth for lensing of giants is greater than for the lensing of main-sequence stars, and that this effect can be quantified. The direct consequence of blending is that the measured lightcurve is not represented by the ideal point-lens-point-source model lightcurve. Di Stefano & Esin (1995) present methods to test whether the deviation from a PSPL-lightcurve can be attributed to blending. They also suggest that the effect of blending can be used to learn more about the lensing event than would be possible otherwise (e.g., it could be that without the blend contribution, this particular star may not have been above the brightness threshhold at baseline and hence not among the list of monitored stars). If blending is neglected, the lens mass distribution will be skewed towards lower masses than the actual underlying distribution of lenses.

Wozniak & Paczyński (1997) point out a strong degeneracy of the fitting procedure for single lensing events between blended and non-blended events. They conclude that it is practically impossible to identify blending by photometric means alone. Some blends might be detected astrometrically, but the majority has to be corrected for statistically.

Alard (1997) analysed the situation in which the lensing event is not on the main star, but rather on an unresolved background star which represents only a 'blended' contribution to the light of the main (giant) star. He showed that such apparently short-duration events can be easily misinterpreted as brown-dwarf lensing events. Furthermore, Alard (1997) points out that there are ways to identify such events: usually, there is a color shift during the event. High resolution, dense, multi-band sampling helps identify such events and to estimate their contribution to the total lensing rates. He identifies OGLE-5 (Udalski et al. 1994) as an obvious such event. Another method to identify and 'deblend' such events was suggested by Goldberg (1998): the shift of the center-of-light due to one of the stars in the seeing disk being magnified produces an astrometric signature which should be measurable in a fair fraction of such events.

Han et al. (1998) found that the contribution of the lens to the blending (suggested by Nemiroff 1997) has a small to moderate effect on the determination of the optical depth (decrease of 20% under the most extreme circumstances) and the Galactic mass distribution. Han & Kim (1999) derived analytical relations between the lensing parameters with and without the effect of blending and investigated the dependence of the derived lensing parameters on the amount of blended light and the impact parameter. In Han et al. (2000), it is shown that the difference image analysis method (Alard & Lupton 1998) is a very efficient way for the astrometric deblending of microlensing events, which was further developed by Gould & An (2002).

**Parallax Effects**

For microlensing events with a duration of many months or longer (i.e. comparable to the orbital period of the Earth) and a relative velocity between source, lens and observer comparable to (or smaller than) the orbital velocity of the Earth, the relative motion cannot be treated as a straight line any more. Rather, the changing observer's position in the course of the microlensing event influences the shape of the lightcurve: the PSPL-lightcurve is modified, it is not symmetric any more. Such events were predicted by Refsdal (1966) and Gould (1992), with the suggested applications to get more information and constraints on mass and transverse velocities of the lenses.

The first such event observed was reported by Alcock et al. (1995). It is the longest of their 45 microlensing events detected towards the Galactic bulge in their first year of observation (Alcock et al. 1997). In Fig. 25, the B-band and R-band lightcurves are shown, together with the best fit assuming only linear motion (dashed line), and the best fit including the motion of the Earth (solid line). Whereas the former clearly shows systematic deviations, the latter provides a very good fit. Since the event is achromatic, there is little doubt that this is a bona fide microlensing event, despite the deviation from the symmetric PSPL-shape.

Alcock et al. (1995) discuss the nature of the deviation from linear motion and emphasize that it is impossible to distinguish between the possibilities that motions of lens, source or observer lead to this modification. However, with the knowledge of the orbital parameters of the Earth, they tried to fit the lightcurve and argued that their reasonable fit with these assumptions is a strong argument in favour of Earth's motion really causing the deviations. This then allows them to compare the projected Einstein ring diameter crossing time with the size of the Earth orbit and hence obtaining a second constraint on the three unknown parameters of a typical microlens situation: lens mass $M$, lens distance $D_d$ and relative transverse velocity $v_t$.

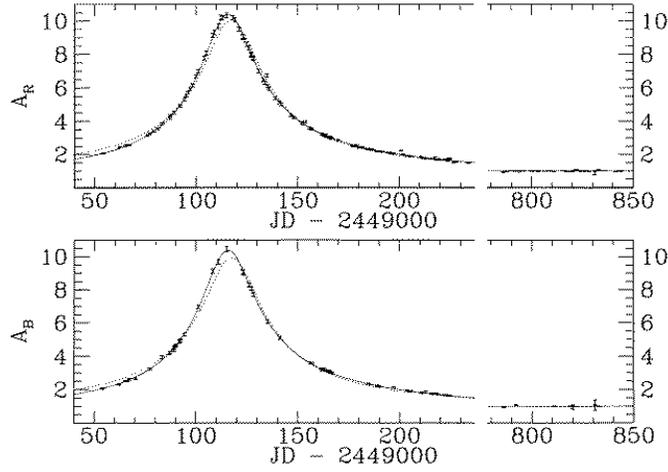

**Fig. 25.** Lightcurve of the first detected parallax lightcurve in R band (top) and B band (bottom); the (linear) magnification is shown as a function of time in days from JD 2,449,000. The dashed curve shows the best linear velocity point-lens-point-source fit, the solid line is the best fit allowing for the parallax effect, the motion of the Earth around the Sun, from Alcock et al. (1995)

In order to include the orbital motion of the Earth, the expression of the impact parameter as a function of time $u(t)$ gets more complicated than the standard form (Alcock et al. 1995):

$$\begin{aligned}
u^2(t) = \quad & u_0^2 + \omega^2(t-t_0)^2 + \alpha^2 \sin^2[\Omega(t-t_c)] \\
& + 2\alpha \sin[\Omega(t-t_c)][\omega(t-t_0)\sin\theta + u_0\cos\theta] \\
& + \alpha^2 \sin^2\beta \cos^2[\Omega(t-t_c)] \\
& + 2\alpha \sin\beta \cos\theta[\Omega(t-t_c)][\omega(t-t_0)\cos\theta - u_0\sin\theta],
\end{aligned} \qquad (15)$$

where $\theta$ is the angle between the velocity vector $v_t$ and the north ecliptic axis, the angular frequency $\omega = 2/\hat{t}$, and $t_c$ is the time at which the Earth is closest to the line connecting Sun and source. The parameters $\alpha$ and $\Omega$ are defined as:

$$\alpha = \frac{\omega(1\mathrm{AU})}{\tilde{v}}\{1 - \epsilon \cos[\Omega_0(t-t_p)]\} \qquad (16)$$

and

$$\Omega = \Omega_0 + \frac{2\epsilon \sin[\Omega_0(t-t_p)]}{(t-t_c)}. \qquad (17)$$

Here $t_p$ is the time of the perihelion, $\tilde{v} = v_t/(1-x)$ is the transverse speed of the lensing object projected to the solar position, $\Omega_0 = 2\pi/yr$ and $\epsilon = 0.017$ is the eccentricity of the Earth motion. This inclusion of the Earth's motion

508

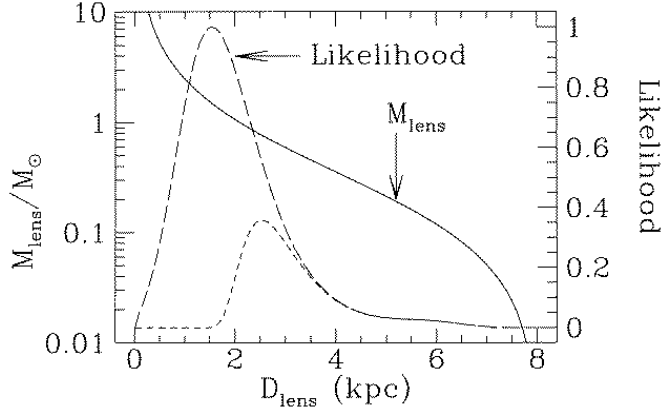

**Fig. 26.** Lens mass versus lens distance (solid line, left scale) and likelihood function for lens distance, using projected velocity and Galactic model (long-dashed curve, right scale) plus upper limit on brightness from a main-sequence lens (short-dashed line, right scale), from Alcock et al. (1995)

into their fitting procedure reduced the $\chi^2$ per degree of freedom from roughly 10 to a value of order unity (for 206 degrees of freedom; more details see in Alcock et al. 1995).

With the additional parameter $v_t = \tilde{v}(1-x)$ replaced in the equation defining the Einstein crossing time, one obtains for the mass of the lens M(x) as a function of the lens distance:

$$M(x) = \frac{1-x}{x} \frac{\tilde{v}^2 \hat{t}^2 c^2}{16 G L}, \tag{18}$$

which is displayed in Fig. 26. This curve shows that the lens could, e.g., be a low mass object (brown dwarf) in the Galactic bulge at large $D_{\text{lens}}$, or an M dwarf in an intermediate distance range of 2 kpc to 6 kpc, or a solar type star (or even higher mass) nearby. However, from the limits on the apparent brightness of the lens (as a massive main sequence star only milli-arcseconds away from the background lensed star, it should contribute a significant amount of light within the seeing disk of the latter) one can constrain the mass at the upper end.

Alcock et al. (1995) tried to use even more constraints, namely on the velocities of lens and source and obtained two likelihood functions (dashed lines in Fig. 26) for the distance of the lens, based on some reasonable velocity limits. The most likely distance of the lens appears to be $D_L = 1.7^{+1.1}_{-0.7}$ kpc, corresponding to a mass range of $M = 1.3^{+1.3}_{-0.6} M_\odot$. With the assumption that the lens is a main sequence star, the constraints are slightly different: $D_{L,MS} = 2.8^{+1.1}_{-0.6}$ kpc and $M = 0.6^{+0.4}_{-0.2} M_\odot$.

For another case, Mao (1999) reported about an ongoing microlensing event toward the Carina spiral arm, discovered by the OGLE team (Udalski et al. 1998): OGLE-1999-CAR-1 (see Fig. 27). He showed that this long duration event exhibits strong parallax signatures, and determines the lens transverse velocity projected onto the Sun-source line to be about 145 km/s.

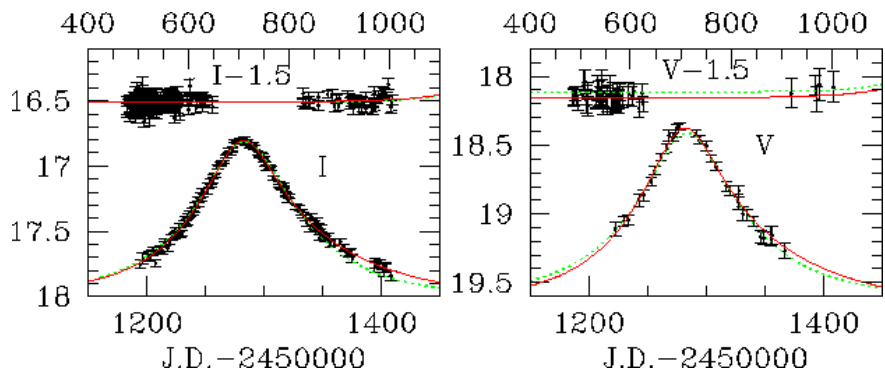

**Fig. 27.** Lightcurves of event OGLE-1999-CAR-1 in I band (left) and V -band (right): dotted lines are best fit PSPL-fits (linear motion), whereas the solid lines are best fits including both parallax and blending (from Mao (2001); more details there)

In a systematic search for parallax signatures among 512 OGLE-II microlensing from 1997-1999, Smith, Mao & Wozniak (2002a) fitted both standard linear motion models and parallax models which included the motion of the earth around the Sun. Using additionally information on the duration of the events, they identified one convincing new candidate, sc33_4505, which is caused by a slow-moving and likely low-mass object, similar to other known parallax events (see Fig. 28). Smith et al. (2002a) emphasize that irregular sampling and gaps between observing seasons hamper the recovery of parallax events.

The first multi-peak parallax event (predicted by Gould 1994) was published by Smith et al. (2002b): the highly unusual microlensing lightcurve of OGLE-1999-BUL-19 (Fig. 29) exhibits multiple peaks which are *not* caustic crossings. The Einstein radius crossing time for this event is approximately 1 yr, which is unusually long. Smith et al. (2002b) show that a simple explanation for these additional peaks in the light curve is the parallax motion of the Earth. The fact that this effective transverse velocity between lens and source is significantly lower than the speed of the orbit of the Earth around the Sun ($v_{\mathrm{Earth}} \approx 30$ km/s) results in a periodic modulation of the impact parameter, superimposed on the linear motion: the motion of the Earth induces these multiple peaks. Smith et al. (2002b) also discuss binary-source signature but conclude that this is a less likely explanation.

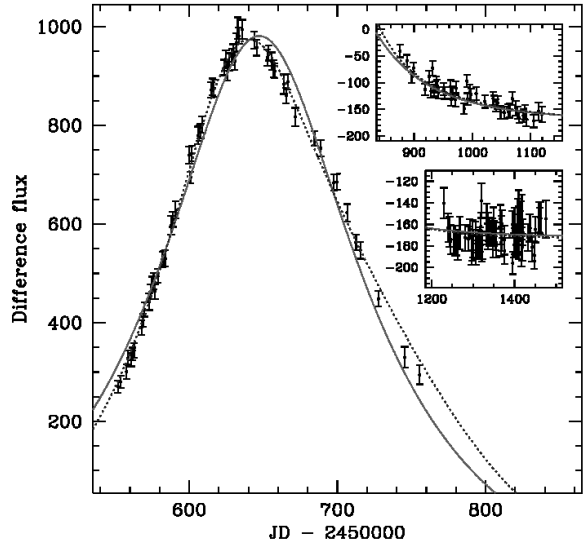

**Fig. 28.** Lightcurve of event OGLE-II event sc33_4505 towards the Carina spiral arm. The solid line is the best linear motion PSPL-fit, whereas the dotted line is the best parallax fit (from Smith et al. (2002a))

In Fig. 29, the lightcurve of the multi-peak event OGLE-1999-BUL-19 is shown (bottom panel). The top panel shows the modulated apparent motion of the lens, projected in the observer plane relative to the observer-source line-of-sight, i.e. the location of the lens with respect to the Earth (denoted by the small cross).

**Short-period Binaries**

In Section 2, lensing by binary stars is considered. There, only the static situation is discussed: the lens configuration is assumed constant during the microlensing event, i.e. the binary period is much larger than the crossing time. However, this assumption will not always be true, there are small-separation binaries with periods of order years, months, or days.

Dominik (1998) investigated this case and discussed three scenarios: the rotating binary lens, rotating binary source, and observer on Earth orbiting the Sun (parallax, see above discussion). The most dramatic effects are expected in the case of a rotating binary lens, because the caustic structure changes with time. In the other two scenarios, the caustic configuration is static, the effect is only a modulation of the straight relative motion (parallax) and/or the superposition of two "static" lightcurves, which might cause

511

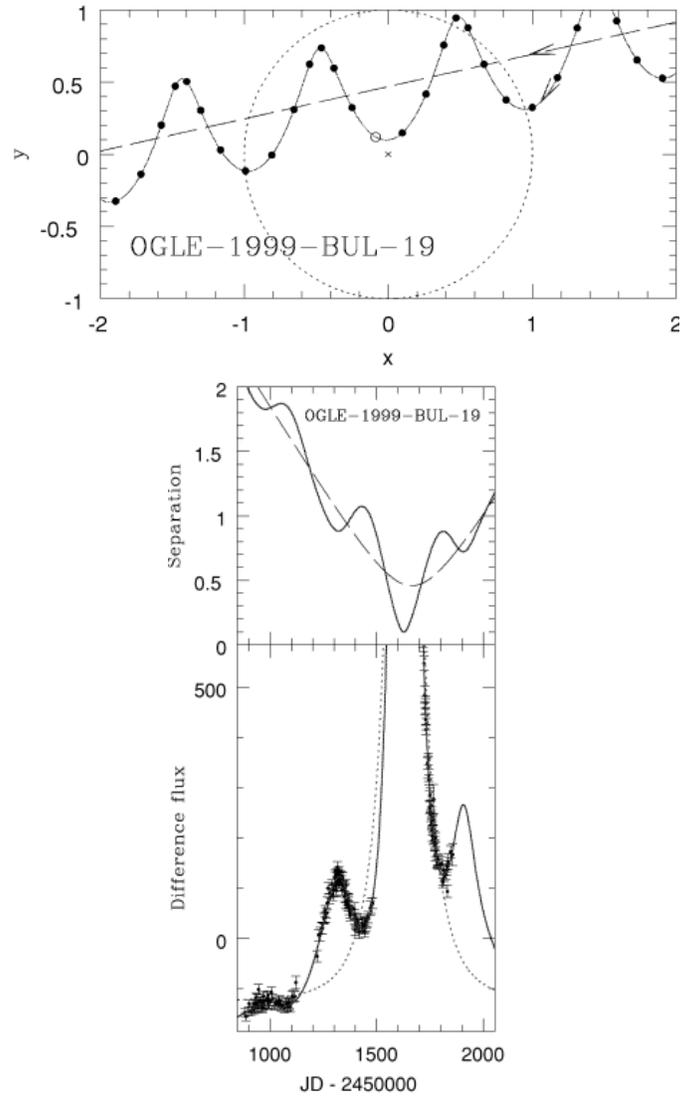

**Fig. 29.** Multi-peak lightcurve OGLE-1999-BUL-19: lens position relative to the observer as a function of time (top panel) and lightcurve plus corresponding impact parameter as a function of time (bottom panel) with best fit linear motion PSPL-curve (dashed) and best fit parallax model (solid), from Smith et al. (2002b)

some color changes as well (cf. Griest & Hu, 1992). In this sub-section, only the rotating lens will be discussed further.

Dominik (1998) shows that the scenario of a rotating binary introduces five additional parameters, compared to a static binary: two rotation angles, the rotation period, the eccentricity and the phase. In Fig. 30, the effect of binary motion is illustrated on a lightcurve comparable to that of the event MACHO-LMC-1: the binary period varies between 365 days and 25 days. It is obvious that the effect of binary rotation is most pronounced for short binary periods, compared to the event duration.

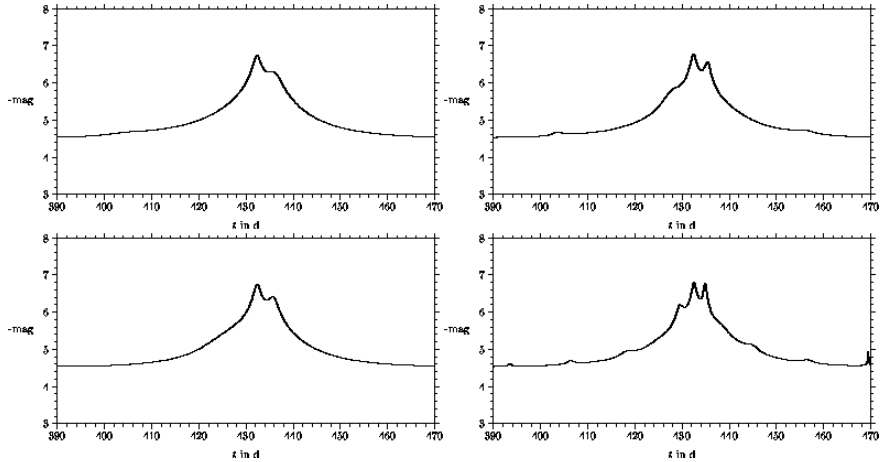

**Fig. 30.** Simulated lightcurves for a rotating binary lens, based on MACHO-LMC-1: binary rotation periods decrease from 365 days (top left) to 100 days (top right), 50 days (bottom left) and 25 days (bottom right), from Dominik (1998)

In 2000, the first detection of a rotating binary lens was published (Fig. 31): the lightcurve of MACHO 97-BLG-41 was the first event with a source crossing two physically distinct caustics (Albrow et al. 2000). Analysing PLANET data for MACHO 97-BLG-41 (46 V-band and 325 I-band observations from five southern observatories), Albrow et al. (2000) showed that this data set is incompatible with a static binary lens. They do find a good model with a rotating binary lens of mass ratio q = 0.34 and angular separation $d = 0.5R_E$. The binary separation changes significantly in size during the 35.17 days between the separate caustic transits. Albrow et al. (2000) use this event to derive the first kinematic estimate of the mass, distance, and period of a binary microlens. The relative probability distributions for these parameters peak at a total lens mass of $M \approx 0.3 M_\odot$, which would imply an M-dwarf binary system. The most likely lens distance is $D_L \approx 5.5$ kpc, and the binary period is $P \approx 1.5$ yr.

What made this model particularly convincing is the following: MACHO/GMAN data covering several sharp features in the light curve which

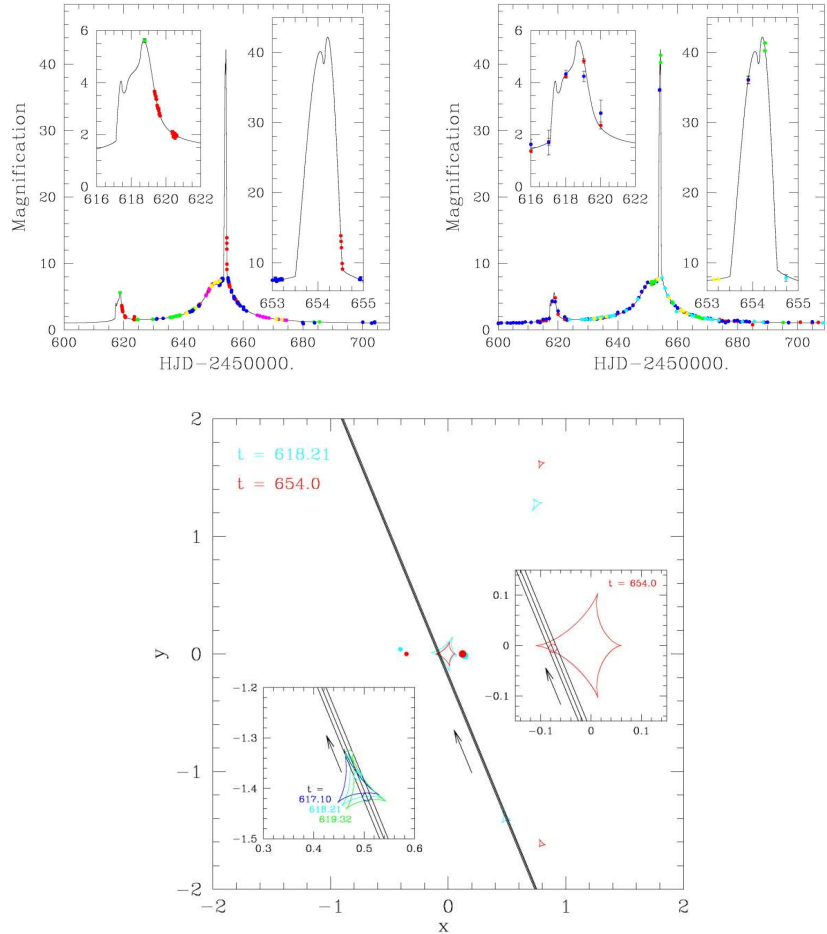

**Fig. 31.** Rotating binary event MACHO 97-BLG-41: a) PLANET data (points) and best-fit rotating binary model (solid line); inset enlarges the two caustic crossing regions (top left); b) Same best-fit line as in a), but here the data points of the MACHO/GMAN collaboration are added which did not enter the modelling procedure (top right); c) Caustic topology of best-fit rotating binary model, shown at time close to first and second caustic crossing. Straight line shows source trajectory. The positions of the two binary lens components are shown as large/small dots. Insets show regions close to the caustics, with the two additional lines indicating the finite source size (bottom) (from Albrow et al. 2000)

are not probed by the PLANET observations and which did not enter the modeling, fall almost perfectly on the best fit lightcurve. This event MACHO 97-BLG-41 (see Fig. 31) had previously been modelled by a static binary lens

514

plus a planetary companion. This much simpler and more robust rotating binary model makes a fit using a third lensing body less plausible.

**Finite Source, Limb Darkening, Star Spots**

When the source size is small compared to the Einstein radius of the lens *and* to the impact parameter, then the point source approximation is justified in a single lens scenario. However, if one of these conditions is violated, then the microlensing lightcurve is affected by the finite source.

Nemiroff & Wickramasinghe (1994) were the first to investigate this. They show that the central peak of the lightcurve is modified if the impact parameter $u_0$ is smaller than the source radius: Fig. 32 (top) shows a circular source with uniform surface brightness crossing the point caustic behind a point lens centrally (top arrow and top lightcurve) and just barely within the source diameter (lower arrow and lightcurve).

The central peak of the microlensing lightcurve is clearly lower and broader than for a point source. Nemiroff & Wickramasinghe (1994) pointed out that an exact determination of the deviation from the PSPL-lightcurve can be used to determine the time it took the stellar disk to cross the central point caustic. With an independent determination of the source radius from knowledge of the stellar type and use of stellar evolution theory, this could be used to determine the transverse velocity. Witt (1995) estimated that at least 3% of all microlensing events in the Galactic bulge will be affected by finite source effects. He defined "being affected" by angular impact parameter being smaller than the angular source radius.

Peng (1997) looked into this question in more detail and presented the effects for a limb-darkened finite source with radius $r = 0.055 R_E$, corresponding to a star with $R = 10 R_\odot$ at $D_S = 9$kpc and a $0.1 M_\odot$-lens at $D_L = 8$kpc. Fig. 32 (bottom) shows lightcurves for such a source with four different impact parameters: $u_0 = 0.00, 0.055, 0.2$ and $0.5$. He finds that the source size can be fitted with reasonable accuracy only if the impact parameter $u_0$ of the event is smaller than the stellar radius.

The first observed finite source effect was reported by Alcock et al. (1997). Their lightcurve of MACHO Alert 95-30 shows significant deviations from the point source lightcurve near the peak (cf. Fig. 33). They could determine the ratio between impact parameter and stellar radius to $u_0/R_* = 0.715 \pm 0.003$. With additional spectroscopic and photometric information they could identify the source as an M4 III star with a radius of $R = (61 \pm 12) R_\odot$ located at the far side of the Galactic bulge at about $D_S \approx 9$ kpc. The lens angular velocity could be determined relative to the source, to $(21.5 \pm 2.9)$ km/sec/kpc. With a likelihood analysis, the lens mass was determined to $m_L = 0.67^{+2.53}_{-0.46} M_\odot$.

Yoo et al. (2004) analysed the short-duration event OGLE-2003-BLG-262, $t_E = (12.5 \pm 0.1)$ day. The lens is identified as a K giant in the Galactic bulge. The finite-source effects are used to measure the angular Einstein radius to

515

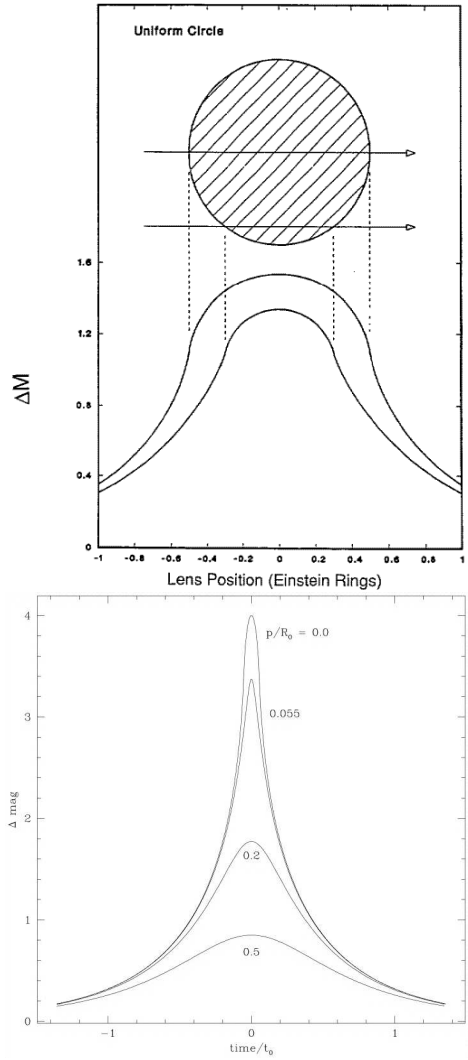

**Fig. 32.** Finite source effect: a) central part for lightcurves with zero or very small impact parameter (left, from Nemiroff & Wickramasinghe 1994); b) lightcurves for a limb darkened source with radius $r = 0.055R_E$, corresponding to star with $R = 10R_\odot$ at $D_S = 9$kpc and a $0.1M_\odot$-lens at $D_L = 8$kpc; four different impact parameters: $u_0 = 0.00, 0.055, 0.2$ and $0.5$ (right, from Peng 1997)

be $\theta_E = (195 \pm 17)\mu$as. The lens mass could be constrained to the FWHM interval $0.08 < M/M_\odot < 0.54$, and the lens-source relative proper motion to $v_{\rm rel} = (27 \pm 2)$ km/s/kpc.

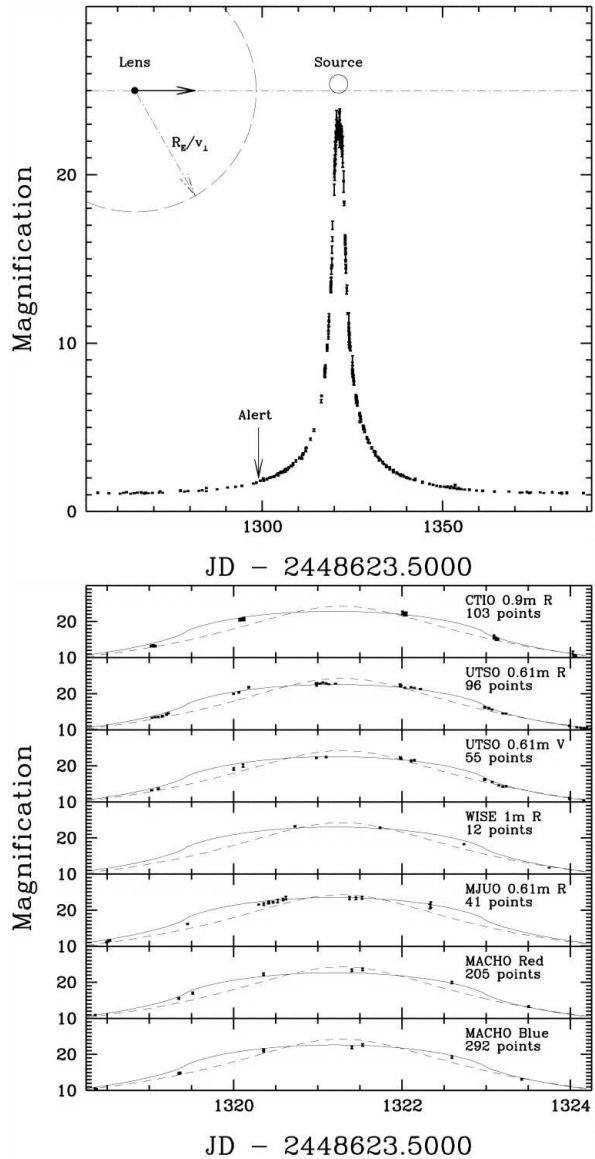

**Fig. 33.** Finite source effect: Full lightcurve of MACHO Alert 95-30 (top) and data close to the central peak (bottom) with point source and extended source fit; the arrow indicates when the alert was sent out (from Alcock et al. 1997)

As described in the context of binary lenses, the lightcurve of a caustic crossing event offers the opportunity to determine the size of the source star, and even the surface brightness profile, i.e. to measure limb darkening of a star

517

that is many kpc away ! This was successfully applied for the first time by the PLANET team on the event MACHO 97-BLG-28 (Albrow et al. 1999). The source star could be spectroscopically identified as a K giant. The observed lightcurve (Fig. 34) was modelled as being due to a cusp crossing of a binary lens caustic (cf. Fig. 35).

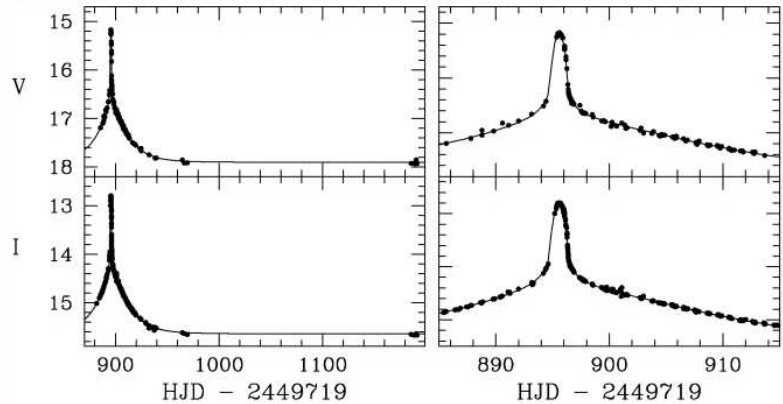

**Fig. 34.** PLANET lightcurve of event MACHO 97-BLG-28, covering a 300 day period (left) and a zoom of 30 days around the maximum (right). The V band lightcurve (top) consists of 155 data points, the I band lightcurve (bottom) comprises 431 data points. (from Albrow et al. 1999)

Modelling of the lens system resulted in a binary with mass ratio $q = 0.23$, and an instantaneous projected separation of $d = 0.69$ (for a lens in the Galactic bulge this corresponds to roughly 1 to 2 AU). The very good coverage of the lightcurve (696 data points in V and I from PLANET observatories in Chile, South Africa and Australia) made it possible to determine the radial surface brightness profile of the source star in the Galactic bulge. In particular the sharp central peak could be monitored with a time resolution of 3 to 30 minutes. The analysis resulted in a determination of the square-root limb darkening coefficient: for the assumed two parameter limb darkening law

$$I_\lambda(\theta) = I_\lambda(0) \left[1 - c_\lambda(1 - \cos\theta) - d_\lambda(1 - \sqrt{\cos\theta})\right], \qquad (19)$$

where $\theta$ is the angle between the normal to the stellar surface and the line-of-sight, and $I_\lambda$ is the intensity for wavelength $\lambda$, the parameters were determined for the two filters to $c_I = 0.40$, $d_I = 0.37$, and $c_V = 0.55$, $d_V = 0.44$. These values are in excellent agreement with the predictions for K giants from numerical modelling of stellar atmospheres. A source profile with a uniform surface brightness could be strongly ruled out.

Another example is event OGLE-1999-BUL-23 (Albrow et al. 2001), a binary lens system as well, for which modelling resulted in a mass ratio $q =$

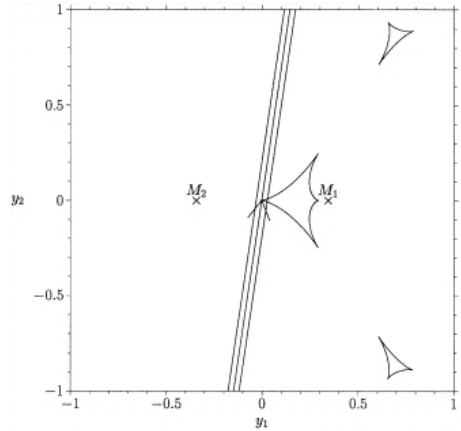

**Fig. 35.** Binary lens configuration (masses $M_1$ and $M_2$), caustic configuration and source track (width corresponds to diameter) of event MACHO 97-BLG-28 (from Albrow et al. 1999)

0.39 and an instantaneous projected separation of $d = 2.42$. The source star is assumed to be a G/K subgiant in the Galactic bulge with an effective temperature of $T_{\text{eff}} \approx 4800K$. The resulting limb darkening coefficients (a different limb darkening law was applied here) are consistent as well with theoretical predictions.

The possibility to detect star spots with microlensing has been explored by Hendry et al. (2002). It turns out that this will indeed be possible, though not in the immediate future. With sufficiently well-sampled lightcurves and high photometric accuracy, stellar microlensing will at least be able to put interesting constraints on the presence or absence of photospheric star spots.

**Direct Lens Detection**

In December 2001, for the first time the direct image of a lensing object in a microlensing event was published. Alcock et al. (2001) reported the photographic image of a second object very close to the source star of the microlensing event LMC-5. The microlensing event had its maximum on February 5, 1993. Due to the relative motion of lensing object and source star, the angular separation was expected to increase with time. If the lens happens to be an ordinary main sequence star (rather than a "dark object"), there is a chance that after some time it will become visible next to the source star. The exact time cannot be predicted, because neither the transverse velocity nor the distance is known.

The MACHO team had successfully proposed for HST time to take very high resolution images of their previous microlensing events in the direction towards the LMC. On May 13, 1999 an image was taken of this particular source star, and 6.3 years after the peak in the microlensing lightcurve, this HST picture (see Fig. 36) revealed a "faint, red object displaced by 0.124 arcsec from the centre of an LMC main-sequence star that, on the basis of previous analysis, is the source star of this event" (Alcock et al. 2001)[12].

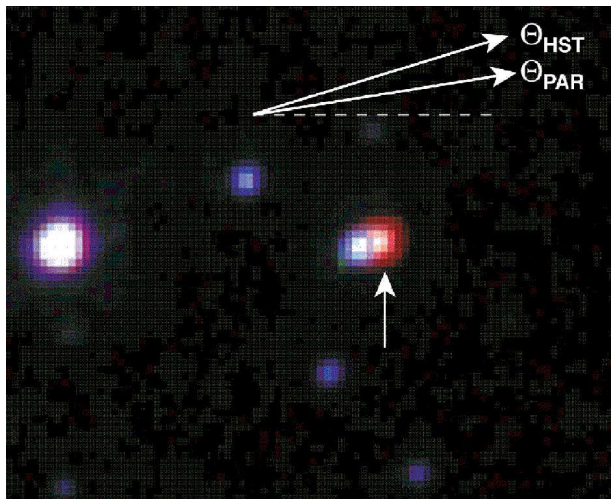

**Fig. 36.** Image of the region near the source star of microlensing event LMC-5 (peaked at February 5, 1993), taken with the HST WFPC2 camera on May 13, 1999. It reveals a faint red object (marked with an arrow) 0.124 arcsec to the top right from the center of the blue source star. The two arrows at the top ($\Theta_{\rm HST}, \Theta_{\rm PAR}$) indicate the direction of the lens motion determined by two methods (from Alcock et al. 2001)

The event LMC-5 had been a very high magnification and hence a very small impact parameter event. So it is safe to assume that at the time of the peak in the microlensing lightcurve, lens and source "coincided". So the relative proper motion of the lens can be easily determined to $\mu_{\rm rel} = (0.0214 \pm 0.0007)$ arcsec/yr.

Alcock et al. (2001) show that the HST and a parallax fit to the lightcurve data (which yields the transverse velocity projected on the observer's plane $\tilde{v} = -18.42^{+1.83}_{-1.91}$AU/yr) can be combined for a complete solution for this lens system. The lens mass is expressed by the observed parameters:

---

[12] The authors emphasize that one normal disk star among the lenses of the 13 to 18 microlensing events towards the LMC is consistent with the number of expected "foreground" events (compared to the majority of the events which may be produced by MACHOs and which should correspond to no visible lens star in the future).

$$m_L = \frac{c^2}{16G}\tilde{v}\hat{t}^2\mu_{\rm rel}. \qquad (20)$$

This yields a value of $M_L = 0.039 M_\odot$, with a $3\sigma$ upper limit of $M_L \leq 0.097 M_\odot$. This is at or below the low end of the stellar mass scale.

In addition, the relation between the relative proper motion ($\mu_{\rm rel}$ in arcsec/yr) with the relative parallax ($\pi_{\rm rel}$, in arcsec) and the transverse velocity ($\tilde{v}$, in AU/yr) provides an estimate for the lens parallax:

$$\tilde{v}/\mu_{\rm rel} = 1/\pi_{\rm rel} \approx 1/\pi_{\rm L}. \qquad (21)$$

The last approximate relation is valid because the lens as a disk star is much closer than the source star in the LMC. The lens parallax gives the distance to the lens: $d_L = \pi_L^{-1} \approx 200^{+40}_{-30}$pc. This leads to an absolute magnitude of $M_V = 16.2^{+0.6}_{-0.5}$. Finally a spectrum taken with the ESO VLT telescope revealed the star to be an M4-5 dwarf in the mass range $M_{\rm M4-5V} \approx (0.095 - 0.13)M_\odot$, inconsistent at the $3\sigma$ level with the parallax determined lens mass ! A photometric distance based on an empirical relation between color and absolute magnitude for M dwarfs leads to a distance determination of $D_{\rm M4-5V} \approx (650 \pm 190)$pc, inconsistent at the $2\sigma$ level with the parallax distance. It was obvious immediately that this apparent conflict should be resolvable with future HST ACS imaging.

The solution of this puzzle came about soon: Drake, Cook & Keller (2004) had used the new HST's Advanced Camera for Surveys (ACS) in order to get very high resolution images of lens and source stars of the MACHO-LMC-5 event. They measure the parallax and determine the distance of the lens star to $D_{\rm L} = 578^{+65}_{-53}$pc, and the proper motion to $\mu = (21.39 \pm 0.04)$mas/yr. They conclude that the lens is an M dwarf which is more likely to be part of the thick disk than the thin disk population. In particular, they confirm Gould's (2004) suggestion that the event MACHO-LMC-5 is a "jerk-parallax" event. Gould had found a second solution to the microlens parallax which was different from the one presented in Alcock et al. (2001).

Furthermore, Nguyen et al. (2004) obtained infrared images of the MACHO-LMC-5 region with the newly launched Spitzer Telescope. Their photometry established an infrared excess, hence confirming that the lens is a M5 dwarf star with a mass of about 0.2 $M_\odot$.

# 6 Astrometric Microlensing

In the point-lens-point-source scenario, there are always two images of a background star. Only when the impact parameter is of order a few Einstein radii, the secondary image very close to the point lens is magnified enough to become important (cf. equation 3). Usually, only the combined magnification of the two images is measured, because the separation of the images is of order milli-arcseconds, unresolvable in most circumstances. Looking into the exact geometrical arrangement of image positions relative to the lens, one sees that the two images, the lens and the source always form a straight line in a point-lens-point-source scenario, as is illustrated in Fig. 38 (top panel). The line rotates by almost 180 degrees in the course of the lensing event. The corresponding center-of-light lies on this straight line as well (Fig. 38, bottom panel).

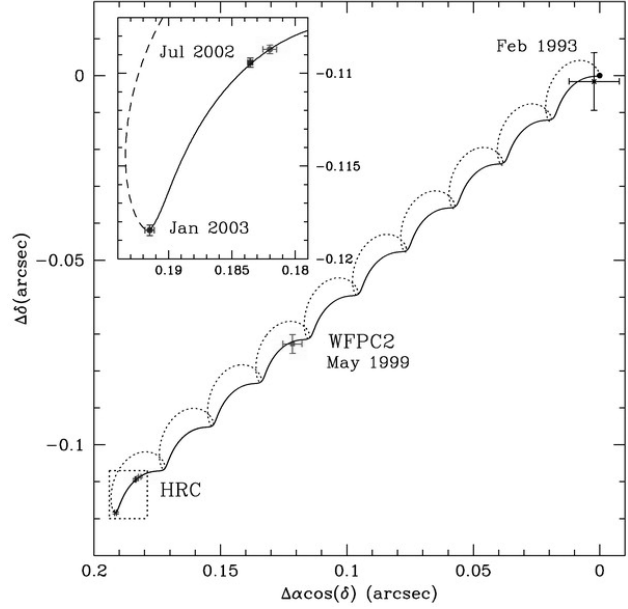

**Fig. 37.** Motion of the lensing star relative to the lensed star at the time of closest approach of event LMC-5. The inset is a zoom and shows the three measurements of the HST HRC camera in 2002. The source star is at location (0.00,0.00) (from Drake et al. 2004)

Paczyński (1998) has shown that the light centroid is displaced relative to the source position by a maximum of

$$\delta\phi_{\max} = 8^{-0.5}\phi_E \approx 0.354\phi_E \quad \text{for} \quad u_{\min} = \sqrt{2}. \tag{22}$$

In Fig. 39, three tracks are shown for $u_{\min} = 0.2$, $\mu_{\max} = 5.07$, $D_S = 50$ kpc, $D_L = 10$, 30 and 45 kpc, and $m = 0.3 M_\odot$.

The fascinating aspect of this center-of-light variation is: if it is measured, the mass of the lens can be determined, the degeneracies can be broken. Combining the definition of the angular Einstein radius

$$\phi_E \approx 0.902 \text{mas} \left(\frac{M}{M_\odot}\right)^{1/2} \left[10\text{kpc}\left(\frac{1}{D_d} - \frac{1}{D_s}\right)\right]^{1/2} \tag{23}$$

with the relative parallactic motion with the angular amplitude (in radians)

522

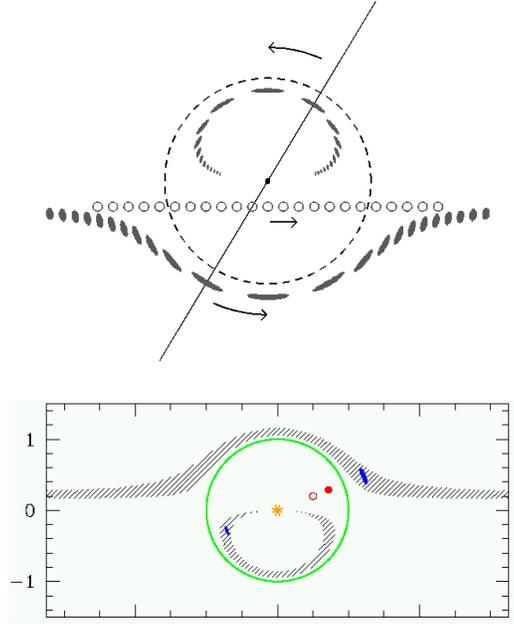

**Fig. 38.** Top: Alignment of the two images with the lens in a point-lens-point-source scenario: the source moves in a straight line in the background from left to right (open circles), the corresponding images (one inside, one outside the dashed Einstein circle), the source position and the lens lie on a straight line which apparently rotates around the lens position, here anticlockwise because the source position is *above* the lens (image courtesy Penny Sackett). Bottom: Similar situation, here the center-of-light is indicated (filled circle) for one particular arrangement (image courtesy Scott Gaudi)

$$\pi_{ds} = 1\mathrm{AU}\ \left(\frac{1}{D_d} - \frac{1}{D_s}\right), \tag{24}$$

(assuming linear relative motion between source and lens), one obtains

$$M = 0.123 M_\odot \frac{\phi_E^2}{\pi_{ds}}, \tag{25}$$

where $\pi_{ds}$, is the measurable parallax of the lens-source !

Currently, the astrometric resolution of the best telescopes is of order 0.1 arcsec. Considering that light centroids can be determined to an accuracy of about 10% of that value, this still leaves us with a 10 milli-arcsec positional accuracy, at least two orders of magnitudes larger than the expected tens or hundreds of micro-arcsec from the astrometric microlensing.

However, this is not a lost case: there are instruments at the horizon which will make such a measurement possible: VLTI (Very Large Telescope Interfer-ometry) and SIM (Space Interferometry Mission). The VLTI uses the combination of the four 8.2m-ESO telescopes in their interferometric mode, making use of the maximum 200m baseline between the unit telescopes. SIM, on the other hand, is a satellite project which will do astrometry with unprecedented accuracy: the specifications state that SIM will be able to do wide-angle astrometry with a nominal accuracy of 4 micro-arcsec, in the narrow-angle mode the accuracy should be as high as 1 micro-arcsec for a 20 mag star ! In addition, parallaxes will be measured with an accuracy of about 5 micro-arcsec, and proper motions down to 2 micro-arcsec/yr. Launch is planned for 2010, according to the web site `http://sim.jpl.nasa.gov`, where also much more information can be found.

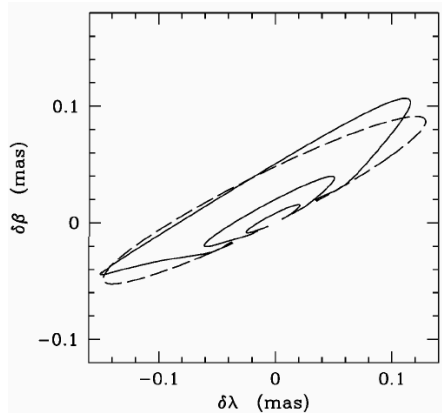

**Fig. 39.** Astrometric displacement (in ecliptic coordinates) caused by three microlensing events described in the text: The source is assumed to be a star in the LMC at a distance of 50 kpc, the lens has a mass of $M = 0.3 M_\odot$, and the three solid lines correspond to three lens distances of $D_L = 10, 30$ and $45$ kpc (the largest lens distance produces the innermost curve). The impact parameter is $u_{min} = 0.2$, corresponding to a maximum magnification of $\mu_{max} = 5.07$. The largest displacement of the curves is $8^{1/2}$ times smaller than the corresponding Einstein radius. The microlensing time scale is $t_0 = 50$ days in all three cases. The dashed curve corresponds to the $D_L = 10$ kpc case, with the effect of Earth's orbital motion suppressed (from Paczyński 1998)

Paczyński (1998) had derived the above relations and values for the SIM mission: SIM can measure the astrometric displacements of the light centroid of microlensing events which are discovered/detected photometrically from the ground. The amplitude of the center-of-light variation can reach a few tenths of the Einstein radius. Such a measurement will make it possible to determine the mass, the distance and the proper motion of almost any star or MACHO, capable of inducing a microlensing event towards the Galactic bulge or the LMC/SMC.

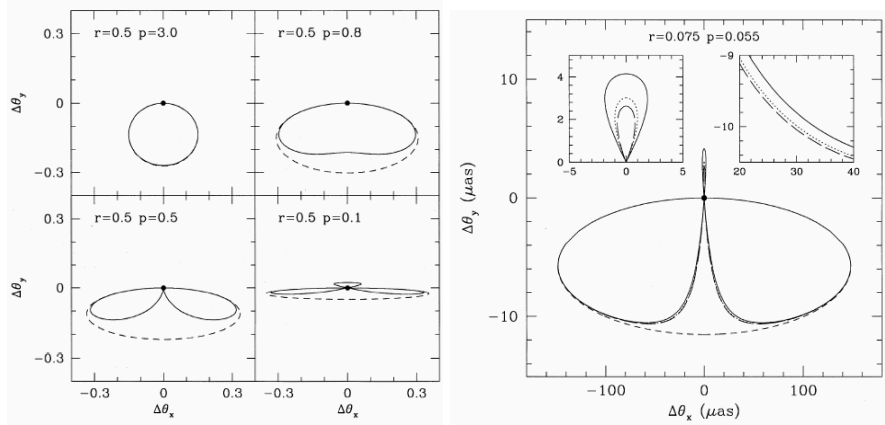

**Fig. 40.** Left panel: Four examples of astrometric trajectories showing finite source size effects: The source size is assumed to be $\theta_s = 0.5\,\theta_E$. In each example, the black dot marks the source position and the lens is moving from $-\infty$ to $+\infty$, parallel to the x-axis. The centroid motion starts at the origin and moves counter-clockwise. The impact parameter (here labelled $p$) of each trajectory is shown at the top left corner in each panel $p = 3.0, 0.8, 0.5, 0.1$. For each example, the solid line shows the trajectory that takes into account the finite source size effect while the dashed lines shows that for a point-source approximation. Right panel: Simulated astrometric trajectories for the first microlensing event (95-BLG-30) that shows photometric extended source effects (parameters taken from Alcock et al. (1997)). The dashed ellipse is the centroid motion for a point source. The solid line shows the trajectory for a source with constant surface brightness, whereas the dotted and long-dashed lines show the predictions for the MACHO R and V bands, respectively. The two insets show magnified views of two regions to indicate the differences between various curves more clearly (both panels from from Mao & Witt 1998)

In addition, Paczyński (1998) pointed out another mode of operation: he suggested to select *lenses* rather than *sources*, in order to get interesting results, namely very high proper-motion stars. These stars are relatively nearby, therefore their angular Einstein radii are relatively large, which means they have a (very) large cross section for astrometric microlensing. As an example, Barnard's star with a parallax of $\pi_\text{Barnard} = 0.522"$, a proper motion of $10.31"/\text{yr}$, and an assumed mass of $0.2\,M_\odot$ has an angular Einstein radius of

$$\phi_{E,Barnard} = 30 mas \frac{M_\text{Barnard}}{0.2 M_\odot}.$$

Since the astrometric lensing effect only falls off with $1/r$, a background star at an angular distance as large as 9 arcsec would still be displaced by 100 microarcseconds, a huge value for SIM. Another nice aspect of this suggestion by Paczyński: these events can be predicted ! There are more than 10,000 stars in

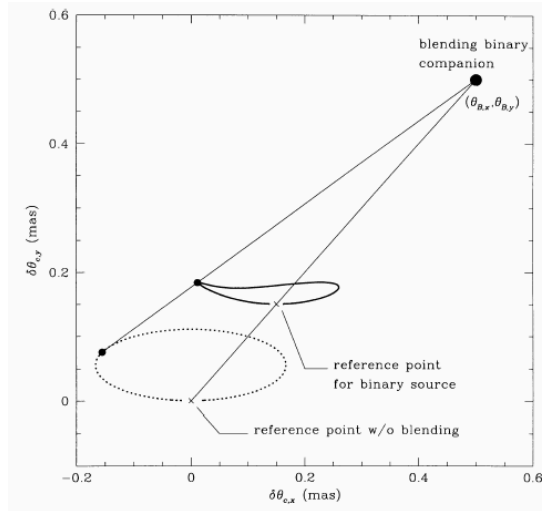

**Fig. 41.** Distortion of the astrometric trajectory of a binary companion: the dotted line is the astrometric shift without blending; with a binary companion a location $(\theta_{B,x}, \theta_{B,y})$, the light centroid will shift towards the companion. The solid line is the resulting trajectory. Parameters used here are time scale $t_E = 15d$, angular Einstein radius $\theta_E = 0.5$mas, contributing light fraction of the blending star is $f = 0.3$ (from Han & Kim 1999)

the Hipparcos catalog with distances under 100pc and proper motion higher than 100 mas/yr !

Mao & Witt (1998) treat finite source effects with respect to astrometric microlensing. They obtain analytically the centroid motion of a source with uniform surface brightness. They conclude that the finite source does affect the centroid shift significantly only when the angular impact parameter is comparable to the angular source size. In that case, the trajectories of the light centroid become clover-leaf like. This offers the exciting possibility to detect stellar radii to very good accuracy. In Fig. 40, four examples of astrometric trajectories with a finite source size of $\theta_s = 0.5\theta_E$ are shown, with impact parameters $u_0 = 3.0, 0.8, 0.5, 0.1$. For comparison, the corresponding point source effect is shown as a dashed line as well. The second panel in this figure is a simulation of how the first microlensing event that showed extended source effects photometrically (MACHO 95-BLG-30) would have looked like astrometrically.

Han & Kim (1999) looked into another degeneracy: can astrometric microlensing help in uncovering blended microlensing events. They found that indeed, due to the high resolution of SIM, many blends will be directly identified: the imaging resolution of SIM is supposed to be 10 milli-arcseconds. But even for very close blends with a separation smaller than that (e.g. physical

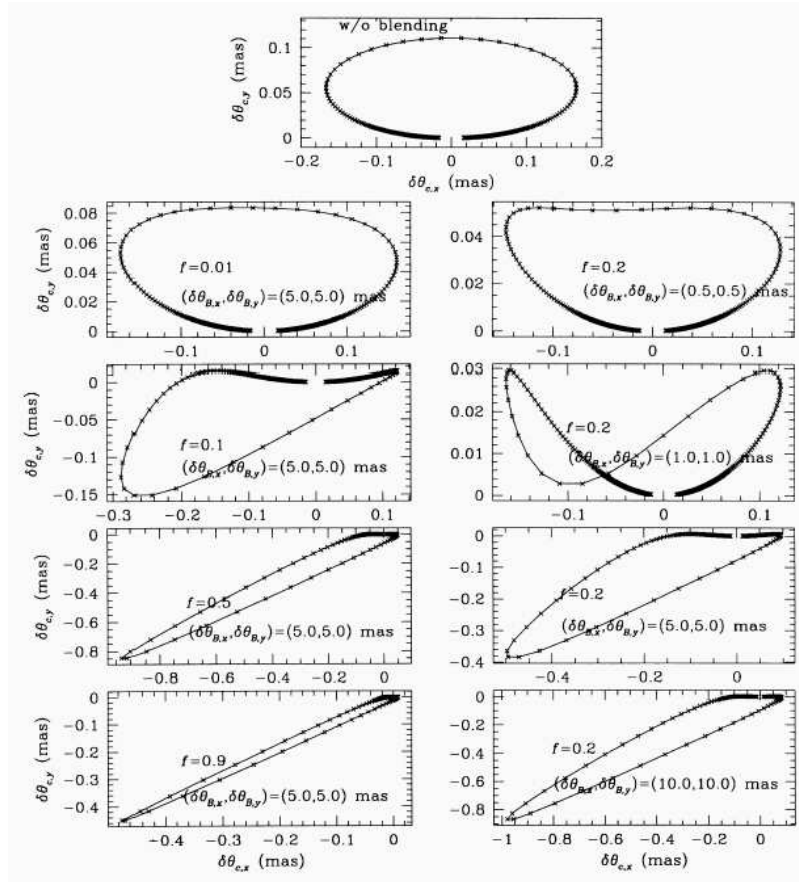

**Fig. 42.** Various forms of centroid shifts by binary-star blending: top panel: no blend contribution; left column: increasing blend contribution from top to bottom; right column: increasing binary separation from top to bottom (from Han & Kim 1999)

binary lenses), the blend signature can be identified in the blend contribution to the astrometric signature. In Fig. 41, the effect of the blended astrometric microlensing is illustrated qualitatively, in Fig. 42, various simulations are shown for increasing blend contribution (left hand column) and increasing binary separation (right hand column).

# 7 Quasar Microlensing

Quasars are affected by gravitational lensing in two ways: the "macrolensing" describes multiply imaged quasars, with angular separations of roughly one

527

arcsecond. These cases are produced by typical galaxy lenses with masses of order $10^{12}$ $M_\odot$. About one out of 500 quasars is multiply imaged. Some 80 such cases are known to date (cf. http://cfa-www.harvard.edu/glensdata, the CASTLES web page), most of them consist of double or quadruple images. Once time delays between the images are determined and the mass distributions of the lenses are modelled properly, these quasar lenses can be used to measure the Hubble constant. Or one can turn this line of reasoning around: assuming one knows the Hubble constant, one can infer the mass profiles of the lensing galaxies (cf. Kochanek et al. 2003).

The second interesting regime is "microlensing": stellar mass lenses affect the apparent brightness of the quasar images. Microlens-induced variability can be used to study two cosmological issues of great interest, the size and brightness profile of quasars on the one hand, and the distribution of compact (dark) matter along the line-of-sight on the other hand. Here a summary of recent observational evidence for quasar microlensing is given, as well as a review of theoretical progress in the field. Particular emphasis is given to the questions which microlensing can address regarding the search for dark matter, both in the halos of lensing galaxies and in a cosmologically distributed form. A discussion of desired observations and required theoretical studies is presented at the end.

## 7.1 Microlensing mass, length and time scales

The lensing effects on quasars by compact objects in the mass range $10^{-6} \leq m/M_\odot \leq 10^3$ are usually called "quasar microlensing". The microlenses can be ordinary stars, brown dwarfs, planets, black holes, molecular clouds, globular clusters or other compact mass concentrations (as long as their physical size is smaller than the Einstein radius). In most practical cases, the microlenses are part of a galaxy which acts as the main (macro-)lens. However, microlenses could also be located in, say, clusters of galaxies or they could even be imagined "free floating" and filling intergalactic space.

The relevant length scale for microlensing is the Einstein radius of the lens in the quasar plane:
$$r_E \approx 4 \times 10^{16} \sqrt{M/M_\odot} \text{ cm},$$
where "typical" lens and source redshifts of $z_L \approx 0.5$ and $z_S \approx 2.0$ are assumed for the numerical value on the right hand side. Quasar microlensing turns out to be an interesting phenomenon, because (at least) the size of the continuum emitting region of quasars is comparable to or smaller than the Einstein radius of stellar mass objects.

The length scale translates into an angular Einstein scale of
$$\theta_E \approx 10^{-6} \sqrt{M/M_\odot} \text{ arcsec}.$$

It is obvious that image splittings on such angular scales cannot be observed directly. What makes microlensing observable anyway is the fact that observer, lens(es) and source move relative to each other. Due to this relative motion, the micro-image configuration changes with time, and so does the total magnification, i.e. the sum of the magnifications of all the micro-images. This change in magnification can be measured over time: microlensing is a "dynamical" phenomenon.

There are two time scales involved: the standard **lensing time scale** $t_E$ is the time it takes the source to cross the Einstein radius of the lens, i.e.

$$t_E = r_E/v_{\perp,\text{eff}} \approx 15\sqrt{M/M_\odot} \ \ v_{600}^{-1} \ \ \text{years},$$

where the same assumptions are made as above, and the effective relative transverse velocity $v_{\perp,\text{eff}}$ is parametrized in units of 600 km/sec: $v_{600}$. This time scale $t_E$ results in discouragingly large values. However, in practice we can expect fluctations on much shorter time intervals. The reason is that the sharp caustic lines separate regions of low and high magnification. Hence, if a source crosses such a caustic line, we can observe a large change in magnification during the **crossing time** $t_{cross}$ it takes the source to cross its own diameter $R_{source}$:

$$t_{cross} = R_{source}/v_{\perp,\text{eff}} \approx 4R_{15} \ \ v_{600}^{-1} \ \ \text{months}.$$

Here the quasar size $R_{15}$ is parametrized in units of $10^{15}$cm.

In microlensing of multiple quasars, the normalized surface mass density is of order unity: $\kappa \approx 1$. This means that at any given time, a whole ensemble of microlenses is affecting the quasar. An illustration is shown in Fig. 43: in the top panel, an example source profile is superimposed on the magnification pattern of a randomly placed ensemble of microlenses for a surface mass density $\kappa = 0.5$. The bottom panel shows the corresponding micro-image configuration, which displays all effects that gravitational lensing can produce: offset of position, distortion, magnification and multiple images !

### 7.2 Early and recent theoretical work on quasar microlensing

Right after the discovery of the first multiply imaged quasar, Chang & Refsdal (1979) suggested that the flux of the two quasar images can be affected by stars close to the line-of-sight. Gott (1981) proposed that a massive galaxy halo could be made of low mass stars and "should produce fluctuations of order unity in the intensities of the QSO images on time scales of 1-14 years." Young (1981) was the first to use numerical simulations in order to explore the effect of quasar microlensing.

Because the optical depth (or surface mass density) at the position of an image is of order unity, microlensing is expected to be going on basically "all the time", due to the relative motion of source, lens(es) and observer. In addition, this means that the lens action is due to a coherent effect of many microlenses, because the action of two or more point lenses whose projected positions is of order of their Einstein radii combines in a very non-linear way

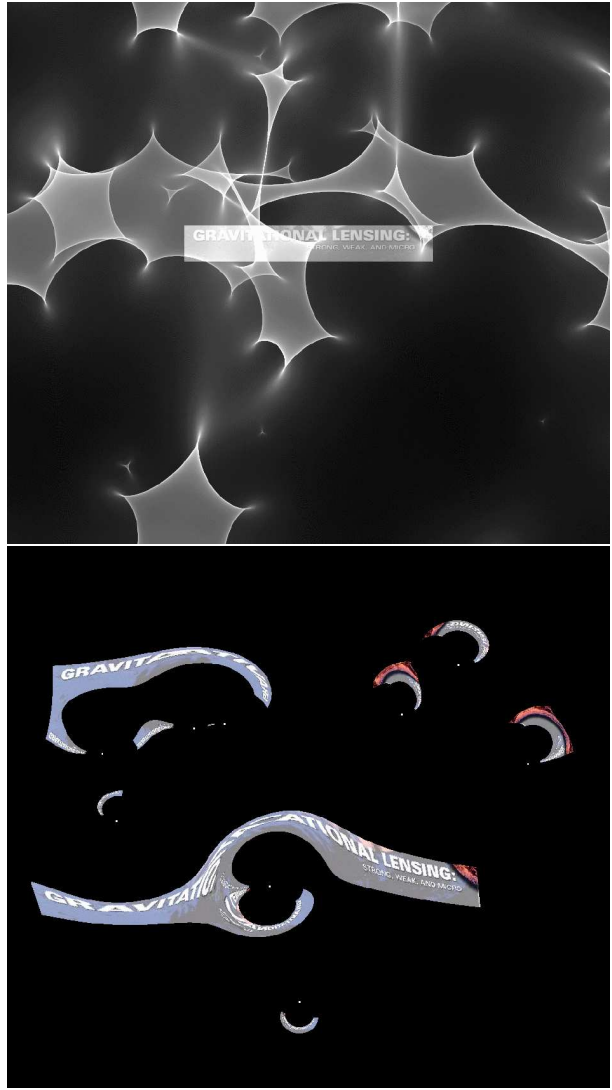

**Fig. 43.** Microlensing effect on an extended source: top: superposition of example source profile and microlensing magnication pattern produced by an ensemble of stellar lenses; bottom: corresponding image configuration

(cf. Wambsganss 1998). An illustration of this coherent action can be found in Figs. 44 and 45:

The magnification distribution produced by an ensemble of lenses is indicated in the quasar plane by different colors. The three dashed lines show the tracks of a quasar. In Fig. 45 the corresponding lightcurves are displayed,

530

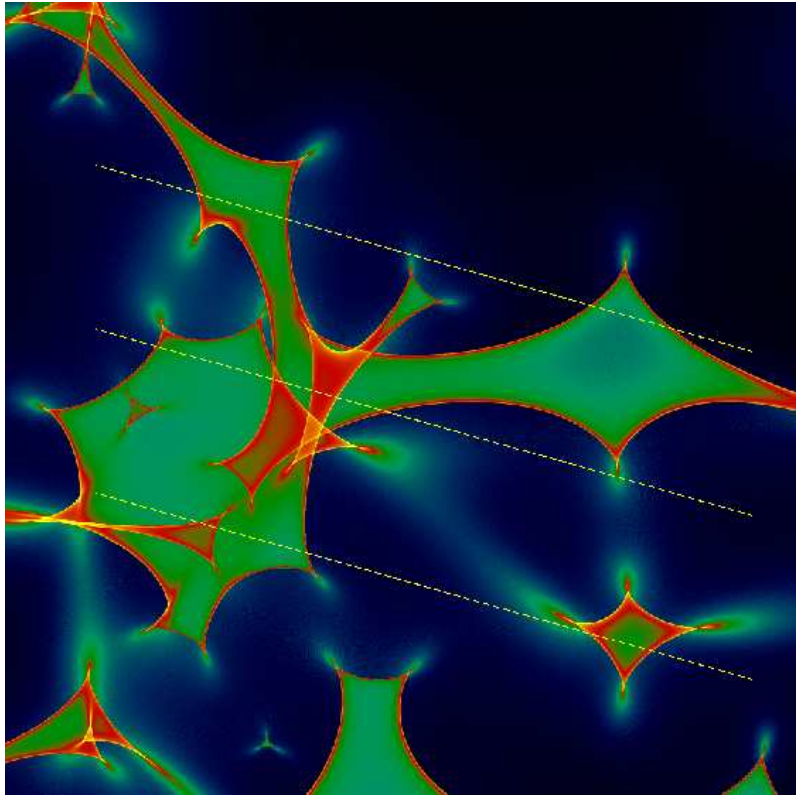

**Fig. 44.** Microlensing magnification pattern produced by stars in a lensing galaxy. The color steps represent different magnifications, with the sharp caustic lines corresponding to the highest magnification. The dashed lines indicates three tracks along which a background quasar moves. The corresponding lightcurves are displayed in Fig. 45

for two different source sizes. If the size of the quasar is small compared to the inter-caustic spacing, each caustic crossing is resolved individually, which results in relatively high maxima in the lightcurves (solid line). For a larger source (dashed line, factor 10 larger than solid line), the peaks are smoothed out, the character of the lightcurve is different.

The lens action of more than two point lenses cannot be easily treated analytically any more. Hence numerical techniques were developed in order to simulate the gravitational lens effect of many compact objects. Paczyński (1986a) had used a method to look for the extrema in the time delay surface. Kayser, Refsdal & Stabell (1986), Schneider & Weiss (1987) and Wambsganss (1990) had developed and applied an inverse ray-shooting technique that produced a two-dimensional magnification distribution in the source plane. An alternative technique was developed by Witt (1993) and Lewis et al. (1993); they solved the lens equation along a linear source track. All the recent theoretical work on microlensing is based on either of these techniques.

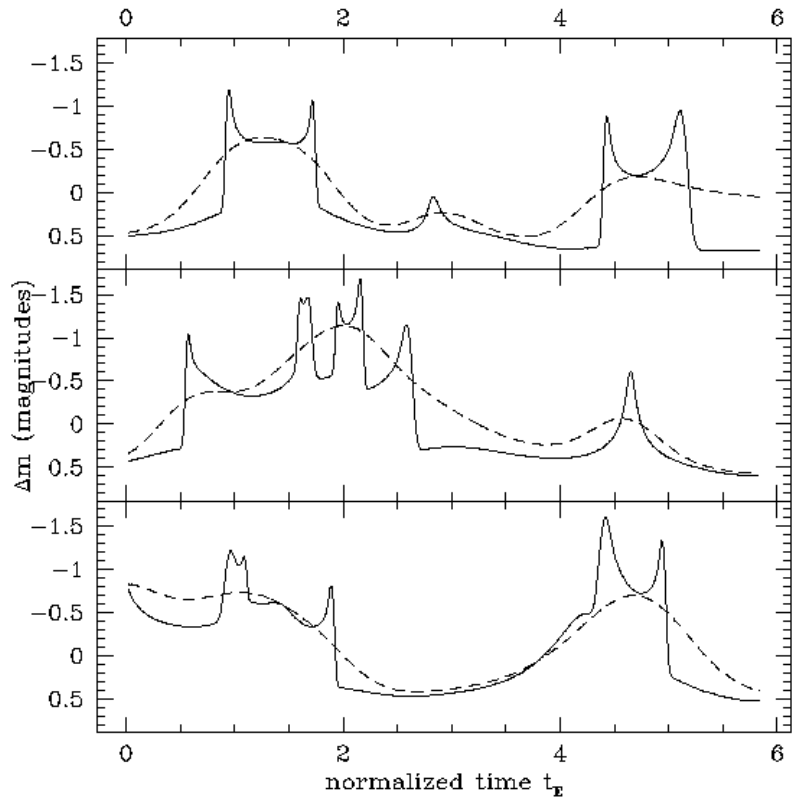

**Fig. 45.** Microlensing lightcurves for the three tracks shown in Fig. 44. The solid line correspond to a small source (Gaussian shape with width of about 3% of the Einstein radius), the dashed line represents a source that is a factor of 10 larger

More recently, Fluke & Webster (1999) explored analytically caustic crossing events for a quasar. Lewis et al. (1998) showed that spectroscopic monitoring of multiple quasars can be used to probe the broad line regions (cf. also Lewis & Belle 1998). Wyithe et al. (2000a, 2000b) investigated and found limits on the quasar size and on the mass function in Q2237+0305.

Agol & Krolik (1999) and Mineshige & Yonehara (1999) developed techniques to recover the one-dimensional brightness profile of a quasar, based on the earlier work by Grieger et al. (1988, 1991): Agol & Krolik showed that frequent monitoring of a caustic crossing event in many wave bands (they used of order 40 simulated data points in eleven filters over the whole electromag-netic range), one can recover a map of the frequency-dependent brightness distribution of a quasar. Mineshige & Yonehara (1999) in a similar approach explored the effect of microlensing on two different accretion disk models. In another paper, Yonehara et al. (1998) showed that monitoring a microlensing event in the X-ray regime can reveal structure of the quasar accretion disk as small as AU-size.

Summarized, the theoretical papers exploring microlensing made basically four predictions concerning the potential scientific results. Microlensing should help us to determine:

1. the existence and effects of compact objects between the observer and the source,
2. the size of quasars,
3. the two-dimensional brightness profile of quasars,
4. the mass (and mass distribution) of lensing objects.

In the following sub-section the observational results to date will be discussed in some detail.

### 7.3 Observational Evidence for Quasar Microlensing

Fluctuations in the brightness of a quasar can have two causes: they can be intrinsic to the quasar, or they can be microlens-induced. For a single quasar (i.e., one that is not multiply imaged), the difference is hard to tell. However, once there are two or more gravitationally lensed (macro-)images of a quasar, we have a relatively good handle to distinguish the two possible causes of variability: any fluctuations caused by intrinsic variability of the quasar show up in all the quasar images, after a certain time delay[13]. So once a time delay is measured in a multiply-imaged quasar system, one can shift the lightcurves of the different quasar images relative to each other by the time delay, correct for the different (macro-)magnification, and subtract them from each other. All remaining incoherent fluctuations in the "difference lightcurve" can be attributed to microlensing. In a few quadruple lens systems we can detect microlensing even without measuring the time delay: in some cases the image arrangement is so symmetrical around the lens that any possible lens model predicts very short time delays (of order days or shorter), so that fluctuations in individual images that last longer than a day or so and are not followed by corresponding fluctuations in the other images, can be safely attributed to microlensing. This is in fact the case in the quadruple system Q2237+0305.

#### The Einstein Cross: Quadruple Quasar Q2237+0305

In 1989, evidence for cosmological microlensing was found by Irwin et al. (1989) in the quadruple quasar Q2237+0305: one of the components showed fluctuations, whereas the others stayed constant. In the mean time, Q2237+0305 has been monitored by many groups (Corrigan et al. 1991; Ostensen et al. 1996; Lewis et al. 1998). The most recent (and most exciting) results (Fig. 46, and Woźniak et al. 2000a,b) show that all four images vary dramatically, going up and down like a rollercoaster in the last three years: $\Delta m_A \approx 0.6$ mag, $\Delta m_B \approx 0.4$ mag, $\Delta m_C \approx 1.3$ mag, $\Delta m_D \approx 0.6$ mag. Comparison of these lightcurves with simulations (cf. Figs. 44 and 45) show that the continuum emitting region of the quasar is relatively small, of order $10^{14}$cm (see, e.g., Wambsganss, Paczyński & Schneider 1990; Wyithe 2000b, Yonehara 2001).

---

[13] This argument can even be turned around: the measured time delays in multiple quasars are the ultimate proof of the intrinsic variability of quasars.

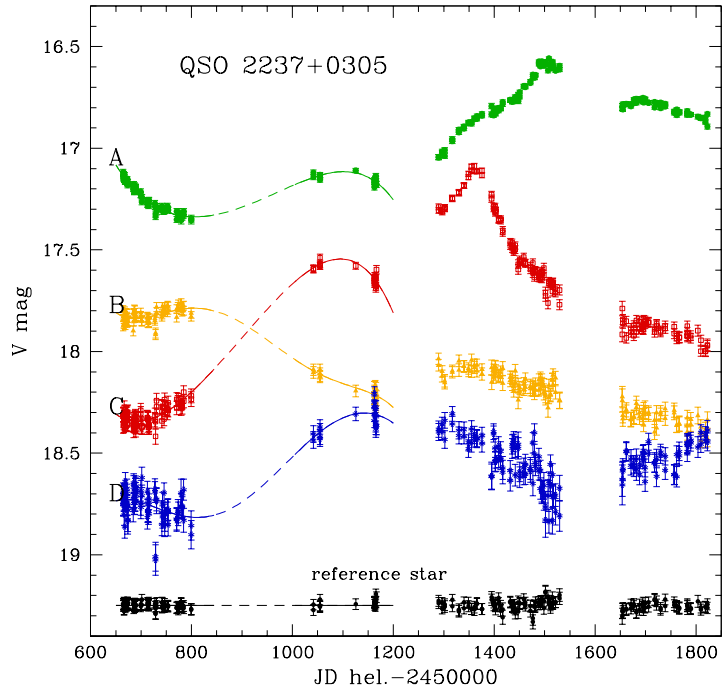

**Fig. 46.** Microlensing lightcurve of the quadruple quasar Q2237+0305, as measured by the OGLE team (Woźniak et al. 2000a,b; see also http://bulge.princeton.edu/∼ogle/ogle2/huchra.html)

### The Double Quasar Q0957+561

The microlensing results for the double quasar Q0957+561 are not quite as exciting. Vanderriest et al. (1989) were the first to put attention on the obser-vational evidence for potential microlensing in early lightcurve of the double quasar. In the first few years after its discovery, there is an almost linear change in the (time-shifted) brightness ratio between the two images (Schild 1996): $\Delta m_{AB} \approx 0.25$ mag over 5 years. But since about 1991, this ratio stayed more or less "constant" within about 0.05 mag, so not much microlensing has been going on in this system recently (Schild 1996; Pelt et al. 1998; Schmidt & Wambsganss 1998).

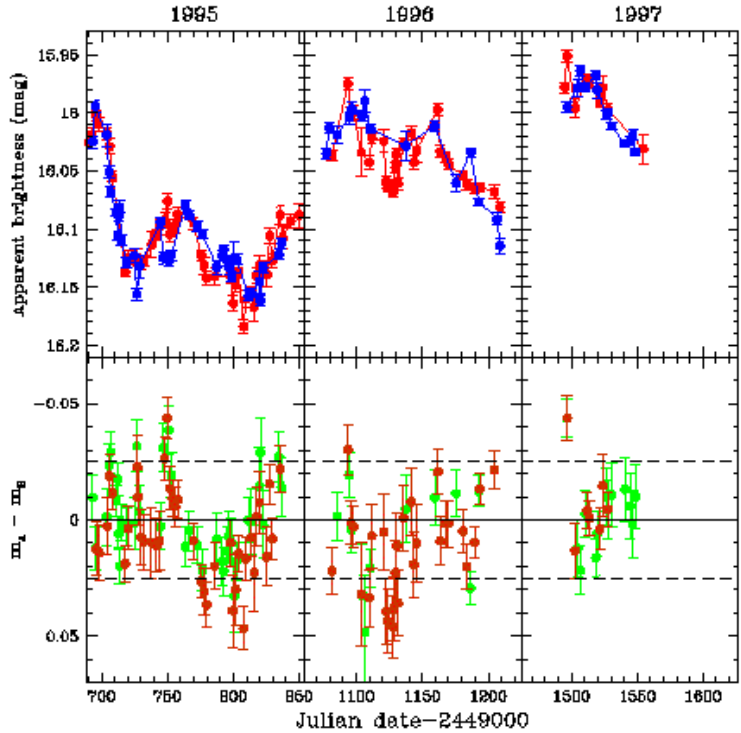

**Fig. 47.** Observed lightcurves of the double quasar Q0957+561; top: superposition of lightcurves of image A and (time shifted and magnitude shifted) image B; bottom: difference lightcurves (Wambsganss et al. 2000)

With numerical simulations and limits obtained from three years of Apache Point monitoring data of Q0957+561 (see Fig. 47), Wambsganss et al. (2000) exclude a whole range of "MACHO" masses as possible dark matter candidates in the halo of the lensing galaxy in 0957+561. They extracted simulated lightcurves according to the timing of the observed ones and evaluated 100000 cases for seven different values for the lens mass (from $m/M_\odot = 10^{-7}$ to $10^0$) and four different quasar sizes ($10^{14}$cm to $3 \times 10^{15}$cm): The small "difference"

535

between the time-shifted and magnitude-corrected lightcurves of images A and B ($|\Delta m_{A-B,Q0957}| \leq 0.05$ mag) excludes a halo of the lensing galaxy made of compact objects with masses of $10^{-7} M_\odot - 10^{-2} M_\odot$ (cf. Figs. 48 and 49).

Refsdal et al. (2000) investigated the microlensing properties of the double quasar as well, using both the original linear change of 0.25 mag over a five year period and the subsequent 8 years of no or very little microlensing. They found constraints on the source size of $R \leq 6 \times 10^{15}$ cm, and the mass of the microlensing objects most likely to be in the range $10^{-6} \leq M/M_\odot \leq 5$.

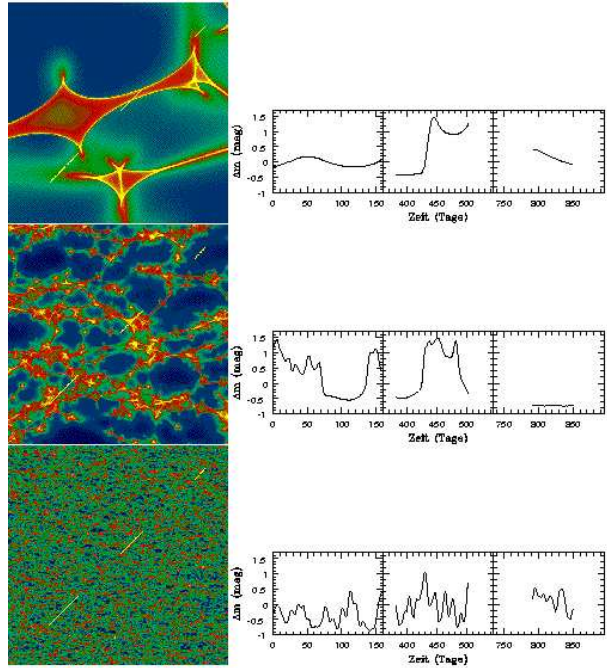

**Fig. 48.** Simulated microlensing lightcurves for the double quasar Q0957+561; Left: Magnification patterns for compact objects in three different mass ranges; the three-part straight line indicates the track of the background quasar. Right: corresponding three-part microlensing lightcurves (Wambsganss et al. 2000)

Recently, the double quasar Q0957+561 was the target of a monitoring campaign particularly searching for short time scale variations. Colley et al. (2003a,b) report the observations and the result: making use of their very precise determination of the time delay in this system: $\Delta t = (417.09 \pm 0.07)$ days, they found no microlensing fluctuations with amplitudes higher than 0.1 mag, and nearly rule out that objects in the mass range about $10^{-5} M_\odot$ make up a large fraction of the dark matter in the lens galaxy. In a further

536

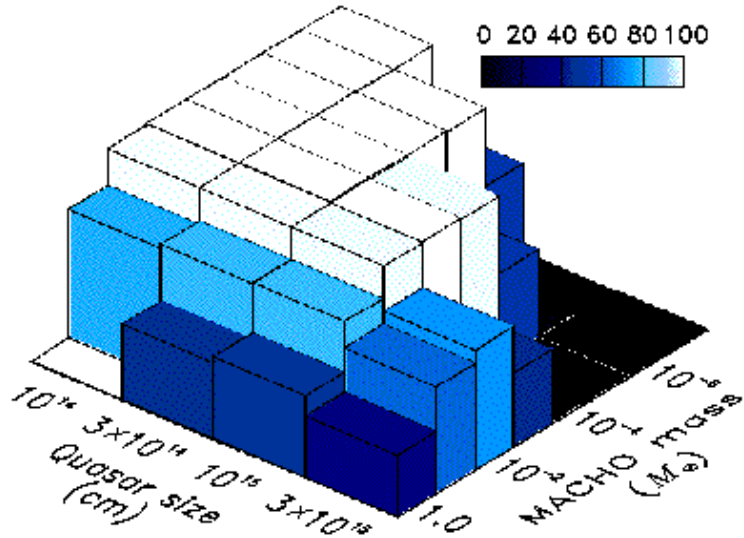

**Fig. 49.** Exclusion diagram: the highest (white) columns indicate values of quasar source size and "macho" mass which are excluded by more than 99.9% probability; the other columns show exlusion probabilities of between 40% and 85% (Wambsganss et al. 2000)

analysis of this data set, Colley & Schild (2003) report a microlensing signal at the 1% level with a time scale of 12 hours. If this result can be confirmed, it does provide a very interesting new window of exploration (and a challenge for theory).

**Other multiple quasars/radio microlensing ?**

A number of other multiple quasar systems are being monitored more or less regularly. For some of them microlensing has been suggested (e.g. H1413+117, Ostensen et al. 1997; or B0218+357, Jackson et al. 2000). In particular the possiblity for "radio"-microlensing appears very interesting (B1600+434, Koopmans & de Bruyn 2000), because this is unexpected, due to the presumably larger source size of the radio emission region. The possibility of relativistic motion of radio jets may make up for this "disadvantage".

**Unconventional Microlensing I: in individual quasars ?**

There were a number of papers interpreting the variability of individual quasars as microlensing (e.g., Hawkins & Taylor 1997, Hawkins 1998). Although this is an exciting possibility and it could help us detect a population of cosmologically distributed lenses, it is not entirely clear at this point whether the observed fluctuations can be really attributed to microlensing. After all, quasars are intrinsically variable, and the expected microlensing in single quasars must be smaller than in multiply imaged ones, due to the lower surface mass density. More studies are necessary to clarify this issue.

### Unconventional Microlensing II: Centroid shifts/Astrometric Microlensing

As in stellar microlensing (cf. Section 6), in quasar microlensing the astrometric signal of the lenses can be used and investigated as well. This was first put forward by Lewis & Ibata (1998), then further investigated by Treyer & Wambsganss (2004). At each caustic crossing, a new very bright image pair emerges or disappears, giving rise to sudden changes in the "center of light" positions (cf. Fig. 50).

The amplitude could be of order 100 micro-arcseconds or larger, which should be observable with the SIM satellite (Space Interferometry Mission), to be launched in 2010. This astrometric microlensing offers the exciting possibility to measure the mass of the lenses (in a statistical way) !

### Unconventional Microlensing III: million solar mass objects or sub-structure: milli-lensing

A decade ago, the idea was popular that dark halos of galaxies could be made of black holes in the mass range of about a million solar masses. Wambsganss & Paczyński (1992) explored this effect on VLBI jets of multiply imaged quasars and suggested that this hypothesis could be tested: High signal-to-noise imaging of the two jet images of Q0957+561 should indicated clear lensing signatures, like kinks, holes, additional milli-images, if a significant fraction of the dark matter in the halo is made of such million solar mass objects (see Fig. 51). Garrett et al. (1994) presented such results and ruled out that the halo of the lensing galaxy in this double quasar consist of such objects.

### Flux ratio anomaly: microlensing or substructure ?

Microlensing may help solve another interesting issue: In a macrolensing scenario producing a quadruple quasar configuration with one close pair – corresponding to the source sitting inside but close to the (macro-)caustic – this image pair should be highly magnified with very similar magnification of the two components. In most of the observed cases, however, this is not the case: close image pairs tend to have quite different magnifications. In almost all cases, the fainter (or demagnified) image seems to be the saddle point image. There are two competing explanations: Substructure in the macro-lens (galaxy) could introduce this flux ratio anomaly (Dalal & Kochanek 2002;

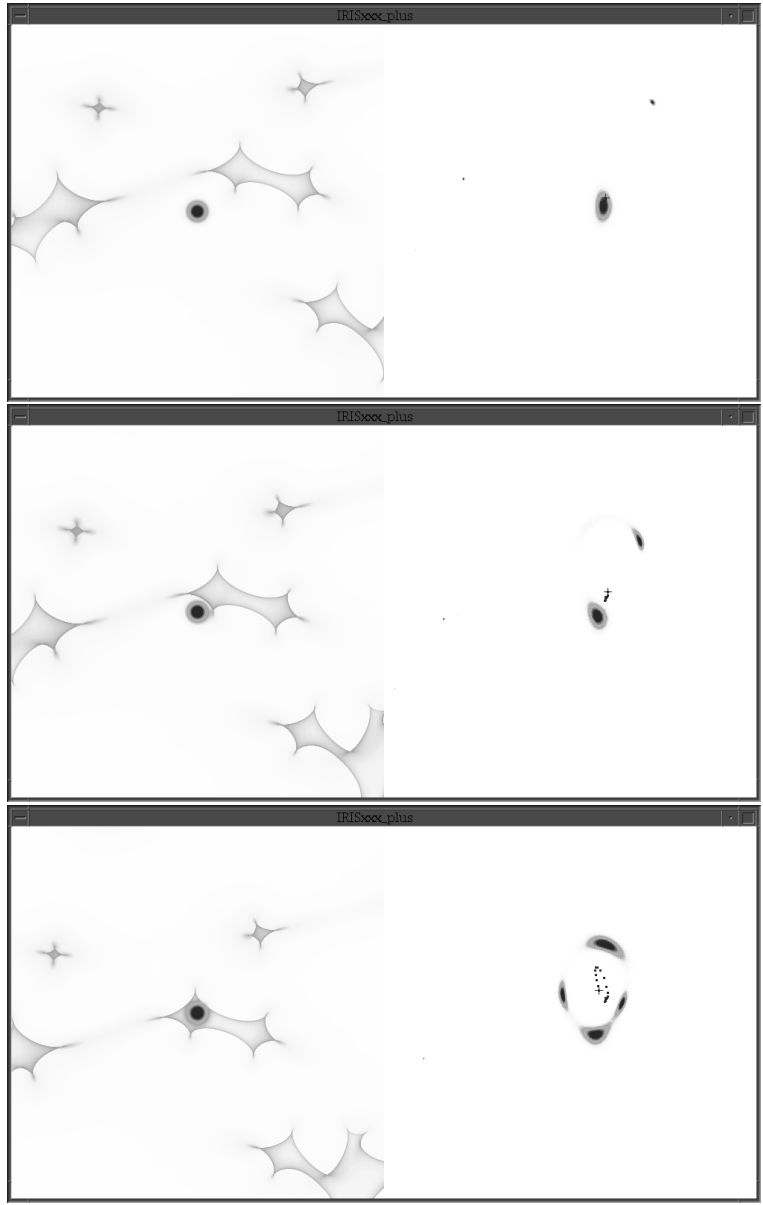

**Fig. 50.** Astrometric microlensing of quasars: due to the relative motion of quasar and foreground stellar lenses, the microimage configuration changes with time. This leads to a change of the light centroid which may be observable. The three panels show three epochs. On the left hand side, the caustics are shown superimposed with the quasar profile. On the right hand side, the micro-image configuration is shown. The plus sign indicates the "center-of-light"; the points next to the plus sign mark the light-centroid of previous epochs, i.e. the motion of the center-of-light for fixed quasar and moving microlenses (after Treyer & Wambsganss 2004)

539

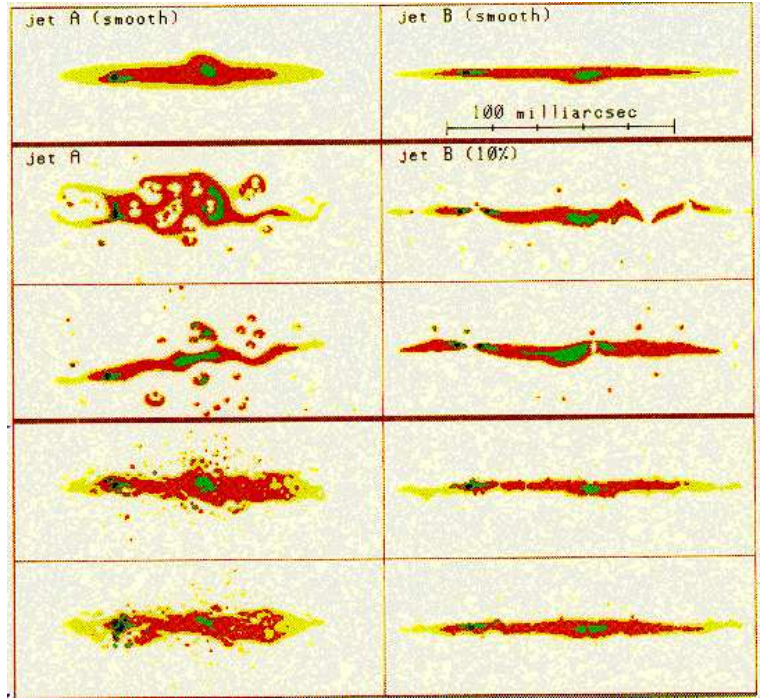

**Fig. 51.** Millilensing by million solar mass black holes affects the VLBI jets of multiply imaged quasars: the top row shows model VLBI jets for images A and B of the double quasar Q0957+561, as produced by the smooth lensing potential. If the halo of the lensing galaxy is made of million solar mass black holes (here: $m/M_\odot = 3 \times 10^5$), then the two jets should be affected by them *differently*, as shown in the four examples below: kinks, holes, additional milli-images should appear in both images uncoherently (Wambsganss & Paczyński 1991)

Metcalf & Madau 2001; Metcalf & Zhao 2002). However, another possibility is microlensing by compact stellar mass objects plus smoothly distributed (dark) matter (Schechter & Wambsganss 2002). A nice thing about the two proposed mechanisms is that they make different predictions: If substructure is the cause for the flux ratio anomaly, then it should act the same way in basically all wave bands, and the flux ratio should be constant in time. If microlensing plus smooth matter is the origin of the discrepancy, then we expect different behaviour in different wavebands, due to the fact, that source size changes as seen in different energy bands. In general, small source sizes (shorter wavelengths) should be affected more drastically then larger sources (which smooth out the microlensing caustics). A second consequence of this microlensing explanation is that the flux ratio should change with time, because the relative positions of the microlenses change over the course of a few years and hence produce fluctuations in the magnification. This issue of microlensing and flux ratio anomalies is discussed in detail in Schechter & Wambsganss (2002).

**7.4 Quasar Microlensing: Now and Forever ?**

Monitoring observations of various multiple quasar systems in the last decade have clearly established that the phenomenon of quasar microlensing exists. There are uncorrelated variations in multiple quasar systems with amplitudes of more than a magnitude and time scales of weeks to months to years. However, in order to get close to a quantitative understanding, much better monitoring programs need to be performed. Summarized, and considering the "early promises" of quasar microlensing, the following can be stated:

1. the existence and effects of compact objects between the observer and the source: has been achieved;
2. the size of quasars: partly fulfilled, some limits on the size of quasars have been obtained;
3. the two-dimensional brightness profile of quasars: we are still (far) away from solving this promise;
4. the mass (and mass distribution) of lensing objects: it is fair to say that the observational results are consistent with certain (conservative) mass ranges.

Looking at the issue today, there are two important questions on the theoretical side: what do the lightcurves tell us about the lensing objects, and what can we learn about the size and structure of the quasar. As a response to the first question, the numerical simulations are able to give a qualitative understanding of the measured lightcurves (detections of microlensing in lightcurves of some multiply imaged quasars, and non-detections in others). The amplitudes and times scales of the events are in general consistent with "conservative" assumptions about the object masses and velocities. But due to the large number of parameters/unknowns (quasar size, masses of lensing objects, transverse velocity) and due to the large variety of lightcurve shapes, no satisfactory quantitative explanation or even prediction could be achieved. So far mostly "limits" on certain parameters have been obtained. The prospects of getting much better lightcurves of multiple quasars, as shown by the OGLE collaboration, should be motivating enough to explore this direction in much more quantitative detail.

The question of the structure of quasars deserves more attention. Here gravitational lensing is in the unique situation to be able to explore an astrophysical field that is unattainable by any other means. Hence more effort should be put into attacking this problem. This involves much more ambitious observing programs, with the goal to monitor caustic crossing events in many filters over the whole electromagnetic spectrum, and to further develop numerical techniques to obtain useful values for the quasar size and profile from unevenly sampled data in (not enough) different filters. Theoretically, Kochanek (2004) attacked this question in a brute force way: simulating microlensing for a large set of parameters and comparing with observed lightcurves for constraints on the input values. So far only a restricted data set of Q2237+0305 was used. This method should clearly be applied to more microlensing lightcurves. On the observational side, the way to go is building one (or more) dedicated telescope(s) for quasar monitoring. Moderate size is sufficient: 1m to 2m class. Excellent site is essential: median seeing better than one arcsecond. The use of robotic telescopes in a time-sharing mode is a much desired first step in this direction.

ACKNOWLEDGEMENTS. It is very great pleasure to thank a large number of people who contributed to the final version of this manuscript. First of all, I have to thank Penny Sackett, who unfortunately could not lecture at the 33rd Saas Fee Advanced Course on Gravitational Lensing, due to the fire at Mt. Stromlo. I have to thank her for providing me with text and figures and other support which were very useful for giving the lectures and for writing them up. I want to thank the organizers of the 33rd Saas Fee lectures, Georges Meylan, Philippe Jetzer and Pierre North, who did a marvellous job in planning, organizing and running this very nice and pleasant course ! It was a lot of fun to interact with my fellow lecturers Peter Schneider and Chris Kochanek, as well as with all the student and postdoc participants of the course. It is a pleasure to thank a number of colleagues who provided me with material for presentations and/or discussed these matters with me, including David Bennett, Ian Bond, Scott Gaudi, Andy Gould, Cheonho Han, Bohdan Paczyński, Nicholas Rattenbury, Andrzej Udalski, Phil Yock, and the PLANET team members. Finally, I want to thank the members of the gravitational lens group in Potsdam: Dijana Dominis, Pascal Hedelt, Janine Heinmüller, Andreas Helms, Susanne Hoffmann, Daniel Kubas, Robert Schmidt, Olaf Wucknitz for many discussions and in particular for reading the manuscript in various stages and providing valuable comments for corrections and improvements. Last but not least I want to thank our secretary in Potsdam, Andrea Brockhaus, who has always provided a very pleasant work atmosphere.